\documentclass[11pt]{article}

\usepackage{setspace}       
\usepackage{titlesec}       
\usepackage{lmodern}        
\usepackage{titling}        

\usepackage[utf8]{inputenc}
\usepackage{stmaryrd}
\usepackage{amsmath,amsthm,amsfonts,amssymb,amscd}
\usepackage{hyperref}
\usepackage{multirow,booktabs}
\usepackage[table]{xcolor}
\usepackage{fullpage}
\usepackage{dsfont}
\usepackage{multicol}
\usepackage{lastpage}
\usepackage{enumitem}
\usepackage{fancyhdr}
\usepackage{mathrsfs}
\usepackage{wrapfig}
\usepackage{setspace}
\usepackage{calc}
\usepackage{multicol}
\usepackage{cancel}
\usepackage[retainorgcmds]{IEEEtrantools}
\usepackage[margin=3cm]{geometry}

\usepackage{empheq}
\usepackage{framed}
\usepackage[most]{tcolorbox}
\usepackage{xcolor}
\colorlet{shadecolor}{orange!15}
\theoremstyle{definition}

\newtheorem{example}{Example}[section]
\newcommand\eref[1]{Eq.~\ref{#1}}
\newcommand\fref[1]{Fig.~\ref{#1}}
\newcommand\exref[1]{Example.~\ref{#1}}
\newcommand\bra[1]{\left\langle #1 \right|}
\newcommand\ket[1]{\left| #1 \right\rangle}
\newcommand\average[1]{\left\langle #1 \right\rangle}
\newcommand\braket[2]{\left\langle #1\middle| #2 \right\rangle}
\newcommand\tr{\mathrm{Tr}\,}
\DeclareMathOperator{\Tr}{Tr}
\newcommand\dd{\mathrm{d}}
\newcommand\ee{\mathrm{e}}
\newcommand\ii{\mathrm{i}}
\newcommand\diag[1]{\mathrm{diag}\left(#1\right)}
\newcommand\ketbra[2]{\left| #1\right\rangle\left\langle #2 \right|}

\newcommand{\dbraket}[2]{\ensuremath{\langle #1 , #2 \rangle}}
\newtheorem{proposition}{Proposition}

\usepackage{chngcntr}
\counterwithin{figure}{section}
\counterwithin{equation}{section}

\newtcolorbox{referencesbox}{
  colback=orange!15,
  colframe=black,
  title=References,
  fonttitle=\bfseries,
  left=1em,
  right=1em,
  top=1em,
  bottom=1em,
  boxrule=0.8pt,
  arc=4pt
}

\begin{document}



\begin{titlepage}
    \centering
    \vspace*{0.5cm}
    
    \includegraphics[width=0.3\textwidth]{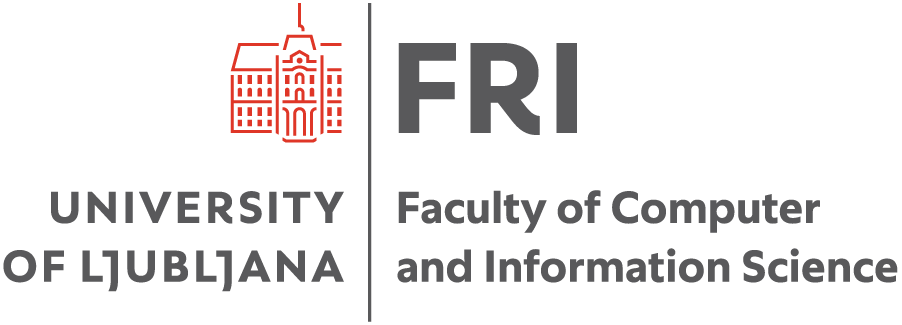} \\[4cm]
    
    {\Huge\bfseries Quantum machine learning \\[0.4cm]}
    
    
    {\Large by \\[0.3cm] \textbf{Bojan Žunkovič}}\\[8cm]
    
    {\large Faculty of computer and information science \\[0.2cm]
    University of Ljubljana \\[2cm]}
    
    {\large \today}
    
    \vfill
    
    
\end{titlepage}

\newpage
\tableofcontents
\newpage
\section{Approaches to QML}
\begin{figure}[!htb]
    \centering
    \includegraphics[width=0.4\textwidth]{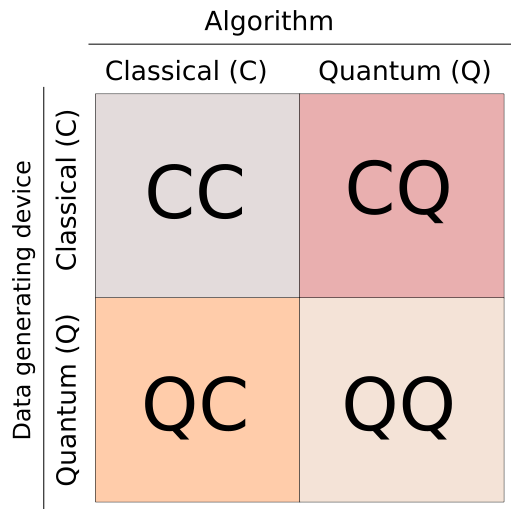}
    \caption{Four approaches to quantum computing based on the data source and the algorithm used to process the data.}
    \label{fig: qml areas}
\end{figure}

We distinguish four subareas of quantum machine learning and adopt the typology based on whether we obtain the data by a classical (C) or quantum device (Q) and if we process it with a classical (C) or quantum (Q) device/algorithm (shown in \fref{fig: qml areas}).

In the case of classical data and devices, we consider approaches that attempt to solve problems in one area using tools developed in another area. Typical examples include applying neural networks to describe complex quantum systems and utilising tensor networks for various machine learning tasks, such as classification, anomaly detection, data embedding, language modelling, and generative modelling. Another important research direction is also the development of quantum-inspired classical algorithms. A prominent example is the dequantization of quantum recommender systems.

In the case of classical data and quantum devices/algorithms, the primary goal is to replace the computationally intensive part of a classical algorithm with a more efficient quantum algorithm. The "hard"/"efficient" might concern computation, generalisation or robustness. The most prominent quantum "algorithm" is a variational quantum circuit, a quantum version of a neural network.

In the case of quantum data and classical algorithms, we typically use machine learning to assist quantum experiments. One possibility is to process the data from quantum measurements with standard machine learning tools. Another use of machine learning in quantum experiments is device calibration, a tedious, error-prone, and critical task. 

The last case of quantum data and a quantum algorithm is relatively unexplored due to the lack of quantum control and relatively short coherence times of current quantum devices. One possible application in this area could be quantum cryptography.

We will explore how quantum computers could help with classical machine learning tasks (the CQ case). We will also mention some quantum-inspired classical algorithms (the CC case).

\begin{referencesbox}
\begin{itemize}
  \item Schuld, M., \& Petruccione, F. (2021). Machine learning with quantum computers (Vol. 676, pp. 163-169). Berlin: Springer. (Chapter 1: Introduction)
\end{itemize}
\end{referencesbox}

\section{Classical and quantum probability}
To understand what quantum computing might mean for machine learning, we must distinguish between and understand the differences between classical and quantum data, as well as classical and quantum processing devices. We will begin by contrasting classical probability with quantum probability. Our approach will be less rigorous and formal, and we intend to provide as much intuition about quantum probability as possible, drawing on our knowledge and experience with classical probability theory. We introduce the main concepts of classical probability and then extend them to the quantum case.
\subsection{Classical probability}
For our purposes, it will be sufficient to define the probability in terms of Kolmogorov axioms summarised as follows: The probability space $(\Omega, F, P)$ is a space with a sample space $\Omega$, event space $F$ and a suitably normalised measure $P(\Omega)=1$. If the sample space is finite, the probability $P$ is a finite set of positive real numbers summing to one. 

There are several interpretations of probability, such as frequencies, propensities, degrees of freedom, and plausibility. Here, we will take an operational approach and not discuss the issue of interpretation, as it does not affect the properties we will discuss in the following sections. The quantum probability inherits a similar interpretation problem, which is even more complicated due to the non-intuitive character of quantum probability.

We will use the following notation
\begin{align*}
    X -& \text{random variable},\\
    x_i -& \text{simple event or a value of the random variable},\\
    N -& \text{number of simple events/values of random variable }X,\\
    P(x_i)=p_i -& \text{probability for the outcome},\\
    p_i>0-& \text{positivity},\\
    |\vec{p}|_1=\sum_{i=1}^{N} p_i=1 -& \text{normalization},\\
\end{align*}
Since we will restrict ourselves to the finite-dimensional setting, we can represent the probability as a vector $\vec{p}=(p_1,p_2,\ldots, p_j)$, where $p_j$ represents probabilities for simple events. In this case, we can represent random variables as a vector where we assume each event is associated with a real number pertaining to the random variable. We represent simple events as diagonal matrices with only one element equal to 1. The representation of a compound event (union of simple events) is equal to the sum of the projectors of simple events. The projection of a probability vector with an event projector is an unnormalized probability vector. Its $L_1$ norm determines the probability for the event, and its normalised version determines the probability vector after we observe the event. We call the change of the probability vector once we observe an event as the collapse of probability. In an observation/experiment, one of the considered events has to occur. Consequently, the projectors of all considered events should sum to identity $\sum_j \Pi_{e_j}=\mathds{1_N}$. With $\mathds{1}_N$, we will denote an $N$ dimensional identity matrix. The last requirement implies that the event probabilities sum to one.

We describe more than one random variable by the joint probability. In the case of two random variables $X$ and $Y$ described by the probabilities $P_1$ and $P_2$ there is joint probability $P_12(X=x_i,Y=y_j)=p^{12}_{ij}$. If $X$ has $N$ events and $Y$ has $M$ events, the joint probability is a set of $NM$ positive numbers summing to one. In general, the joint probability $p^{12}_{ij}$ is not a product of probabilities $p^1_i$ and $p^2_j$. If we can write the joint probability of $K$ random variables as 
\begin{align}
    P^{12\ldots K}_{ij\ldots k} = p^1_ip^2_j\ldots p^K_k
\end{align}
we say that the random variables are independent. Otherwise, they are dependent. We obtain the marginal distribution of a random variable by summing over all remaining random variables
\begin{align}
    p^1_i = \sum_{j\ldots k}p^{12\ldots K}_{ij\ldots k}.
\end{align}
\begin{shaded}
\begin{example}
\label{ex: classical collapse}
Imagine we have two coins. The first (coin A) is unbiased, and the second (coin B) has a 3/4 probability of getting heads. We first throw coin A. If we observe a head, we throw coin A again; if we get tails, we throw coin B. We want to describe the probability vector during the game. Let us denote simple events as follows $e_0=(heads,heads), e_1=(heads,tails), e_2=(tails,heads), e_3=(tails,tails)$. For the described game the probability vector is given by $\vec{p}=(\frac{1}{4},\frac{1}{4},\frac{3}{8},\frac{1}{8})$ and the projectors for the simple events are
\begin{align*}
    \Pi_{e_0}=\mathrm{diag}(1,0,0,0),\\
    \Pi_{e_1}=\mathrm{diag}(0,1,0,0),\\
    \Pi_{e_2}=\mathrm{diag}(0,0,1,0),\\
    \Pi_{e_3}=\mathrm{diag}(0,0,0,1),\\
\end{align*}
the $\text{diag}$ denotes a diagonal matrix with provided elements on the diagonal. Now, imagine we throw the first coin. We formally write the compound events as $e_0 \cup e_1$ and $e_2 \cup e_3$. Their projectors sum to identity and are given by 
\begin{align*}
    \Pi_{e_0\cup e_1}=\mathrm{diag}(1,1,0,0),\\
    \Pi_{e_2\cup e_3}=\mathrm{diag}(0,0,1,1).\\
\end{align*}
The unnormalised projections of the probability vector with the event projectors are 
\begin{align*}
    \vec{p}_{\rm H}=\vec{p}\cdot\Pi_{e_0 \cup e_1} &= {\rm diag}(\frac{1}{4},\frac{1}{4},0,0),\quad p(e_0\cup e_1)=|\vec{p}_{\rm H}|_1=\frac{1}{2},\\ 
    \vec{p}_{\rm T}=\vec{p}\cdot\Pi_{e_2 \cup e_3} &= {\rm diag}(0,0,\frac{3}{8},\frac{1}{8}),\quad p(e_0\cup e_1)=|p_{\rm T}|_1=\frac{1}{2}.
\end{align*}
As expected, both events are equally likely. If we observe heads, the probability vector is updated to 
\begin{align*}
    \vec{p}\rightarrow \vec{p}_{\rm H}/|\vec{p}_{\rm H}|_1={\rm diag}(\frac{1}{2},\frac{1}{2},0,0),
\end{align*}
and if we observe tails, the probability vector becomes 
\begin{align*}
    \vec{p}\rightarrow \vec{p}_{\rm T}/|\vec{p}_{\rm T}|_1 ={\rm diag}(0,0,\frac{3}{4},\frac{1}{4}).
\end{align*}
We describe the marginal probabilities for the second throw to get heads or tails by the following marginal probability vectors
\begin{align*}
    \mbox{we observed heads} &\rightarrow \vec{p}^{\,2}_{\rm H}=(\frac{1}{2},\frac{1}{2}),\\
    \mbox{we observed tails} &\rightarrow \vec{p}^{\,2}_{\rm T}=(\frac{3}{4},\frac{1}{4}).
\end{align*}
The outcome of the first throw determines the marginal probability vector for the second throw. However, we need to know which outcome occurred when throwing either coin A or coin B. A similar situation will appear in quantum mechanics, only that the collapse of the probability will be independent of the knowledge of the first observation, which is a distinct feature of quantum probabilities.
\end{example}
\end{shaded}

\begin{figure}[!htb]
    \centering
    \includegraphics[width=0.6\textwidth]{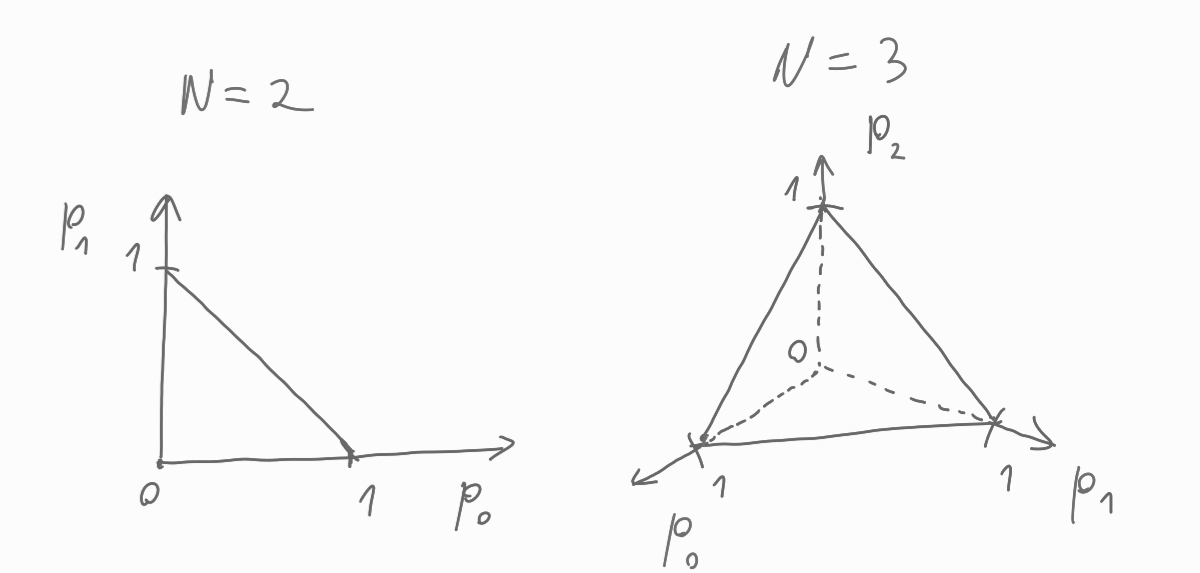}
    \caption{The space of all probabilities. In the case of $N=2$ (left), it is a line; in the case of $n=3$, it is a triangle.}
    \label{fig: probability simplex}
\end{figure}

\paragraph{Probability simplex}
In the following, we will not be strict and sometimes use probability to denote a probability vector, not its $L_1$ norm. Let us now consider the space of all possible probabilities with $N$ outcomes. We define pure probabilities (or pure probability vectors) as probabilities with only one specific simple event. The space of simple events is thus isomorphic to the space of pure probabilities. We can obtain any other probability as a convex combination of pure probabilities. The space of all possible probabilities with $N$ outcomes is a simplex. In \fref{fig: probability simplex}, we show the probability simplex for the cases $N=2$ and $N=3$.

We want to compare the probabilities and find simple functionals describing their properties. The most common concepts are majorisation, Shanon entropy, relative entropy, and Fisher-Rao metric. We shall introduce them in the following.

\begin{figure}[!htb]
    \centering
    \includegraphics[width=0.6\textwidth]{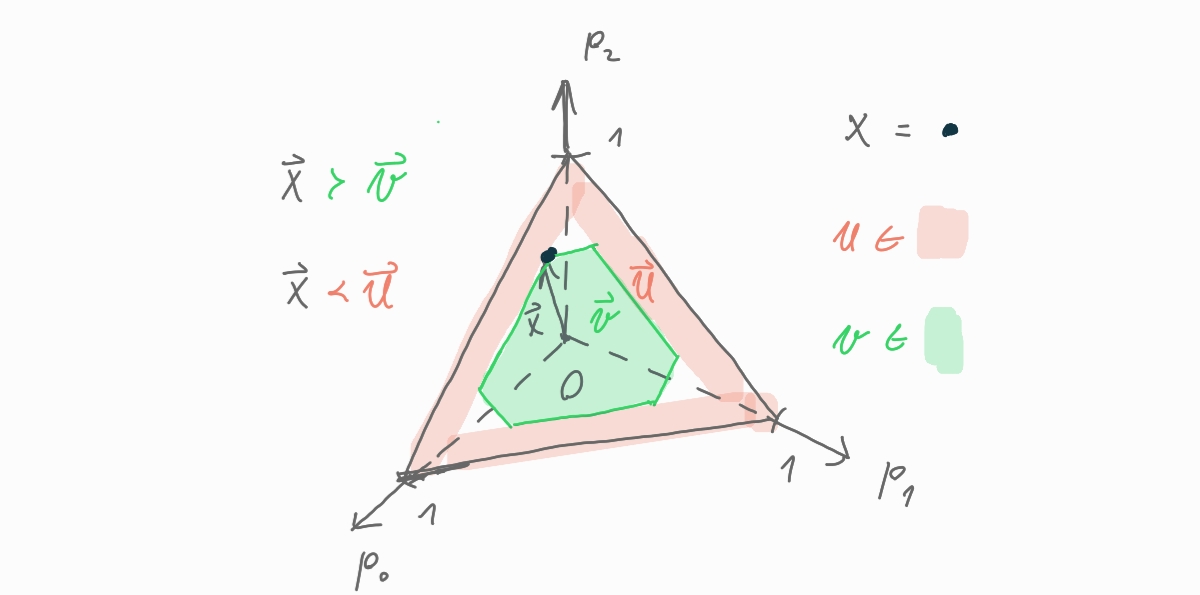}
    \caption{Majorisation relation in the $N=3$ probability simplex. The vectors in the red region majorise the vector $\vec{x}$, and the vectors in the green region are majorised by the vector $\vec{x}$.}
    \label{fig: majorisation}
\end{figure}

\paragraph{Majorisation}
The simplest concept with which we can compare two probability distributions is majorisation. Consider two vectors with $N$ real elements $\vec{x}$ and $\vec{y}$. The vector $\vec{x}$ majorises the vector $\vec{y}$ (written as $\vec{x}\succeq \vec{y}$) iff
\begin{align}
    \sum_{i=1}^k x^{\downarrow}_i\geq \sum_{i=1}^k y^{\downarrow}_i,\quad\mbox{ for all } \quad k=1,2,\ldots N.
\end{align}
The vectors $x^{\downarrow}$ and $x^{\downarrow}$ are such permutations of the original vectors that the entries appear in decreasing order (e.g. $x^\downarrow_1\geq x^\downarrow_1\geq \ldots \geq x^\downarrow_N$). Majorisation is transitive, i.e. $\vec{x}\succeq \vec{y}$ and $\vec{y}\succeq \vec{z}$ implies $\vec{x}\succeq \vec{z}$. All pure probabilities majorise all probabilities. Similarly, the uniform probability is majorised by all probabilities. Therefore, we can think of majorisation as determining which probability is more uniform. There can be probabilities that are unrelated by majorisation as shown in \fref{fig: majorisation} for the case $N=3$.

\begin{shaded}
\begin{example}
Let $\vec{x}=(4,1,1)$, $\vec{y}=(2,1,2)$ and $\vec{z}=(3,1,3)$
\begin{center}
\begin{tabular}{c|c|c|c}
    k & $\sum_{i=1}^k x^\downarrow_i$ & $\sum_{i=1}^k y^\downarrow_i$ & $\sum_{i=1}^k z^\downarrow_i$ \\
    \hline
    1 & 4 & 2 & 3\\
    2 & 5 & 4 & 6\\ 
    3 & 6 & 5 & 7\\
\end{tabular}
\end{center}
We have $\vec{x}\succeq \vec{y}$ and $\vec{z}\succeq \vec{y}$, but there is no majorisation relation between $\vec{x}$ and $\vec{z}$.
\end{example}
\end{shaded}

\paragraph{Shanon entropy} The most well-known probability functional is the Shanon entropy defined as 
\begin{align}
    S(P) = -\sum_{i=1}^N p_i\log(p_i).
\end{align}
Although the entropy is associated with a random variable, we write $S=S(P)$ since the probability is the only property we need from the random variable. We interpret and use the Shanon entropy in many different ways. We can interpret it as the uncertainty about an outcome according to a known probability distribution $P$. It can quantify the amount of information we need to specify an outcome. It can quantify the compression limits of a long string. It can quantify the probabilities of long sequences of independent identically distributed (IID) random variables. And more. In the following, we provide three examples that elucidate the use of Shanon entropy in different contexts.

\begin{shaded}
\begin{example} First, we provide without a proof (which is simple) a fundamental theorem, namely the asymptotic equipartition property (AEP) theorem. If $X_1, X_2,\ldots $ are IID  with $P(X)$, then 
\begin{align}
    -\frac{1}{n}\log p(X_1,X_2,\ldots X_n)\rightarrow S(P), \quad \mbox{in probability.}
\end{align}
Defining the typical set as a set of sequences $x_1,x_2,\ldots x_n$ such that
\begin{align}
    2^{-n(S(P)+\epsilon)}\leq p(x_1,x_2,\ldots x_n)\leq2^{-n(S(P)-\epsilon)}
\end{align}
we have, as a consequence of the AEP, the following properties. The typical set has a probability close to 1, all elements of the typical set are almost equally likely, and the number of elements in the typical set grows as $\exp(n\,S(p))$. 
\end{example}
\end{shaded}

\begin{shaded}
\begin{example}
Imagine we want to find the smallest expected number of binary questions to determine an event if we know the underlying probability $P$. For concreteness, we consider that we have four events associated with the probability vector $\vec{p}=(\frac{1}{2},\frac{1}{4},\frac{1}{8},\frac{1}{8})$. Our task is to find the sequence of YES/NO questions that result in the shortest average number of questions needed to determine the event. In our case, we obtain the shortest average number of questions by the following sequence: 1) "Is the event $e_0$?", 2) if NO "Is the event $e_1$?", 3) if NO, "Is the event $e_3$?" We calculate the expected number of questions as follows
\begin{align*}
    L = 1 p(e_0) + 2p(e_1) + 3p(e_2) + 3*p(e_3) = \frac{1}{2}+\frac{2}{4} + \frac{3}{8} + \frac{3}{8} = 1\frac{3}{4}
\end{align*}
It is not difficult to see that the answer is the Shanon entropy $S(p)$ if calculated with the logarithm of base 2. Therefore, we can regard the $-\log(p_i)$ of the probability of simple events as the amount of information we obtain by measuring the event if the underlying probability distribution is $P$. 
\end{example}
\end{shaded}

\begin{figure}[!htb]
    \centering
    \includegraphics[width=0.6\textwidth]{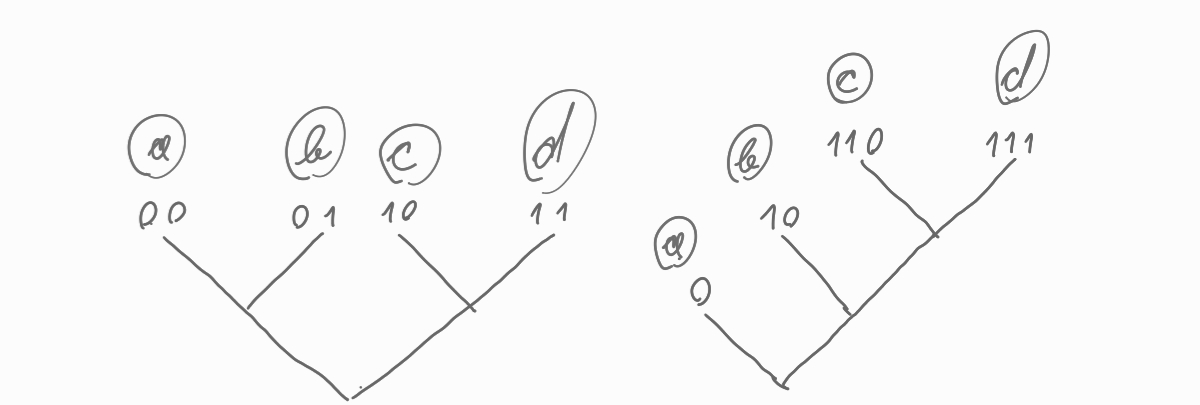}
    \caption{Binary code trees for four characters. The right tree has a shorter expected length if $p_1>p_2+p_3$.}
    \label{fig: code trees}
\end{figure}

\begin{figure}[!htb]
    \centering
    \includegraphics[width=0.8\textwidth]{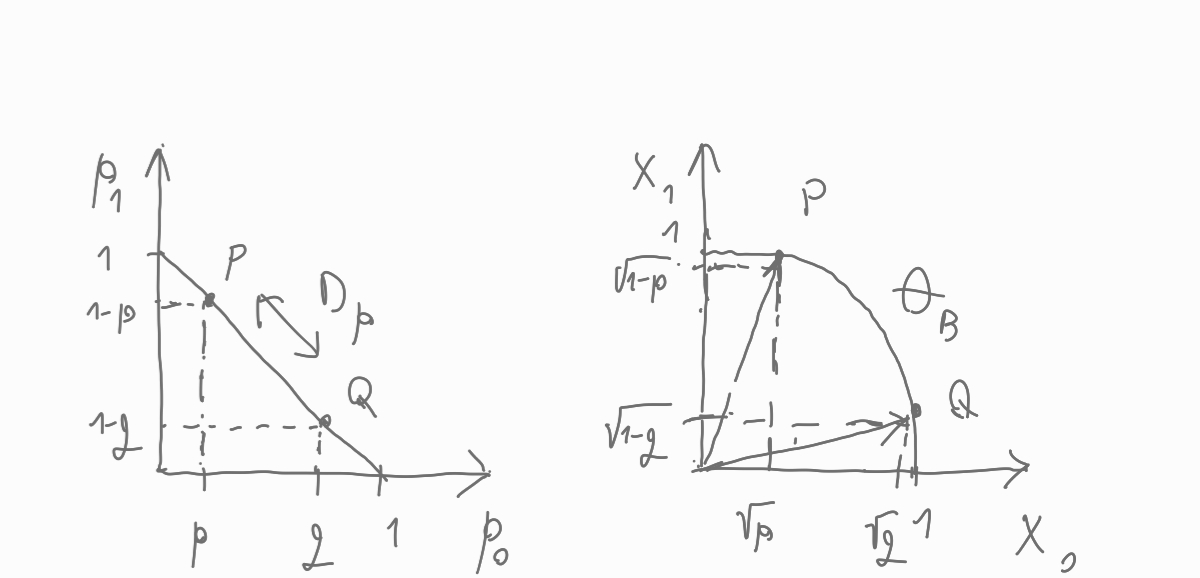}
    \caption{The left plot represents the standard probability simplex for $N=2$ and the $L_1$ distance between the probabilities $P$ and $Q$. The right plot shows the deformed simplex in new coordinates with a flat Fisher-Rao metric and a geometric distance between the probabilities given by the angle between the square root probability vectors.}
    \label{fig: battacharayya}
\end{figure}

\begin{shaded}
\begin{example}
\label{example: code trees}
Let us now consider the problem of compression. We assume that we have a long text where the four possible characters "a", "b", "c", and "d" appear according to the probability $\vec{p}=(\frac{1}{2},\frac{1}{4},\frac{1}{8},\frac{1}{8})$. Our task is to specify a binary prefix code so that the expected length of the text will be the shortest. We calculate the expected length of the code as 
\begin{align}
    L = l(a)p(a)+l(c)p(c)+l(b)p(b)+l(d)p(d).
\end{align}
Let us consider the codes given by the left and the right code tree in \fref{fig: code trees}. 
In the left case, we get
\begin{align*}
    L_{\rm left} = 2\frac{1}{2} +2\frac{1}{4}+ 2\frac{1}{8}+ 2\frac{1}{8} = 2 
\end{align*}
and in the right case, we get
\begin{align*}
    L_{\rm right} = \frac{1}{2}+\frac{2}{4} + \frac{3}{8} + \frac{3}{8} = 1\frac{3}{4}.
\end{align*}
One can show that the right case is the optimal one. Also, the Shanon entropy again gives the shortest expected length, which is, in this context, related to data compression.
\end{example}
\end{shaded}
\paragraph{Relative entropy} The second important probability functional that we will discuss is the relative entropy between two random variables $X$ and $Y$ with probabilities $P(X)$ and $Q(Y)$ defined as
\begin{align}
    S(P||Q) = \sum_{i=1}^Np_i\log\left(\frac{ p_i}{q_i}\right).
\end{align}
Relative entropy also has many applications and interpretations. It describes the convergence of samples to the asymptotic probability. We can also use it as an inefficiency measure in our compression, question strategy,\ldots, if assuming the wrong distributions. We further elaborate on both meanings with some examples
\begin{shaded}
\begin{example} \textbf{Sanov theorem (informal):} Assume we are drawing $\mathcal{N}$ samples from a probability distribution $Q$. The probability to get the frequency count $P^*$ is close asymptotically given by 
\begin{align}
    \mathcal{P}\sim\mathrm{e}^{-\mathcal{N}S(P^*||Q)}.
\end{align}
\end{example}
\end{shaded}
\begin{shaded}
\begin{example}
    Let us continue with the code trees example from the discussion of the Shanon entropy (\exref{example: code trees}) and assume that we think all characters are equally likely. Namely, we assume that the probability of the characters is $q=(\frac{1}{2},\frac{1}{2},\frac{1}{2},\frac{1}{2})$ whereas the real probability is still $\vec{p}=(\frac{1}{2},\frac{1}{4},\frac{1}{8},\frac{1}{8})$. The best encoding for our assumed probability $q$ is the left code tree in \fref{fig: code trees}. The difference between the assumed optimal length and the real optimal length ($L_{\rm left}$ - $L_{\rm right}$ from the example \exref{example: code trees}) is equal to the relative entropy
    \begin{align*}
        S(p||q) = \frac{1}{2}\log(\frac{4}{2}) + \frac{1}{4}\log(\frac{4}{4}) + \frac{1}{8}\log(\frac{4}{8}) + \frac{1}{8}\log(\frac{4}{8}) = \frac{1}{2} + 0 -\frac{1}{8} - \frac{1}{8} = \frac{1}{4}
    \end{align*}
\end{example}
\end{shaded}
\paragraph{Fisher-Rao metric} The Relative entropy determines a probabilistic "distance" between two probabilities. However, it is not a true distance in a mathematical sense since it is not symmetric. If we expand the relative entropy around the first or the second argument, we get
\begin{align}
    S(P||P+{\rm d}P) \approx \sum_{i=1}^N p_i\log\left(\frac{p_i}{p_i+{\rm d}p_i}\right)=\frac{1}{2}\sum_{i=1}^N \frac{{\rm d}p^i{\rm d}p^i}{p^i}.
\end{align}
We get the same result if we expand $S(P+{\rm d}P||P)$. The infinitesimal relative entropy is symmetric. The metric determining the infinitesimal distance between probability distributions is the Fisher-Rao metric defined as
\begin{align}
    g_{ij}=\frac{1}{4}\frac{\delta_{ij}}{p_i}.
\end{align}
We can introduce new coordinates to make the metric flat
\begin{align}
    X^i=\sqrt{p^i} \Rightarrow {\rm d}X^i = \frac{{\rm d}p^i}{\sqrt{p^i}}.
\end{align}
It follows that 
\begin{align}
    {\rm d}s^2 = \frac{1}{4}\sum_{i=1}^N \frac{{\rm d}p^i{\rm d}p^i}{p^i} = \sum_{i=1}^N {\rm d}X^i{\rm d}X^i.
\end{align}
In the new flat space, the ellipsoids around the probability vector become circles, and the normalisation condition becomes $|\vec{p}|_1 = |\vec{X}|_2 = 1$. In the newly introduced coordinates with a flat metric, the probabilities live on a positive part of a sphere. The geodesic distance between probabilities is the angle between the square roots of the probabilities (called the Battacharayya angle $\theta_{\rm B}$) determined by the scalar product between square roots of the probabilities
\begin{align}
    \cos \theta_{\rm B} = \sum_{i=1}^N \sqrt{p^iq^i}.
\end{align}
The \fref{fig: battacharayya} provides a simple visualization in the case $N=2$.
The Fisher-Rao metric also has a notable operational and practical meaning. To see that, we first consider a submanifold of the probability space parametrised by coordinates $\theta^i$. The Fisher-Rao metric induces a metric in this manifold as
\begin{align}
    g_{ab}=\sum_{i,j=1}^Ng_{ij}\frac{\partial p^i}{\partial \theta^a}\frac{\partial p^j}{\partial \theta^b}.
\end{align}
The above metric provides a fundamental constraint on the estimation uncertainty of the parameters $\theta^i$ from the samples of some random variable. If we assume that we have an unbiased estimator $\xi^i$, i.e. $\langle \xi^i\rangle = \theta^i$, we have the following covariance bound
\begin{align}
    \langle\xi^a\xi^b\rangle-\langle\xi^a\rangle\langle\xi^b\rangle \geq \frac{1}{4}g_{ab}.
\end{align}
More pragmatically, the Fisher-Rao metric also defines the natural gradient. If efficiently calculable, the natural gradient improves the training with stochastic gradient descent since it considers the curvature of the probability space. 

Here, we conclude our short introduction to classical probability. Introduced concepts will be a starting point for discussing quantum probability in the following sections.

\begin{referencesbox}
\begin{itemize}
  \item Bengtsson, I., \& Życzkowski, K. (2017). Geometry of quantum states: an introduction to quantum entanglement. Cambridge University Press. (Chapter 2: Geometry of probability distributions -- 2.1, 2.2, 2.3, 2.4, 2.5)
  \item Cover, T. M. (2006 - second edition). Elements of information theory. John Wiley \& Sons. (Chapter 2: Entropy, relative entropy and mutual information)
\end{itemize}
\end{referencesbox}

\subsection{Quantum probability}
This section introduces quantum probability with minimal changes to the classical probability formalism. The goal is to preserve most of the classical intuition/interpretation described in the previous section. We will introduce quantum probability in two steps: 
\begin{itemize}
    \item Extension of the formalism for classical probability from vectors to (diagonal) matrices
    \item Removing the restriction to diagonal matrices
\end{itemize}
The following table serves as a dictionary of the extended formalism of classical probability described in the previous section. We describe the probability in the extended formalism by a diagonal matrix and not by a vector.
\small{
\begin{center}
\begin{tabular}{l|c|c|c}
     Object& Classical (vector) & Extended (matrix)& Example ($N=2$)  \\
     \hline \hline
     Probability &$\vec{p}=(p_1,p_2,\ldots p_N)$ & $\rho={\rm diag}(p_1,p_2,\ldots p_N)$ &  $\rho=\begin{pmatrix}
p_1 & 0 \\
0 & p_2 
\end{pmatrix}$ \\
random variables &$\vec{a}=(a_1,a_2,\ldots a_N)$ & $A={\rm diag}(a_1,a_2,\ldots a_N)$ &  $A=\begin{pmatrix}
a_1 & 0 \\
0 & a_2 
\end{pmatrix}$ \\
simple events & $\Pi_j={\rm diag}(0_1,\ldots 1_j\ldots 0_N)$ & $\Pi_j$ (remains the same) &  $\Pi_1=\begin{pmatrix}
1 & 0 \\
0 & 0 
\end{pmatrix}$ \\
    \hline 
     real probabilities& $p_j\in \mathds{R}$ & $\rho^\dag=\rho$& $p_1=p_1^*$, $p_2=p_2^*$ \\
     real random variables& $a_j\in \mathds{R}$ & $A^\dag=A$& $a_1=a_1^*$, $a_2=a_2^*$ \\
     positivity & $p_j>0$ & $\rho>0$& $p_1>0$, $p_2>0$  \\
     normalisation&$\sum_{i=1}^Np_i=1$ &${\rm Tr}\,{\rho}=1$ & $p_1+p_2=1$ \\
     sum of simple events&$\sum_{i=1}^N\Pi_i = \mathds{1}_N$ &$\sum_{i=1}^N\Pi_i = \mathds{1}_N$  & $\Pi_1+\Pi_2 =\begin{pmatrix}
1 & 0 \\
0 & 1 
\end{pmatrix}$  \\
    event probability & $p_j=|\vec{p}\cdot\Pi_j|_1$ & $p_j=\mathrm{Tr}\,(\Pi_j\rho\Pi_j)$  & $\mathrm{Tr}\,(\Pi_1\rho\Pi_1)=p_1$ \\
    averages & $\langle a\rangle=\sum_{ij=1}^Np_ia_ib_i$& $\langle A\rangle={\rm Tr}\,\rho A$ & ${\rm Tr}\,\rho A=a_1p_1+a_2p_2$\\
    Independent $P$ and $Q$ & $\vec{p}^{\,12}=\vec{p}^{\,1}\otimes\vec{q}^{\,2}$ &  $\rho^{12}=\rho^1\otimes\rho^2$&  $\rho^{12}=\begin{pmatrix}
p_1q_1 & 0 & 0 & 0\\
0 & p_1q_2 & 0 & 0\\
0 & 0 & p_2q_1 & 0\\
0 & 0 & 0 & p_2q_2
\end{pmatrix}$\\
    marginals &$p^1_i=\sum_{j=1}^Np^{12}_{ij}$ &$\rho^1={\rm Tr}_2\,\rho$ & $\rho^1={\rm Tr}_2\rho^{12}=\begin{pmatrix}
p_1 & 0 \\
0 & p_2 
\end{pmatrix}$ \\
prob. update & $\vec{p}^{\rm UN}_{\Pi_j} = \Pi_j \vec{p}$ & $ \rho_{\Pi_j}^{\rm UN} =\Pi_j\rho\Pi_j$ & 
$\rho_{\Pi_1}^{\rm UN}=\begin{pmatrix}
p_1 & 0 \\
0 & 0 
\end{pmatrix}$
\end{tabular}
\end{center}
}
With $\vec{p}^{UN}$ and $\rho^{\rm UN}$ we denote the unnormalized probabilities. With the restriction to the diagonal matrices given in the first three rows of the above table, the standard and the extended columns are equivalent. We arrive at quantum probability if we remove this restriction and allow all objects that satisfy the rest of the equations in the extended column. It is standard to call the quantum probability a density matrix. In the following, we shall describe, similarly to the classical case, the essential properties of the quantum probability/density matrix.
\begin{shaded}
\begin{example}
\label{ex: qubit}
\textbf{Qubit} Let us start with the simplest case $N=2$ (qubit). In this case, we can parametrise the most general density matrix satisfying the hermiticity and normalisation as
\begin{align}
    \rho  = \frac{1}{2}\begin{pmatrix}
1+z & x-{\rm i}y \\
x+{\rm i}y & 1-z 
\end{pmatrix}.
\end{align}
The positivity condition is equivalent to the positivity of the eigenvalues of the above density matrix. After simplification, we obtain 
\begin{align}
    x^2+y^2+z^2\leq 1
\end{align}
The probabilities to get $a 0$ or $a 1$ are 
\begin{align}
    p_0={\rm Tr}\,\Pi_0\rho \Pi_0 =\frac{1}{2}(1+z),\quad\mbox{and}\quad p_1={\rm Tr}\,\Pi_1\rho \Pi_1 =\frac{1}{2}(1-z),
\end{align}
where
\begin{align}
    \Pi_0=\begin{pmatrix}
1 & 0 \\
0 & 0 
\end{pmatrix},\quad \mbox{and } 
\quad \Pi_1=\begin{pmatrix}
0 & 0 \\
0 & 1 
\end{pmatrix}.
\label{eq: qubit projectors}
\end{align}
\end{example}
\end{shaded}

\paragraph{Is the mapping to quantum probability unique?}
We now consider the following problem. Given a classical probability $\vec{p}$ and a set of "simple" projectors $\{\Pi_j\}$, find a corresponding quantum probability. One possibility is just the diagonal matrix with $p_j$ on the diagonal. However, this is not the only possibility. Any density matrix having the same diagonal will produce the same classical probability for the diagonal projectors.
\begin{shaded}
\begin{example}
In our qubit example, all density matrices of the form
\begin{align*}
    \eta = \begin{pmatrix}
p_0 & \sqrt{p_0p_1}e^{{\rm i }\phi} \\
e^{{\rm -i }\phi}\sqrt{p_0p_1} & p_1 
\end{pmatrix}
\end{align*}
reproduce the same classical probability given by the projectors $\Pi_0, \Pi_1$ defined in \eref{eq: qubit projectors}.
\end{example}
\end{shaded}

\paragraph{Determines the quantum probability as a unique classical probability?}
We have seen that many quantum probabilities are consistent with one classical probability and a set of projectors. The converse is also true. A single quantum probability (density matrix) $\rho$ determines many classical probabilities associated with different sets of projectors. Since we do not have the restriction to diagonal matrices, we can transform one set of projectors $\{\Pi_j\}$ into another valid set of projectors by a unitary transformation $\{U\Pi_jU^\dag\}$. The second set will typically result in different classical probabilities for different unitaries. 
\begin{shaded}
\begin{example}
In our qubit example, we define the following set of projectors
\begin{align*}
\Pi_0(\varphi)&=\begin{pmatrix}
\cos\varphi   \\
\sin\varphi   
\end{pmatrix} \begin{pmatrix}
\cos\varphi  &\sin\varphi   
\end{pmatrix}=  \begin{pmatrix}
\cos\varphi^2  &\cos\varphi\sin\varphi \\
\cos\varphi\sin\varphi & \sin\varphi^2
\end{pmatrix}\\
\Pi_1(\varphi)&=\begin{pmatrix}
-\sin\varphi   \\
\cos\varphi   
\end{pmatrix} \begin{pmatrix}
-\sin\varphi  &\cos\varphi   
\end{pmatrix}=  \begin{pmatrix}
\sin\varphi^2  &-\cos\varphi\sin\varphi \\
-\cos\varphi\sin\varphi & \cos\varphi^2
\end{pmatrix}
\end{align*}
It is clear that $\Pi_0(\varphi)+\Pi_1(\varphi) = \mathds{1}_2$. Since we have $||(\cos\varphi,\sin\varphi)||_2=||(-\sin\varphi,\cos\varphi)||_2=1$ we also have $\Pi_j(\varphi)^2=\Pi_j(\varphi)$. Finally, we get the following classical probabilities associated with the qubit density matrix $\rho$ from \exref{ex: qubit} and the set of projectors $\{\Pi_j(\varphi)\}$
\begin{align*}
    p_0&={\rm Tr}\, \Pi_0(\varphi) \rho \Pi_0(\varphi) = \frac{1}{2}(1+z(\cos\varphi-\sin\varphi)+x\cos\varphi\sin\varphi),\\
    p_1&={\rm Tr}\, \Pi_1(\varphi) \rho \Pi_1(\varphi) = \frac{1}{2}(1-z(\cos\varphi-\sin\varphi)-x\cos\varphi\sin\varphi).
\end{align*}
\end{example}
\end{shaded}
\paragraph{Pure and mixed states} We have seen that we do not have a one-to-one mapping between quantum and classical probabilities. The set of quantum probabilities seems larger than the set of classical probabilities. In the following, we will make the above statement more precise.

The quantum probability (density matrix) is a Hermitian matrix. Hence, we can diagonalise it by a unitary transformation. Due to positivity and normalisation, its eigenvalues are positive real numbers that sum to one. We thus have
\begin{align}
    \rho=U\Lambda U^\dag,\quad U^\dag U=\mathds{1}_N,\quad \Lambda={\rm diag}(\lambda_1,\ldots \lambda_N),\quad \lambda_j\geq0,\quad \sum_{i=1}^N\lambda_i=1
\end{align}
We use the convention that the eigenvalues are ordered $\lambda_1\geq\lambda_2\geq\ldots\lambda_N$. Density matrices with only one non-zero eigenvalue $\lambda_1=1$ are called pure quantum probabilities. If the quantum probability is not pure, we call it a mixed quantum probability. We have
\begin{align}
    \rho^2=\rho&,\quad \mbox{pure quantum probability}\\
    \rho^2<\rho&,\quad \mbox{mixed quantum probability}.
\end{align}
Pure quantum probabilities are analogues of pure classical probabilities. Every pure classical probability is also a pure quantum probability. 

Pure quantum probabilities are the central object in quantum computation and machine learning. Therefore, we introduce a new notation. We write $\rho_{\rm pure}=\ket{\psi}\bra{\psi}$. We denote the "square root" of pure quantum probability with a \textit{ket} $\ket{\psi}$, which in our case represents a column vector. We represent the row vector a \textit{bra} $\bra{\psi}$, where we also take complex conjugation. \textit{Bra} and \text{ket} represent adjoint vectors. We can combine them and construct operators (matrices) or scalars (complex numbers). Pure states are related to the eigenvectors of the density matrix with the eigenvalue 1. The averages over pure quantum probabilities can be calculated as $\average{A}_\psi={\rm Tr}\,A\rho =\bra{\psi}A\ket{\psi}$. The normalisation of quantum probabilities implies the $L_2$ normalisation of the "square root" of pure quantum probabilities, namely $||\psi||_2=\braket{\psi}{\psi}=1$. 

To summarise, pure quantum probabilities are vectors in a complex Hilbert space, i.e. $\psi \in \mathcal{H}$, and the random variables (observables) are operators in that space. 

In the following, we will use the standard terminology and refer to the "square root" of the pure quantum probability as a quantum state or a pure quantum state. We will imply that we can write it as a matrix or a vector. We will refer to mixed quantum probabilities as mixed quantum states. The term density matrix refers to the matrix representation of a pure or mixed quantum state.

We observe that $\vec{p}\in\mathds{R}^N, ||\vec{p}||_1=1$ while $\ket{\psi}\in\mathds{C}^N, ||\psi||_2=1$. We also see that while there is a finite number of pure classical probabilities, we have infinitely many pure quantum states.

\begin{figure}[!htb]
    \centering
    \includegraphics[width=0.8\textwidth]{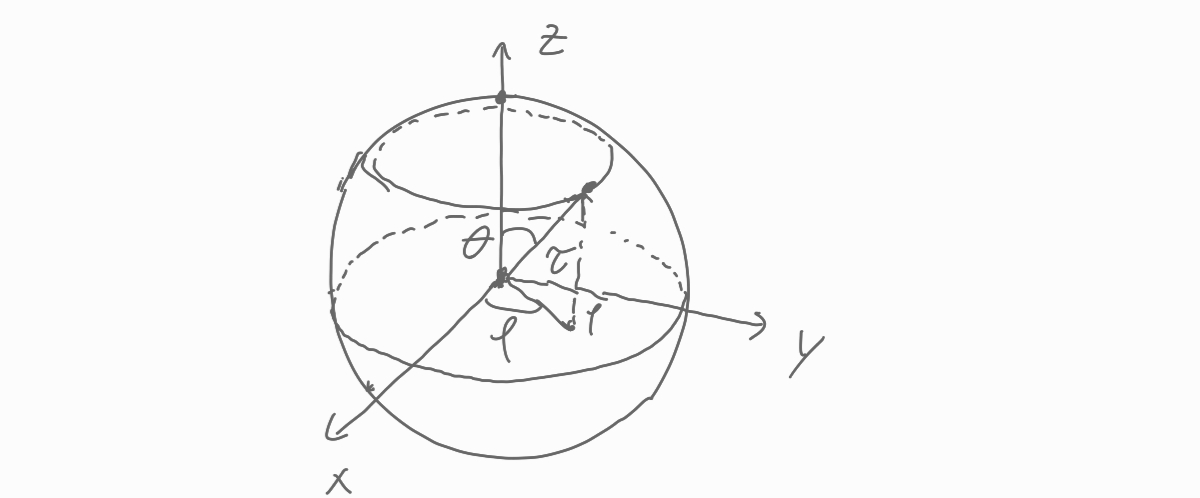}
    \caption{We show a representation of the Bloch ball. Pure quantum states live on the surface of the unit ball (Bloch sphere), whereas mixed quantum states live inside the sphere. The uniform distribution is at the origin.}
    \label{fig: bloch sphere}
\end{figure}

\begin{shaded}
\begin{example} \textbf{(Qubit)} Let us revisit the qubit example \exref{ex: qubit} and write the density matrix in terms of the ubiquitous Pauli matrices $\sigma^x,\sigma^y,\sigma^z$:
\begin{align}
    \rho = \frac{1}{2}(\mathds{1}+\vec{\tau}\cdot\vec{\sigma}),\quad \vec{\tau}=(x,y,z),\quad \vec{\sigma}=(\sigma^x,\sigma^y,\sigma^z).
\end{align}
The qubit is in a pure state iff $||\vec{\tau}||_2=1$. In this case, we can write
\begin{align}
    \rho = \ket{\psi}\bra{\psi},\quad \ket{\psi} = \begin{pmatrix}
\cos\theta/2   \\
{\rm e}^{-{\rm i} \varphi}\sin\theta/2,
\end{pmatrix}
\end{align}
The angles $\theta$ and $\varphi$ parametrise a general pure qubit state. The density matrix of this state is given by
\begin{align}
    \rho = \begin{pmatrix}
\cos^2\theta/2 &{\rm e}^{-{\rm i} \varphi}\cos\theta/2\sin\theta/2    \\
{\rm e}^{{\rm i} \varphi}\cos\theta/2\sin\theta/2 & \sin^2\theta/2,
\end{pmatrix}=\frac{1}{2}\begin{pmatrix}
1+\cos\theta &\sin\theta(\cos\varphi -{\rm i}\sin\varphi)\\
\sin\theta(\cos\varphi -{\rm i}\sin\varphi) & 1-\cos\theta.
\end{pmatrix}
\end{align}
We can now express the elements of the vector $\vec{\tau}$ in terms of the angles $\theta$ and $\varphi$
\begin{align}
    x &= \sin\theta \cos\varphi,\\
    y &= \sin\theta \sin\varphi,\\
    z &= \cos\theta.
\end{align}
As shown in the \fref{fig: bloch sphere}, the $N=p$ probability manifold is a unit ball called the Bloch ball. The pure states live on its surface, known as the Bloch sphere. This simple example illustrates the geometric difference between quantum and classical probability. In the $N=2$ case, the classical probability is simply a line with pure states being its boundary points. The quantum case is more complicated as all quantum states live in a unit ball, and there are infinitely many pure states.
\end{example}
\end{shaded}
\paragraph{Superposition and $\ket{Schroedinger}$} For a pure quantum state, one can always choose a set of projectors {$\Pi_j$} such that the corresponding classical probability is pure. In this case, the outcome is deterministic, and we know the quantum state exactly. However, in contrast to the classical case, we can also perform observations for any other set of projectors. In most other cases, the obtained classical probability will not be pure, and the outcome will be random. The ambiguity of classical measurements for a pure quantum state is the origin of the Schr\"{o}dinger cat thought experiment, where we "evolve" a cat into a state of being alive and dead. Here, we interpret a pure quantum state as a deterministic object, the same way as we can with a pure classical state. We only go this far with interpreting the superposition, which is operationally only a statement that a pure quantum state expressed in a chosen basis can have more than one non-zero component. 
\begin{shaded}
\begin{example}
Let us consider the following density matrix
\begin{align*}
    \rho=\frac{1}{2}\begin{pmatrix}1&1\\ 1&1 \end{pmatrix}.
\end{align*}
Together with projectors \eref{eq: qubit projectors} this defines a uniform classical probability $\vec{p}=(\frac{1}{2},\frac{1}{2})$. However, we also notice that the density matrix is a pure quantum state given by
\begin{align*}
    \ket{\psi}=\frac{1}{\sqrt{2}}\begin{pmatrix}1\\1\end{pmatrix}=\frac{1}{2}(\ket{0}+\ket{1}).
\end{align*}
Since a pure quantum state is also a projector, any set of projectors containing the state will have a deterministic outcome for our chosen state. For example 
\begin{align*}
        \Pi_0=\frac{1}{2}\begin{pmatrix}1&1\\ 1&1 \end{pmatrix}, \quad     \Pi_1=\frac{1}{2}\begin{pmatrix}1&-1\\ -1&1 \end{pmatrix}.
\end{align*}
The above projectors determine a pure classical probability with $p_0=1$ and $p_1=0$.
\end{example}
\end{shaded}

\paragraph{Collapse of the probability} What happens with the quantum state when we observe an event? The answer to this question is the same as in the classical case. If we throw a fair coin initially, the probability is $\vec{p}=\frac{1}{2}(1,1)$. After we observe either 0 or 1, we update the probability to $\vec{p}=(1,0)$ or $\vec{p}=(0,1)$. In the quantum case, it is the same. The quantum states become states we obtain if we first project the observed state onto the projector associated with the outcome and then normalise it. We express this in equations as follows
\begin{align}
    \rho\xrightarrow{\Pi_j} \frac{\Pi_j\rho\Pi_j}{{\rm Tr} \Pi_j\rho\Pi_j}.
    \label{eq:measurement}
\end{align}

\paragraph{Majorisation of quantum probabilities}
In the classical case, we introduced the majorisation order with non-increasing permuted probability vectors. As we have seen, a quantum probability $\rho$ determines many classical probabilities through projective measurements $\{\Pi_j\}$. However, we can use a unique probability to generalise majorisation, namely the probability determined by the eigenvectors of the quantum probability $\rho$. Therefore, we define the majorisation of quantum probabilities by 
\begin{align}
    \label{eq: rho majorisation}
    \rho\succeq\sigma\quad\Leftrightarrow\vec{\lambda}(\rho)\succeq\vec{\lambda}(\sigma).
\end{align}
We can show that the probability vector $\vec{\lambda}(\rho)$ majorises any other probability vector determined by a projective measurement $\{\Pi_j\}$. In other words, we determine the least uniform (most pure) probability by the projectors on the eigenspace of the quantum probability $\rho$.
\begin{figure}[!htb]
    \centering
    \includegraphics[width=0.4\textwidth]{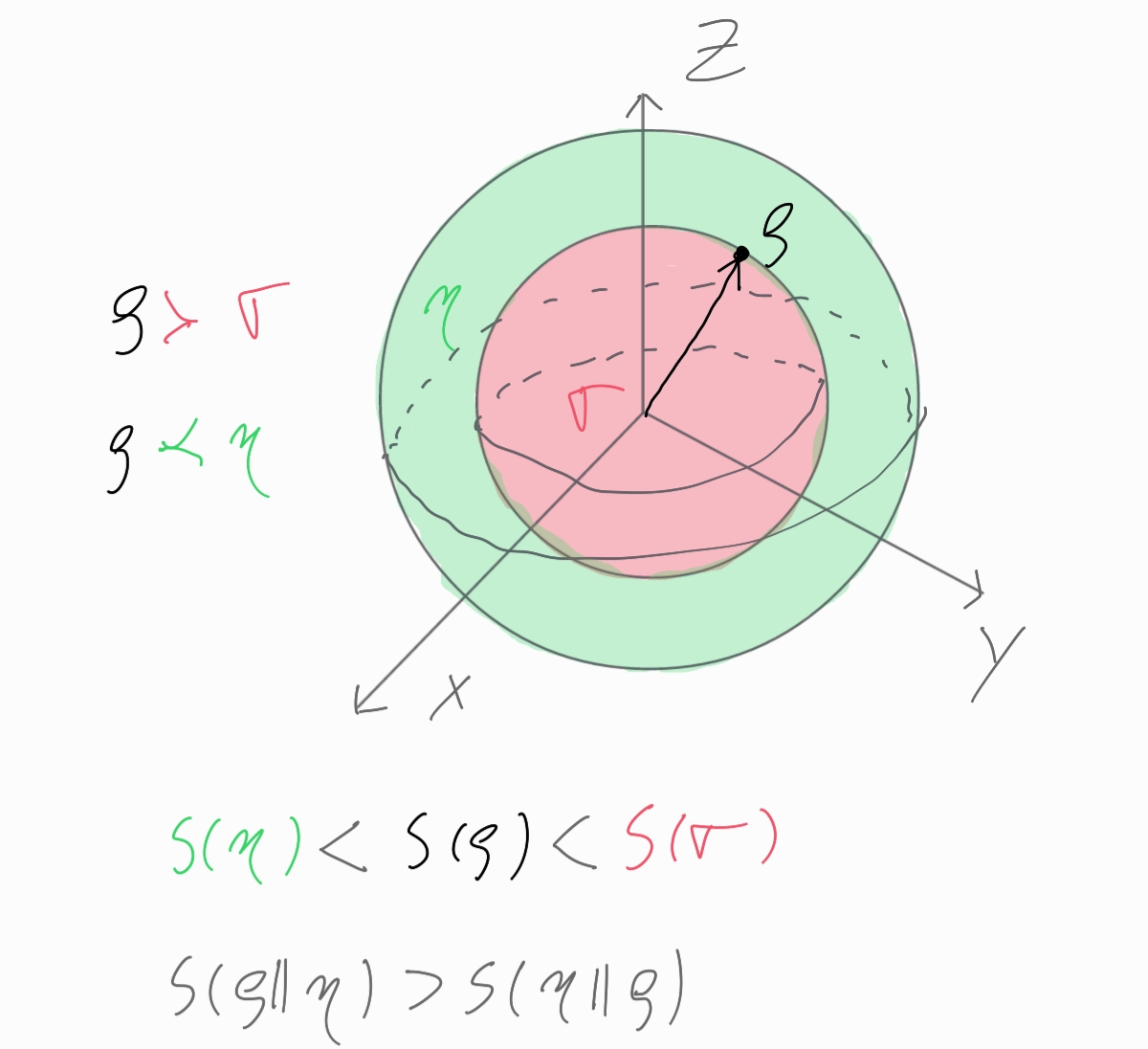}
    \caption{Majorization and entropies in a qubit case. The green quantum probabilities majorize the red quantum probabilities. Related inequalities hold for von Neumann entropy and quantum relative entropy.}
    \label{fig: quantum majorization}
\end{figure}

\paragraph{Von Neumann entropy} The most important probability functional in the classical case is the Shanon entropy. Its generalisation to the quantum case should recover the same result in the limit of diagonal density matrices. Namely
\begin{align}
    S(\rho_c)=\sum_{i=1}^{N} \lambda_i\log\lambda_i, \rho_c={\rm diag}(\lambda_1,\ldots,\lambda_N).
\end{align}
We can write this expression in a basis-independent form 
\begin{align}
    S(\rho) = {\rm Tr}\rho\log\rho,
\end{align}
which is valid for any density matrix. It already represents the correct quantum analogue of the Shanon entropy, called von Neumann entropy. However, this is not the only choice we could make. For example, every complete set of projectors defines a classical probability, which then has an associated Shanon entropy
\begin{align}
    S(\vec{p})=\sum_{i=1}^N p_i\log p_i,\quad p_i={\rm Tr}\,\rho \Pi_i.
\end{align}
One can show that von Neumann's entropy determines the minimum of the above entropies over all possible measurements. In this sense, the von Neumann entropy determines the minimum amount of uncertainty we can have when observing a quantum state, and it is a measure of purity. This agrees with the generalisation of majorisation discussed above and is explicitly shown in \fref{fig: quantum majorization}.
\paragraph{Quantum relative entropy} Similarly as before, we define the quantum relative entropy by classical analogy 
\begin{align}
    D(\rho||\eta)={\rm Tr}\rho(\log\rho-\log\eta).
\end{align}
The relative entropy quantifies the distinguishability of two quantum states by the best possible observation/measurement.

In the qubit case, we explicitly calculate the relative entropy $S(\rho_a||\rho_b)$ as
\begin{align}
    \label{eq: rel ent qubit}
    S(\rho_a||\rho_b)=\frac{1}{2}\ln\left(\frac{1-\tau_a^2}{1-\tau_b^2}\right)+\frac{\tau_a}{2}\ln\left(\frac{1-\tau_a}{1+\tau_a}\right)-+\frac{\tau_a}{2}\ln\left(\frac{1-\tau_b}{1+\tau_b}\right),
\end{align}
where $\rho_{a,b}=\frac{1}{2}(\mathds{1}+\vec{\tau}_{a,b}\cdot\vec{\sigma})$ and $\tau_{a,b}=\lVert\vec{\tau}_{a,b}\rVert_2$. From the \eref{eq: rel ent qubit} we explicitly see that $S(\rho_a||\rho_b)>S(\rho_b||\rho_a)$ iff $\tau_a<\tau_b$ (see also \fref{fig: quantum majorization}).

\paragraph{Quantum statistical distance (Fisher-Study distance)}
Let us consider pure states $\ket{\psi}$ and $\ket{\varphi}$ and a set of projectors $\{\Pi_1,\Pi_2\ldots \Pi_N\}$. They define two classical probabilities $\vec{p}=(p_1,p_2\ldots p_N)$ and $\vec{q}=(q_1,q_2\ldots q_N)$ with 
\begin{align}
    p_j=\bra{\psi}\Pi_j\ket{\psi}\quad\mbox{and}\quad q_j=\bra{\varphi}\Pi_j\ket{\varphi}.
\end{align}
We also write $\Pi_j=\ket{j}\bra{j}$. The geodesic distance between the two classical probabilities $\vec{p}$ and $\vec{q}$ is given by the Battacharayya angle $\theta_{\rm B}$
\begin{align}
    \cos\theta_{\rm B} = \sum_{i=1}^N\sqrt{p_iq_i} = \sum_{i=1}^N|\braket{\psi}{j}||\braket{\varphi}{j}|
    \label{eq: quantum distance}
\end{align}
We want to find the maximum possible distance for all allowed sets of measurements. Therefore, we must minimise the right-hand side of \eref{eq: quantum distance}. By using the inequality 
\begin{align}
    \sum_{i=1}^N|\braket{\psi}{j}||\braket{\varphi}{j}|\geq |\braket{\psi}{\varphi}| = \cos \theta_{\rm FS}
\end{align}
we obtain the probabilistic distance between two quantum states. The distance between two quantum states is the largest possible distance between two classical probabilities defined by the states and a set of projectors $\{\Pi_j\}$. 

\paragraph{Multiple quantum random variables: Entanglement} We will now discuss the properties of quantum probabilities with many random variables. Consider two uncorrelated pure quantum states with states $|\psi_1\rangle$ and $|\psi_2\rangle$.
A pure quantum state 
\begin{align}
    |\psi\rangle = |\psi_2\rangle\otimes|\psi_2\rangle.
    \label{eq: product state}
\end{align}
describes the joint quantum probability of the states

\begin{shaded}
\begin{example}
Let us consider a simple example of a two-qubit state. The first qubit is described by the state $|\psi_1\rangle=|0\rangle$ and the second by the state $|\psi_2\rangle=|1\rangle$. We write the joint probability distribution as $|\psi\rangle=|0\rangle\otimes|1\rangle$ or simply $|\psi\rangle=|01\rangle$. If we choose the basis $|0\rangle=\begin{pmatrix}1\\0\end{pmatrix}$, $|0\rangle=\begin{pmatrix}0\\1\end{pmatrix}$ we can express the quantum states as vectors
\begin{align}
    |\psi_1\rangle = \begin{pmatrix}1\\0\end{pmatrix},\quad |\psi_2\rangle = \begin{pmatrix}0\\1\end{pmatrix},\quad\mbox{and}\quad |\psi\rangle=|\psi_1\rangle\otimes|\psi_2\rangle=\begin{pmatrix}0\\1\\0\\0\end{pmatrix}.
\end{align}
\end{example}
\end{shaded}
However, we can not write all states as a tensor product of two states (as in \eref{eq: product state}).
We call such states entangled states. A principal example of an entangled state is a Bell state 
\begin{align}
\ket{\psi^+}=\frac{1}{\sqrt{2}}(\ket{00}+\ket{11})
\end{align}
We can express the state in a standard basis as 
\begin{align}
    \ket{\phi^+}=\frac{1}{\sqrt{2}}\begin{pmatrix}1\\0\\0\\1\end{pmatrix}.
\end{align}
To show that the state $\ket{\phi^+}$ is entangled, we consider the most general two-qubit state that we can  write as a tensor product of one qubit states
\begin{align}
    \ket{\psi_1}=\begin{pmatrix}a\\b\end{pmatrix},\quad \ket{\psi_1}=\begin{pmatrix}c\\d\end{pmatrix}.
\end{align}
We express the joint state in a standard basis as
\begin{align}
    \ket{\psi}=\ket{\psi_1}\otimes\ket{\psi_2}=\begin{pmatrix}ac\\ad\\bc\\bd\end{pmatrix}.
\end{align}
The equation $\ket{\psi}=\ket{\phi}$ shows that neither $a,b,c,d$ can be zero. At the same time, one of $ a$ and $ d$ and one of $ b$ and $ c$ should be zero. Consequently, we can not write the state $\ket{\phi^+}$ as a product of two one-qubit states, i.e., the state is entangled. The marginal probabilities of entangled states are mixed states, and the marginal probabilities of non-entangled states are pure states.
\begin{shaded}
\begin{example}
    Let us calculate the marginal probabilities of the Bell state $\ket{\phi^+}$. We calculate them by the partial trace operation as
    \begin{align}
        \rho=\frac{1}{2}\begin{pmatrix}1&0&0&1\\0&0&0&0\\0&0&0&0\\1&0&0&1\end{pmatrix},\quad \rho_1={\rm Tr}_2\,{\rho}=\frac{1}{2}\begin{pmatrix}1&0\\0&1\end{pmatrix},\quad \rho_2={\rm Tr}_1\,{\rho}=\frac{1}{2}\begin{pmatrix}1&0\\0&1\end{pmatrix}.
    \end{align}
\end{example}
\end{shaded}

\paragraph{Schmidt decomposition} An important tool to understand entangled states is the Schmidt decomposition
\begin{align}
    \ket{\psi}=\sum_{i=1}^{K}\sqrt{\lambda_i}\ket{\varphi_i}\ket{\theta_i},
    \label{eq:schmidt}
\end{align}
where $\ket{\varphi_i}$ and $\ket{\theta_j}$ denote orthonormal bases of the two parts of the quantum state, $\lambda_i>0$, and $\sum_{i=1}^K\lambda_i=1$.
\begin{proof}
In general we can a twopartite quantum state as $\ket{\psi}=\sum_{i,j}C_{i,j}\ket{i}\ket{j}$, where $i=1,\ldots N$ and $j=1,\ldots M$ and $\ket{i}$, $\ket{j}$ are orthonormal basis states in the respective subsystems. We can view the coefficients $C$ as an $N\times M$ dimensional complex matrix and expand it via the singular value (SVD) decomposition 
\begin{align}
    C = U\diag{\sqrt{\lambda_1},\sqrt{\lambda_2},\ldots\sqrt{\lambda_{K}}}V,\quad U^\dag U=VV^\dag=\mathds{1}_K,\quad K=\min(N,M).
\end{align}
Inserting the SVD decomposition in the general expression for the state, we obtain
\begin{align}
    \ket{\psi}=\sum_{i=1}^{N}\sum_{j=1}^{M}\sum_{i=1}^{K}U_{ik}\sqrt{\lambda_k}V_{k,j}\ket{i}\ket{j}.
\end{align}
We now define new basis states $\ket{\varphi_i}$ and $\ket{\theta_j}$ as
\begin{align}
    \ket{\varphi_k}=\sum_{i=1}^NU_{i,k}\ket{i},\quad \ket{\theta_k}=\sum_{j=1}^MV_{k,j}\ket{j}. 
\end{align}
Finally, we must show that the new basis states $\ket{\varphi_i}$ and $\ket{\theta_j}$ are orthonormal. We see this by using the unitarity of $U$ and $V$
\begin{align}
    \braket{\varphi_{k'}}{\varphi_k}&=\sum_{i',i=1}^N \overline{U}_{i',k'}U_{i,k}\braket{i'}{i}=\sum_{i=1}^N \overline{U}_{i,k'}U_{i,k}=\delta_{k',k},\\ 
    \braket{\theta_{k'}}{\theta_k}&=\sum_{j',j=1}^N \overline{V}_{k',j'}V_{k,j}\braket{j'}{j}=\sum_{j=1}^M \overline{V}_{k',j'}V_{k,j}=\delta_{k',k}.
\end{align}
\end{proof}
An important consequence of the Schmidt decomposition is that we can write any quantum state (pure or mixed) as a marginal of a larger pure state.
\begin{align}
    \rho_{12}&=\ketbra{\psi}{\psi}=\sum_{k',k=1}^{K}\sqrt{\lambda_{k'}\lambda_k}\ket{\varphi_k\theta_k}\bra{\varphi_{k'}\theta_{k'}},\\
    \rho_1&={\rm Tr}_2\,\rho_{12}=\sum_{k=1}^K\braket{\theta_k}{\psi}\braket{\psi}{\theta_k}=\sum_{k=1}^K\lambda_k\ketbra{\varphi_k}{\varphi_k}.
\end{align}
Since in the above equations, $\ket{\psi}$ is an arbitrary bipartite quantum state, also $\lambda_k>0$, and $\ket{\varphi_k}$ can take any allowed values. Hence, any $\rho_1$ can be written as a marginal of a larger pure state $\ket{\psi}$.

\paragraph{Completeness of the reduced density matrix} We will now consider a probability $\vec{p}=(p_1,p_2,\ldots p_N)$ determined by projectors $\{\Pi_j\otimes \mathds{1}_M\}$ acting nontrivially only on the first part of the system
\begin{align}
    p_j=\tr\left(\rho_{12}\Pi_j\otimes\mathds{1}_M\right)
\end{align}
Since the projectors act nontrivially only on the first part of the system, we can write
\begin{align}
    p_j&=\sum_{i=1}^N\sum_{j=1}^M\bra{i}\bra{j}\rho_{12}\Pi_j\otimes\mathds{1}_M\ket{i}\ket{j}\\ \nonumber
    &=\sum_{i=1}^N\sum_{j=1}^M\bra{i}\bra{j}\rho_{12}\ket{j}\Pi_j\ket{i} \\ \nonumber
    &=\sum_{i=1}^N\bra{i}\sum_{j=1}^M\bra{j}\rho_{12}\ket{j}\Pi_j\ket{i} \\ \nonumber
    &=\sum_{i=1}^N\bra{i}\rho_{1}\Pi_j\ket{i} \\ \nonumber
    &=\tr\rho_1\Pi_j \nonumber
\end{align}
The first line in the above derivation is just an explicit representation of the trace. To get the second line, we used the fact that the projectors are trivial on the second space; hence, they do not change the state $\ket{j}$, and we can pull it through and leave out the tensor product since the remaining projector acts only on the first part of the system. In the third line, we commuted the second (finite) sum with the "bra" of the first part of the system $\bra{i}$. In the fourth line, we introduced the reduced density matrix as a partial trace over the second system $\rho_1={\rm Tr}_2\rho_{12}$. In the last line, we compactly wrote the equation as a trace of the first system. The meaning of the obtained equation is that to determine any probability on the part of the system, we only need to know the reduced density matrix of the considered part.

\paragraph{Reduced density matrix of a classical probability} In classical probability, marginal distributions do not have any information about the joint distribution. On the other hand, quantum marginal probabilities (or reduced density matrices) still preserve just enough information to reconstruct the joint pure state. To see this, we first use the Schmidt decomposition of the joint distribution and write the reduced densities with the Schmidt vectors $\ket{\varphi_k}$ and $\ket{\theta_k}$
\begin{align}
    \rho_{12}&=\ketbra{\psi}{\psi}=\sum_{k',k=1}^{K}\sqrt{\lambda_{k'}\lambda_k}\ket{\varphi_k\theta_k}\bra{\varphi_{k'}\theta_{k'}},\\
    \rho_1&={\rm Tr}_2\,\rho_{12}=\sum_{k=1}^K\braket{\theta_k}{\psi}\braket{\psi}{\theta_k}=\sum_{k=1}^K\lambda_k\ketbra{\varphi_k}{\varphi_k},\\
    \rho_2&={\rm Tr}_1\,\rho_{12}=\sum_{k=1}^K\braket{\varphi_k}{\psi}\braket{\psi}{\varphi_k}=\sum_{k=1}^K\lambda_k\ketbra{\theta_k}{\theta_k}.
\end{align}
Since the $\ket{\varphi_k}$ and $\ket{\theta_k}$ are orthonormal in the respective Hilbert spaces, they are eigenvectors of the reduced density matrices $\rho_1$ and $\rho_2$, respectively. We also see that the reduced density matrices have equal spectra, which are given by $\lambda_1,\lambda_2,\ldots \lambda_N$. Therefore, we can calculate the vectors $\ket{\varphi_k}$, $\ket{\theta_k}$ and the values $\lambda_k$ by diagonalising the reduced densities $\rho_1$ and $\rho_2$. Then, we can reconstruct the joint probability with the help of the Schmidt decomposition. This construction only works if the joint quantum probability is a pure state. 
\begin{shaded}
\begin{example}
\label{ex: probability reconstruction}
Let us consider a small example and study two quantum probabilities which determine the same classical probability by the projectors to the computational basis {$\ket{00},\ket{01},\ket{10},\ket{11}$}. The classical probability should be $\vec{p}=(\frac{1}{2},0,0,\frac{1}{2})$. 

\textbf{Classical probability} The classical probability can be written as a density matrix as
\begin{align}
    \rho_{12}=\begin{pmatrix}\frac{1}{2}&0&0&0\\
    0&0&0&0\\0&0&0&0\\0&0&0&\frac{1}{2}\end{pmatrix}.
\end{align}
Then the marginals are 
\begin{align}
    \rho_1={\rm Tr}_1\,\rho_{12}=\frac{1}{2}\begin{pmatrix}1&0\\0&1\end{pmatrix},\quad 
    \rho_2={\rm Tr}_2\,\rho_{12}=\frac{1}{2}\begin{pmatrix}1&0\\0&1\end{pmatrix}
\end{align}
Besides $\rho_{12}$, there are many density matrices with the above marginals $\rho_1$ and $\rho_2$. For example 
\begin{align}
    \eta=\rho_1\otimes\rho_2 = \frac{1}{4}\begin{pmatrix}1&0&0&0\\
    0&1&0&0\\0&0&1&0\\0&0&0&1\end{pmatrix}.
\end{align}
In other words, classical marginal probabilities have no information about the correlations in the joint probability.

\textbf{Quantum probability} Another possibility to encode a classical probability with a quantum probability is to use the pure state
\begin{align}
    \ket{\psi}&=\frac{1}{\sqrt{2}}(\ket{00}+\ket{11})\\
    \rho_{12} &= \ketbra{\psi}{\psi} = \frac{1}{2}\begin{pmatrix}1&0&0&1\\
    0&0&0&0\\0&0&0&0\\1&0&0&1\end{pmatrix}.
\end{align}
We reproduce the correct classical probability since we have $\diag{\rho}=\vec{p}$. The marginals are now given by 
\begin{align}
    \rho_1 = \frac{1}{2}\begin{pmatrix}1&0\\0&1\end{pmatrix},\quad \rho_2 = \frac{1}{2}\begin{pmatrix}1&0\\0&1\end{pmatrix}.
\end{align}
The marginals are the same as in the classical case! How can they now have more information about the joint probability? The difference is that, in this case, the joint probability is a pure state. Which we can reconstruct with the prescription from the main text. From $\rho_1$ and $\rho_2$ we first get the the eigenvectors $\ket{\varphi_k}$, $\theta_k$, and the eigenvalues $\lambda_k$. Since both reduced densities are diagonal we have $\lambda_1=\lambda_2=\frac{1}{2}$, and $\ket{\varphi_1}=\ket{0}$, $\ket{\varphi_2}=\ket{1}$, $\ket{\theta_1}=\ket{0}$, $\ket{\theta_2}=\ket{1}$. Now we reconstruct the initial joint probability via the Schmidt decomposition
\begin{align}
    \sum_{k=1,2}\sqrt{\lambda_k}\ket{\varphi_k}\ket{\theta_k}=\frac{1}{\sqrt{2}}\ket{0}\ket{0}+\frac{1}{\sqrt{2}}\ket{1}\ket{1} =\frac{1}{\sqrt{2}}(\ket{00}+\ket{11}) \ket{\psi}.
\end{align}
Note that even this procedure is not unique. However, if we know that the initial joint probability describes a classical distribution and has positive phases, then the reconstructed joint probability is unique.
\end{example}
\end{shaded}

\paragraph{Spooky action at a distance}
In the example \exref{ex: classical collapse}, we considered correlated coin flips. We showed that the marginal probability for the second coin flip changes when we observe the outcome of the first coin flip. We observe a similar phenomenon with quantum probability, with a crucial difference that no information transfer between the system's first and second parts is necessary to reproduce the correlations. We sometimes refer to this change in the marginal probability of a distant system upon observing the first system as "spooky action at a distance". For concreteness, we consider the standard maximally entangled Bell state in \exref{ex: spooky action}.
\begin{shaded}
\begin{example}
\label{ex: spooky action}
We start our experiment with a maximally entangled Bell state $\ket{\phi^+}=\frac{1}{\sqrt{2}}(\ket{00}+\ket{11})$. We will observe the change in the marginal probabilities of the second cubit upon observing the first qubit. Before any observation, the marginal probability of the second qubit is 
\begin{align}
    \rho_2=\frac{1}{2}\begin{pmatrix}1&0\\0&1\end{pmatrix}.
\end{align}
Then we perform an observation of the first qubit and observe with probability $\frac{1}{2}$ the state $\ket{0}$ and with probability $\frac{1}{2}$ the state $\ket{1}$. We express the updated joint probability after the collapse as
\begin{align}
    \ket{\phi^+}\xrightarrow{\mbox{observe 0}} \ket{00},\\
    \ket{\phi^+}\xrightarrow{\mbox{observe 1}} \ket{11}.
\end{align}
The marginal probability changes accordingly
\begin{align}
    \rho_2\xrightarrow{\mbox{observe 0} } \ketbra{0}{0},\\
    \rho_2\xrightarrow{\mbox{observe 1} } \ketbra{1}{1}.
\end{align}
In other words, if we observed 0 for the first qubit, we will observe 0 also for the second qubit, and if we observed 1 for the first qubit, we will observe 1 for the second qubit. In the classical example, we had to transfer some information about the outcome of the first observation, but this is not the case here. 
\end{example}
\end{shaded}
\paragraph{Transformations of quantum states}
We transform quantum states by linear transformation\footnote{Non-linear transformations lead to non-physical effects, e.g. superluminal information transfer.}. In the finite-dimensional case, we represent linear transformations with matrices
\begin{align}
    \ket{\psi}\rightarrow M\ket{\psi}.
\end{align}
Since the transformed state $M\ket{\psi}$ has to be a quantum state as well, it needs to be normalised
\begin{align}
    |\psi|_2=1,\quad |M\psi|_2=1
\end{align}
Norm-preserving matrices are unitary matrices. Hence, we represent transformations of quantum states by unitary matrices
\begin{align}
    \ket{\phi}&\rightarrow U\ket{\phi},\quad U^\dag U = UU^\dag = \mathds{1}_N,
\end{align}
and in the language of density matrices 
\begin{align}
    \rho\rightarrow U\rho U^\dag.
\end{align}

Therefore, we write the most general transformation of a reduced density matrix as
\begin{align}
    \rho_1\rightarrow&{\rm Tr}_2U\rho_1\otimes\ketbra{\nu}{\nu}U^\dag\\ 
    &=\sum_{\mu=1}^K \bra{\mu}U\ket{\nu}\rho_1\bra{\nu}U^\dag\ket{\mu}\\
    &=\sum_{\mu=1}^KA_\mu \rho_1A_\mu,\quad \mbox{ where }\quad A_\mu=\bra{\mu}U\ket{\nu}.
\end{align}
Since $U$ is a unitary matrix, we have
\begin{align}
    \sum_{k=1}^K A^\dag A = \sum_{\nu=1}^K\bra{\nu}U\ketbra{\mu}{\mu}U^\dag\ket{\nu}=\mathds{1}_N
    \label{eq: Kraus condition}
\end{align}
Therefore, the most general transformation of a density matrix can be written with arbitrary operators $A_\nu$ that satisfy the condition \eref{eq: Kraus condition} and is called the Kraus map. 
\begin{shaded}
\begin{example}
Imagine we have two unitaries $U_1$ and $U_2$ to transform our quantum state $\rho$. We apply the unitary $U_1$ with probability $p_1$ and the unitary $U_2$ with probability $p_2=1-p_1$. Our statistical description of the transformed density matrix is then
\begin{align}
    \rho\rightarrow\rho'=p_1U_1\rho U_1^\dag + p_2U_2\rho U_2^\dag.
\end{align}
The transformed matrix is still a valid density matrix since it is a convex combination of density matrices and $p_1+p_2=1$. By introducing $A_j=\sqrt{p_j}U_j$, we write the transformed density matrix with a Kraus map
\begin{align}
    \rho'=\sum_{j=1,2}A_j\rho A_j^\dag.
\end{align}
The operators $A_j$ satisfy the condition \eref{eq: Kraus condition} 
\begin{align}
    \sum_{j=1,2}A_j^\dag A_j =\sum_{j=1,2}p_j U_j^\dag U_j=\sum_{j=1,2}p_j\mathds{1}_N = \mathds{1}_N. 
\end{align}
We can interpret the Kraus map as describing quantum operations with classical noise.
\end{example}
\end{shaded}
\paragraph{General measurements/events} We introduced quantum events by analogy with the classical description. They are described by a set of projectors $\{\Pi_j\}$ that sum to the identity $\sum_{j}\Pi_j=\mathds{1}_N$. The probability $p_j$ to observe a particular event $\Pi_j$ is given by the trace $p_j=\tr{\rho\Pi_j}$. We determine the state after the observation by the collapse rule
\begin{align}
    \rho\xrightarrow{\Pi_j}\frac{\Pi_j\rho\Pi_j}{\tr\Pi_j\rho\Pi_j}.
\end{align}
Similarly, as we did in the case of the transformations, we can study how the measurement generalises by studying a joint probability
\begin{align}
    \rho_1\xrightarrow{\Pi_j} &\frac{{\rm Tr}_2 \left(\Pi_j\rho_1\otimes\ketbra{\nu}{\nu}\Pi_j\right)}{\tr\left( \Pi_j\rho_1\otimes\ketbra{\nu}{\nu}\Pi_j\right)}.
\end{align}
We find that 
\begin{align}
    {\rm Tr}_2 \left(\Pi_j\rho_1\otimes\ketbra{\nu}{\nu}\Pi_j\right)=\sum_{\mu=1}^K\bra{\mu}\Pi_j\ket{\nu}\rho_1\bra{\nu}\Pi_k\ket{\mu}=\sum_{\mu=1}^K A_{\mu,j}\rho_1A^\dag_{\mu,j},\quad \mbox{ where }\quad A_{\mu,j}=\bra{\mu}\Pi_j\ket{\nu}.
\end{align}
Since the projectors sum to identity, we have
\begin{align}
    \sum_{\nu,j}A^\dag A=\sum_{\mu,j}\bra{\nu}\Pi_j\ketbra{\mu}{\mu}\Pi_j\ket{\nu}=\mathds{1}_N
\end{align}
\begin{shaded}
\begin{example}
We have seen that a measurement determined by a set of projectors $\{\Pi_j\}$ does not uniquely determine the quantum state. However, there exists a generalised measurement described by operators $\{A_k\}$, which determines $\rho$ uniquely. We call such a measurement an informationally complete (IC) measurement. In the case of a qubit, an important IC measurement is given by projectors $\Pi_j=\ketbra{\psi_j}{\psi_j}$ where
\begin{align}
    \ket{\psi_1} &= \ket{0},\\
    \ket{\psi_2} &= \frac{1}{\sqrt{3}}\ket{0}+\sqrt{\frac{2}{3}}\ket{1},\\
    \ket{\psi_3} &= \frac{1}{\sqrt{3}}\ket{0}+\sqrt{\frac{2}{3}}{\rm e}^{{\rm i}\frac{2\pi}{3}}\ket{1},\\
    \ket{\psi_4} &= \frac{1}{\sqrt{3}}\ket{0}+\sqrt{\frac{2}{3}}{\rm e}^{{\rm i}\frac{4\pi}{3}}\ket{1}.
\end{align}
The projector $\Pi_j$ also has the property that 
\begin{align}
    \tr\Pi_i\Pi_j=\frac{\delta_{i,j}N+1}{N+1},
    \label{eq: sic}
\end{align}
where $N=2$ is the dimension of the Hilbert space. IC measurements that satisfy the condition \eref{eq: sic} are called symmetric IC measurements and are important in quantum tomography and cryptography. \footnote{We still do not know if SIC exists in any dimension.}
\end{example}
\end{shaded}

\begin{referencesbox}
\begin{itemize}
  \item Bengtsson, I., \& Życzkowski, K. (2017). Geometry of quantum states: an introduction to quantum entanglement. Cambridge University Press. (Chapter 2: Geometry of probability distributions -- 2.1, 2.2, 2.3, 2.4, 2.5, Chapter 5: Outline of quantum mechanics)
  \item Nielsen, M. A., \& Chuang, I. L. (2010). Quantum computation and quantum information. Cambridge University Press. (Chapter 2: introduction to quantum mechanics, Chapter 11: Entropy and information)
\end{itemize}
\end{referencesbox}

\section{Quantum computation}
\subsection{Quantum many-body problem}
In this section, we will connect the structure of quantum probability with the physical descriptions of a system. Then, we will discuss the quantum many-body problem and the prospects of quantum machine learning.
\paragraph{Dynamical and statistical description of a quantum system}
So far, we have seen that we represent transformations of quantum states by unitary matrices. However, we still do not know how we can generate those unitaries. To see this, we must introduce the main object determining a quantum system's behaviour, the Hamiltonian (typically denoted by $H$). In general, the Hamiltonian is a Hermitian matrix and is connected to the unitary transformations of quantum states by the Schr\"odinger equation
\begin{align}
    \frac{\dd \ket{\psi}}{\dd t}=-{\rm i}H\ket{\psi}.
\end{align}
If the Hamiltonian $H$ is time-independent, then we have 
\begin{align}
    \ket{\psi(t)}={\rm e}^{-{\rm i}Ht}\ket{\psi}=U\ket{\psi}
\end{align}
For a given unitary, we can always find the generating Hamiltonian. 

The Hamiltonian is also important for the statistical description of the system. If we have a quantum system at a certain temperature $T$, we write the density matrix that describes its statistical properties as
\begin{align}
    \rho_T = {\rm e}^{-H/T}/Z =\frac{1}{Z}\sum_{j=1}^N{\rm e}^{-E_k/T}\ketbra{\xi_k}{\xi_k},
\end{align}
where we introduced the normalization $Z=\tr{\rm e}^{-H/T}$, and the Hamiltonian eigenvector decomposition $H=\sum_{k=1}^NE_k\ketbra{\xi_k}{\xi_k}$. We observe that the eigenvectors follow an exponential (softmax) distribution. If we lower the temperature, the states with the lowest energies become exponentially more likely, and at zero temperature, the system is in the ground state, i.e. the state with the smallest energy. We describe the quantum system as being in a mixed state at higher temperatures.
\begin{shaded}
\begin{example}
To get an intuition about the Hamiltonian, we will first consider a classical problem. Then, we will rewrite it using the quantum formalism. Finally, we will generalise it to the quantum case. 

Imagine we have N magnets arranged on a line. Let us assume that we can place the magnets either with the north pole up or the south pole up. We associate a scalar $\sigma_j\in\{-1,+1\}$ to each of the states. If the first magnet is in the state $\sigma_1=+1$, the next magnet prefers to be in the $\sigma_{2}=-1$. We can describe this interaction by a scalar product $H=\sum_j\sigma_j\sigma_{j+1}$. Further, we can tune the strength of the interaction between the magnets by controlling the size of the wall between two magnets. In this case, we write $H=\sum_jJ_j\sigma_j\sigma_{j+1}.$ Finally, we can add a bigger magnet below each magnet, which always has the north pole up. The final Hamiltonian describing the energy of the system is  
\begin{align}
    H_{\rm Ising} = \sum_{j=1}^{n-1} J_j\sigma_{j}\sigma_{j+1} +\sum_{j=1}^n h_j\sigma_j.
    \label{eq:ising}
\end{align}
We now rewrite the classical Ising Hamiltonian \eref{eq:ising} with the quantum formalism by introducing the standard embedding $\Phi(1)=\ket{0}$ and $\Phi(-1)= \ket{1}$. The classical Ising Hamiltonian can then be written with the $\sigma^{\rm z}$ Pauli matrix Since we have $\sigma^{\rm z}\ket{0}=+1 \ket{0}$. We have $\sigma^{\rm z}\ket{1}=-1 \ket{1}$, and 
\begin{align}
    H_{\rm Ising}^{\rm q} = \sum_{\langle i,j\rangle}J_{i,j}\sigma^{\rm z}_i\sigma^{\rm z}_{j} + \sum_{j}h_j\sigma^{\rm z}_j.
\end{align}
The notation $\langle i,j\rangle$ in the sum means we have a sum over the nearest neighbours. The above equation is the quantum formulation of the classical Ising model since we have 
$H_{\rm Ising}(\sigma_1,\sigma_2,\ldots \sigma_n)=\Phi^\dag(\sigma_1)\otimes\Phi^\dag(\sigma_2)\otimes \ldots \Phi^\dag(\sigma_n)H^{\rm q}_{\rm Ising}\Phi(\sigma_1)\otimes\Phi(\sigma_2)\otimes \ldots \Phi(\sigma_n)$, where $\Psi^\dag(1)=\bra{0}$ and 
$\Phi^\dag(-1)=\bra{1}$. 
We see that operators acting on a single qubit encode an external field, and the operators acting on two or more qubits encode the interaction between the qubits. 

Finding the ground state, i.e. the state with the lowest energy, of the Ising model directly corresponds to quadratic unconstrained binary optimisation (QUBO)
\begin{align}
    \min \sum_{\langle i,j\rangle}w_{i,j}x_ix_j + \sum_{j}b_j x_j
\end{align}
where $x_j\in \{0,1\}$. The QUBO problem is NP-hard, translating into the Ising model's long relaxation times. We already see an intimate connection between interesting computational problems and paradigmatic physical models. Therefore, we can use observable properties of actual physical systems to solve interesting computational problems.  

The model $H_{\rm Ising}^{\rm q}$ is classical since all operators in the sums commute. We must add a term that does not commute with $\sigma^{\rm z}_j$ to introduce quantum effects. The standard choice is a local field in the $x$ direction modelled by the Pauli matrix $\sigma^{\rm x}_j$. The resulting model is called the transverse field Ising model (TFIM)
\begin{align}
    H_{\rm TFIM}= \sum_{\langle i,j\rangle}J_{i,j}\sigma^{\rm z}_i\sigma^{\rm z}_{j} + \sum_{j=1}(h^{\rm z}_j\sigma^{\rm z}_j +h^{\rm x}_j\sigma_j^{\rm x}). 
\end{align}
As we will see later, the TFIM is still insufficient to perform universal quantum computation, but it is sufficient to solve interesting NP-hard problems, e.g., the travelling salesman problem. 
\end{example}
\end{shaded}

\paragraph{Why is quantum computing hard?}
Exact solutions to quantum problems are rare. The reason is that the size of the Hilbert space of a many-body quantum problem grows exponentially with the number of qubits, i.e., $2^n$. Therefore, even numerically, we can find exact solutions for problems up to 40 qubits. Nevertheless, we can describe much larger systems in many important special cases. We typically exploit some of the models' symmetry or unique properties in these cases, e.g., low entanglement, local conservation laws, or small parameters. Accordingly, we developed many different methods to describe these special cases:
\begin{itemize}
    \item \textbf{Analytical methods}: perturbation theory, renormalisation group, integrability, quadratic systems,
    \item \textbf{Numerical methods}: Monte Carlo, tensor networks, numerical renormalisation group.
\end{itemize}
Special cases are important since they often contain essential information about more realistic physical systems. However, they cover a negligible part of the Hilbert space, representing two opportunities from the quantum machine learning perspective. First, we can use standard machine learning tools to describe many-body quantum systems and cover more Hilbert space with new numerical methods. Second, we can use the classically inaccessible part of the Hilbert space to improve machine learning algorithms. Here, we will focus mainly on the latter. 

\paragraph{Quantum advantage} A natural framework to study how quantum computation can assist machine learning and quantify the quantum advantage is complexity theory. We will distinguish computational complexity, sample complexity, and model complexity.

Computational (quantum) complexity is the most common approach to finding the benefits of quantum computing. The main tool is the asymptotic runtime complexity, which denotes the asymptotic functional dependence of the runtime with respect to the input size. A quantum speedup means the quantum algorithm is faster than the classical one in asymptotic complexity. We distinguish five different classes of quantum speedup
\begin{enumerate}
    \item \textit{Provable quantum speedup:} we have proof that no classical algorithm can perform better than the quantum algorithm, Grover's search (quadratic speedup)
    \item \textit{Strong quantum speedup:} the quantum algorithm is better than the best classical algorithm, Shor's factoring  (exponential speedup)
    \item \textit{Common quantum speedup:} the quantum algorithm is better than the best available classical algorithm
    \item \textit{Potential quantum speedup:} the quantum algorithm is better than a specific classical algorithm
    \item \textit{Limited quantum speedup:} quantum algorithm is better than the corresponding classical algorithm (useful when comparing quantum and classical annealing)
\end{enumerate}

Sample complexity is related to generalisation. It determines how many samples we need to generalise from data. Most evidence considering quantum and classical sample complexity suggests they are polynomially equivalent. However, upon introducing noise, the separation becomes more drastic. The quantum algorithm can still generalise in some cases, while the classical problem is considered unlearnable. 

Finally, model complexity refers to the expressiveness of the models, i.e. the classes of functions we can express with a given number of parameters. Due to overfitting, expressive models are not necessarily better than "slim" models. In this context, the main avenue of research is experimental since it is not apparent which quantum models have a suitable bias to capture patterns in specific datasets.

\begin{referencesbox}
\begin{itemize}
  \item Schuld, M., \& Petruccione, F. (2021). Machine learning with quantum computers (Vol. 676, pp. 163-169). Berlin: Springer. (Chapter: Quantum Computing, Chapter: Potential Quantum Advantages )
\end{itemize}
\end{referencesbox}

\subsection{Models of quantum computation}
\begin{figure}[!htb]
    \centering
    \includegraphics[width=0.6\textwidth]{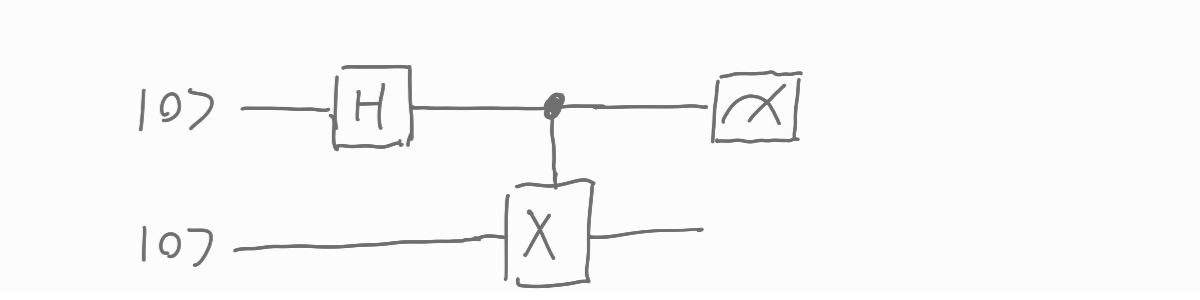}
    \caption{Example quantum circuit. We start in a state $\ket{00}$, then transform it with a Hadamard gate and a ${\rm CNOT}$ gate to the state $\ket{\phi^+}=\frac{1}{\sqrt{2}}(\ket{00}+\ket{11})$, and finally we measure the first qubit.}
    \label{fig: example circuit}
\end{figure}

There are several competing models of quantum computation: gate model, adiabatic quantum computation, measurement-based quantum computation and continuous variable quantum computation. They are equivalent to a polynomial overhead.
\paragraph{Gate model of quantum computation}
\label{sec: gate model}
The most common approach to quantum computing is the gate model of quantum computation. Here, we generate complex quantum transformations by composing smaller, simple-to-implement quantum transformations called quantum gates. The two main components that ensure the scalability of the gate model are fault tolerance and universality. Fault tolerance means we can perform logical transformations even if some of our physical qubits are corrupted. It was shown that it is possible to implement fault-tolerant schemes for a finite set of gates. Conveniently, universality states that it is possible to implement any quantum transformation with a finite set of gates. More formally, the Solovay-Kitaev theorem shows that for any gate $U$ on a single qubit,
and given any $\epsilon>0$, it is possible to approximate $U$ to a precision $
\epsilon$ using $\mathcal{O}(\log c (1/\epsilon))$ gates from a ﬁxed ﬁnite set, where $c$ is a small constant approximately equal to $2$. Combining those two results, we can deduce that we can implement any algorithm with a finite set of universal quantum gates. Importantly, the number of gates we must use for fault-tolerant computation is only polylogarithmically larger than the number of simple transformations in the original algorithm. Finally, there are infinitely many universal sets of quantum gates. The most common universal set is: $\mathrm{CNOT}$, Hadamard gate $H$, phase gate $S$ and the $T$ gate:
\begin{align*}
    \mathrm{CNOT} &= \begin{pmatrix}1&0&0&0\\ 0&1&0&0\\ 0&0&0&1\\ 0&0&1&0 \end{pmatrix} &
    H &=\frac{1}{\sqrt{2}}\begin{pmatrix}1&1\\1&-1\end{pmatrix}\\
    S &=\begin{pmatrix}1&0\\0&\ii\end{pmatrix} &
    T &=\begin{pmatrix}1&0\\0&{\rm e}^{\ii\pi/4}\end{pmatrix}\\
\end{align*}
\begin{shaded}
\begin{example}
We typically represent quantum circuits with diagrams as shown in \fref{fig: example circuit}. We generally draw the diagrams from left to right. We start by writing the initial states of the qubits $\ket{0}$. To each of the states, we associate a wire. If we see only a wire, the qubit will not change during the computation. We draw the unitary transformations as boxes with names inside that denote the type of the transformation. Connecting the corresponding wires to the box indicates the qubits on which the unitary acts. Finally, on the right, we draw a special box to denote a measurement of one or more qubits. 
\end{example}
\end{shaded}
Practically, several challenges limit the applicability of the gate model. The main problem is that the current implementation of gates is noisy and that we can apply only up to several 100 transformations. Another critical issue that we should consider in this setting is that not all qubits are connected. Sometimes, we have to use more than two qubits and additional gates to apply a desired interaction between the two qubits.
\paragraph{Adiabatic quantum computation}
Adiabatic quantum computing builds on two observations/statements. I) Optimisation problems can be formulated such that a quantum state with minimum energy solves the problem determined by a Hamiltonian $H_1$. II) A quantum system remains in its instantaneous eigenstate if a given perturbation is acting on it slowly enough. The idea of adiabatic quantum computing is then the following. Start with a ground state of a simple Hamiltonian (e.g. $H_0=\sum_{i=1}^N\sigma^{\rm x}_i$) and then gradually change it to the final Hamiltonian (e.g. $H_1=\sum_{\langle i,j\rangle}J_{i,j}\sigma_i^{\rm z}\sigma_j^{\rm z} +\sum_ih_i\sigma_i^{\rm z}$) which encodes the solution of a desired problem in its lowest energy state. The caveat is that there is a speed limit on how fast the transition from one $H_0$ to $H_1$ can be. The transition time is lower bounded by the square of the inverse of the difference between the two lowest energy states. 

We can avoid the speed limit problem of adiabatic quantum computing by relaxing the condition of being in the best possible state and doing many tries of the same protocol. If the probability of staying in the ground state is large enough, we will eventually get the best possible solution to the problem. This is the quantum annealing approach to quantum computation.

As in the gate model, there are also some practical obstacles. The main ones are finite temperature and interference from measurement devices. Since they disturb the system, they increase the transition probability to a higher energy state (i.e. a sub-optimal solution).  

\paragraph{Measurement-based and continuous variable quantum computation} Measurement-based quantum computing consists of two steps. First, we prepare a highly entangled quantum state (a cluster state). Then, we perform conditional single-qubit measurements. We use the results of the measurements as conditions for the following measurements. The result of the computation is either the state of unmeasured qubits or the result of the last measurement. Though we have translated many important algorithms into this framework, it is unclear how they can present an advantage for NISQ devices.

So far, we have covered quantum computation with finite-dimensional systems. Continuous-variable quantum computing, on the other hand, manipulates quantum states whose matrix representations are infinite-dimensional. Contrary to measurement-based quantum computing, continuous-variable quantum computing seems beneficial for machine learning algorithms due to a large local Hilbert space that we can use to store and manipulate information.

\paragraph{Implementations}
There are several approaches to building a quantum computer. Their detailed description is out of the scope of the lectures. We only list a few important properties and obstacles:

\textbf{Superconducting circuits}
\begin{itemize}
    \item They are silicon-based; hence, we can use a lot of existing infrastructure
    \item Typical operating temperature is $\sim 10mK$
    \item Short control time $\sim 10ns$
    \item 2D connectivity. Therefore, we can not implement all possible two-body interactions.
\end{itemize}
\textbf{}
\textbf{Trapped ions}
\begin{itemize}
    \item Implemented by particles in magnetic traps
    \item Interactions are controlled by lasers
    \item Long coherence time
    \item Slow control time
\end{itemize}

\textbf{Photonic-based systems}
\begin{itemize}
    \item Based on light, hence work at room temperature
    \item The main implementation problem is photon loss
    \item Difficult to store
    \item It is not clear how they can scale
\end{itemize}

\begin{referencesbox}
\begin{itemize}
  \item Schuld, M., \& Petruccione, F. (2021). Machine learning with quantum computers (Vol. 676, pp. 163-169). Berlin: Springer. (Chapter: Quantum Computing)
  \item Nielsen, M. A., \& Chuang, I. L. (2010). Quantum computation and quantum information. Cambridge University Press. (Chapter 7: Quantum computers: physical realization)
\end{itemize}
\end{referencesbox}
\subsection{Fault-tolerant quantum computing: basic algorithms}
In this section, we consider the gate model of quantum computation. We will introduce it by translating essential quantum experimental protocols into quantum circuits. Then, we will provide some examples that show quantum speedup. Finally, we will discuss several types of algorithms for fault-tolerant quantum computers.

\paragraph{Quantum gates}
As we already discussed, the main building block of the gate model of quantum computing is a finite set of quantum gates that act only on a few qubits. Typically, we extend the universal set of quantum gates given in the previous section to be able to implement quantum protocols with a lesser number of gates. The following tables provide the standard set of gates implemented on current NISQ devices.

\begin{table}[!htb]
    \centering
    \begin{tabular}{c|c|c|c}
        Gate & Circuit & Matrix & Dirac \\ 
        \hline
         $X$ & \includegraphics[width=0.08\textwidth]{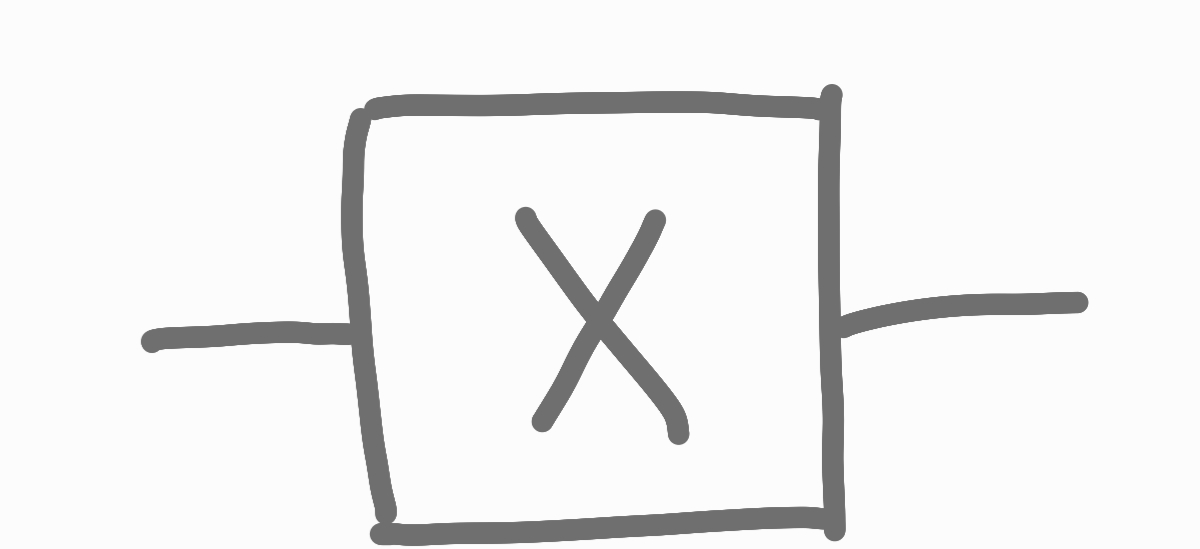} & $\begin{pmatrix}0&1\\1&0\end{pmatrix}$  &  $\ketbra{0}{1} +\ketbra{1}{0}$\\
         $Y$ & \includegraphics[width=0.08\textwidth]{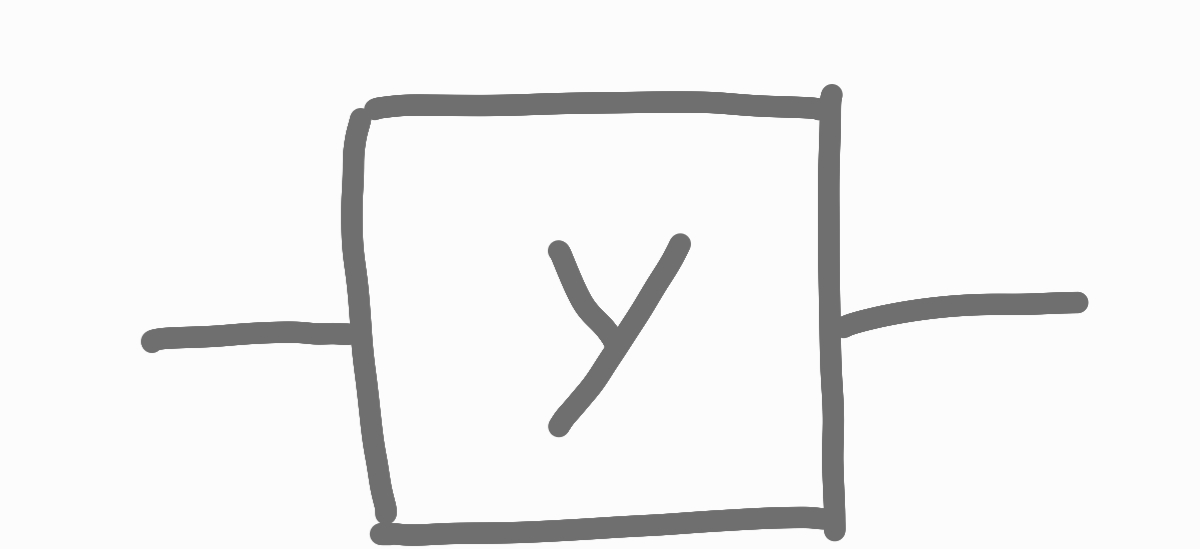} & $\begin{pmatrix}0&-\ii\\ \ii&0\end{pmatrix}$ & $\ii(\ketbra{1}{0}-\ketbra{0}{1})$  \\
         $Z$ & \includegraphics[width=0.08\textwidth]{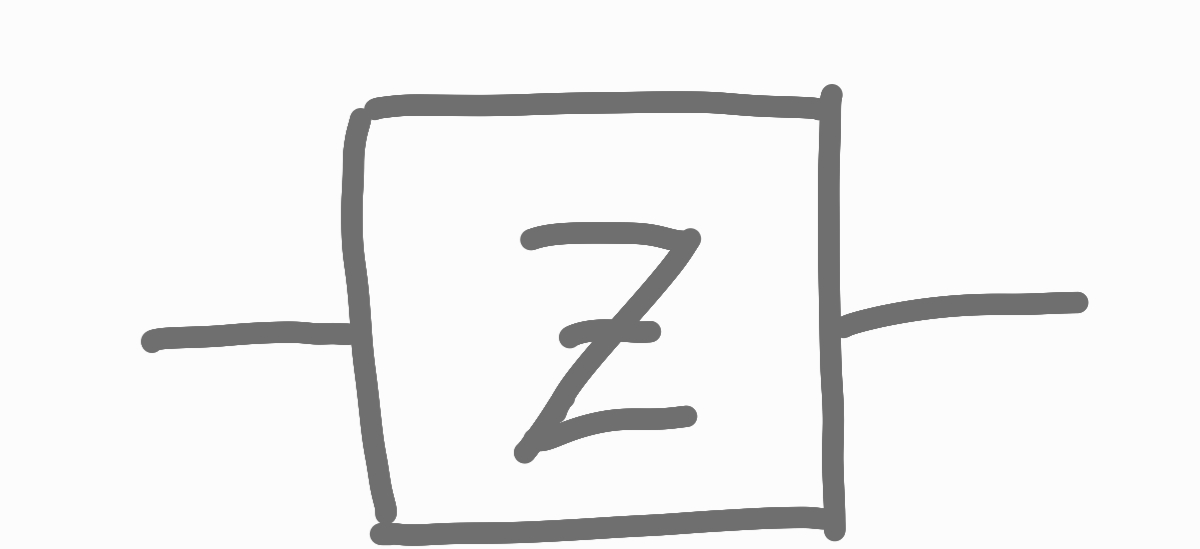} & $\begin{pmatrix}1&0\\0&-1\end{pmatrix}$ & $\ketbra{0}{0}-\ketbra{0}{0}$  \\
         $H$ & \includegraphics[width=0.08\textwidth]{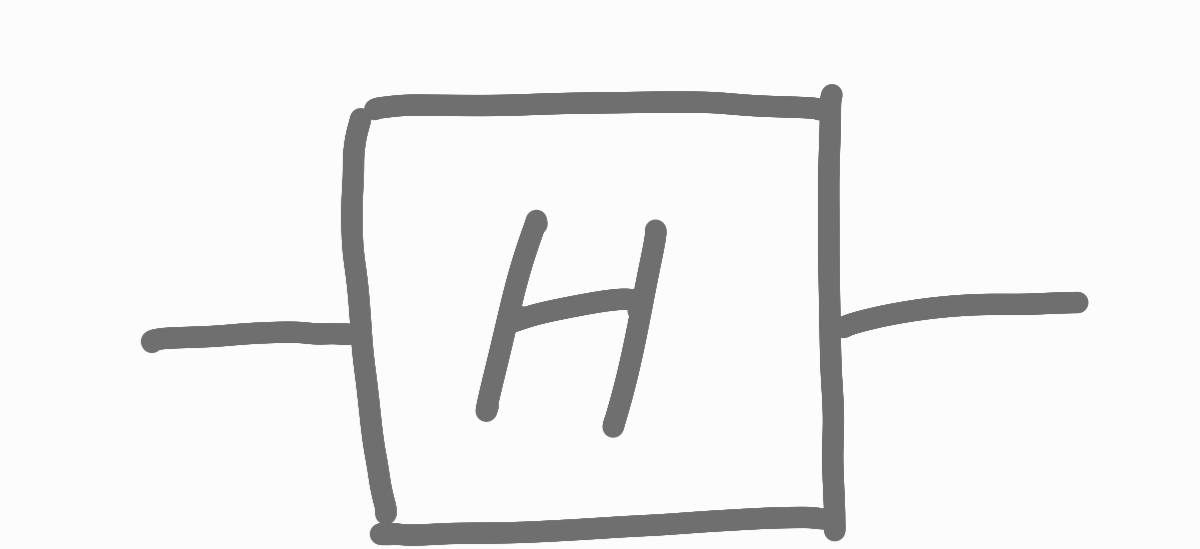} & $\frac{1}{\sqrt{2}}\begin{pmatrix}1&1\\1&-1\end{pmatrix}$ &$\frac{1}{\sqrt{2}} (\ketbra{0}{0}+\ketbra{0}{1} +\ketbra{1}{0} - \ketbra{1}{1})$ \\
         $S$ & \includegraphics[width=0.08\textwidth]{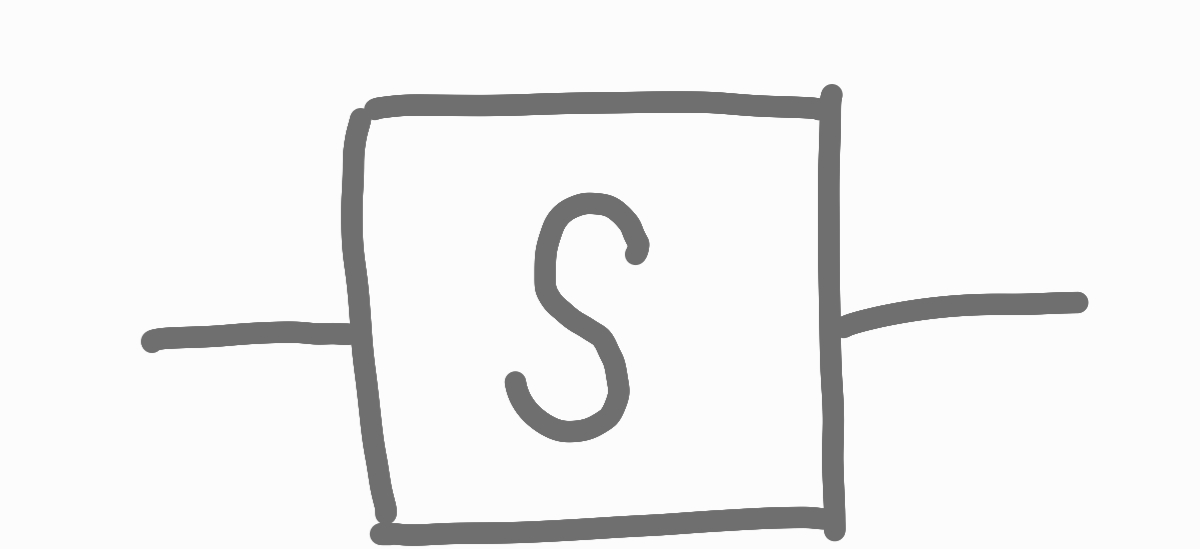} & $\begin{pmatrix}1&0\\0&\ii\end{pmatrix}$ & $\ketbra{0}{0} +\ii \ketbra{1}{1}$  \\
         $T$ & \includegraphics[width=0.08\textwidth]{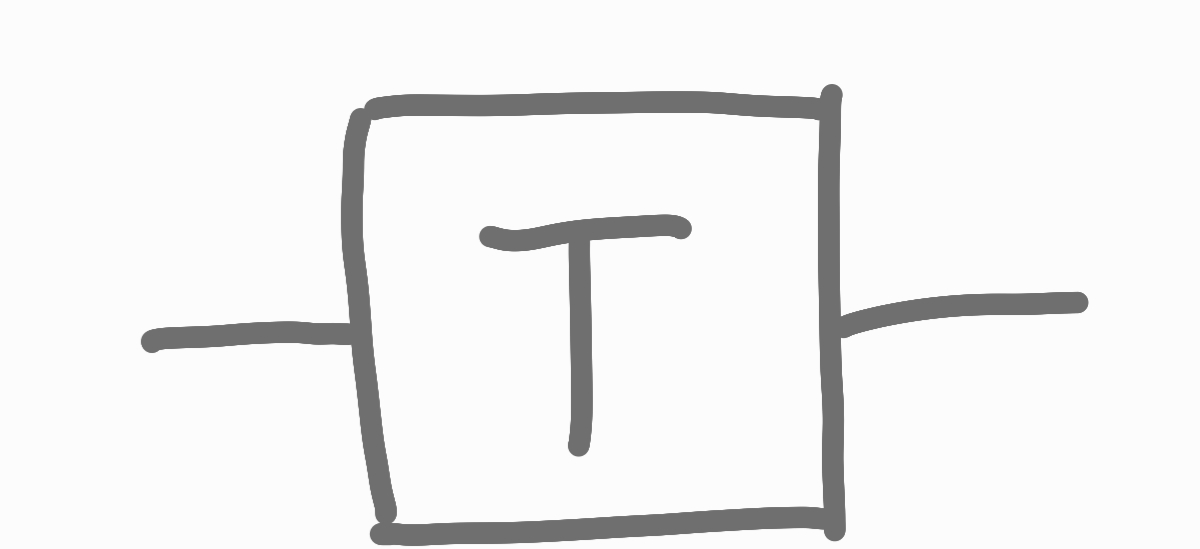} & $\begin{pmatrix}1&0\\0&{\rm e}^{\ii\frac{\pi}{4}}\end{pmatrix}$ & $\ketbra{0}{0} + {\rm e}^{\ii\frac{\pi}{4}}\ketbra{1}{1}$  \\
    \end{tabular}
    \caption{Common one-qubit quantum gates}
    \label{tab: qubit gates}
\end{table}

\begin{table}[!htb]
    \centering
    \begin{tabular}{c|c|c|c}
        Gate & Circuit & Matrix & Dirac \\ 
        \hline
         ${\rm CNOT}$ & \includegraphics[width=0.08\textwidth]{Gatex.jpg} & $\begin{pmatrix}1&0&0&0\\0&1&0&0\\0&0&0&1\\0&0&1&0\end{pmatrix}$  &  $\ketbra{00}{00} +\ketbra{01}{01}+\ketbra{10}{11}+\ketbra{11}{10}$\\
         ${\rm SWAP}$ & \includegraphics[width=0.08\textwidth]{Gatey.jpg} & $\begin{pmatrix}1&0&0&0\\0&0&1&0\\0&1&0&0\\0&0&0&1\end{pmatrix}$ & $\ketbra{00}{00} +\ketbra{10}{01}+\ketbra{01}{10}+\ketbra{11}{11}$ 
    \end{tabular}
    \caption{Common two-qubit quantum gates}
    \label{tab:2qubit gates}
\end{table}

\begin{table}[!htb]
    \centering
    \begin{tabular}{c|c|c}
        Gate & Circuit & Description \\ 
        \hline
        qubit & \includegraphics[width=0.08\textwidth]{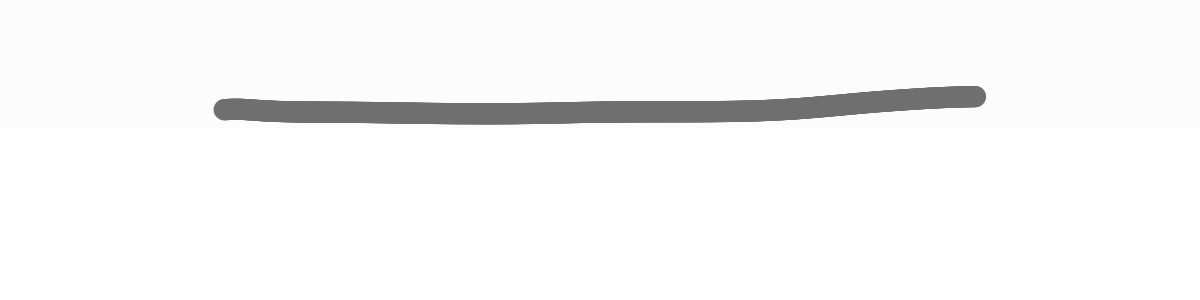} & wire with one qubit\\
         n qubits & \includegraphics[width=0.08\textwidth]{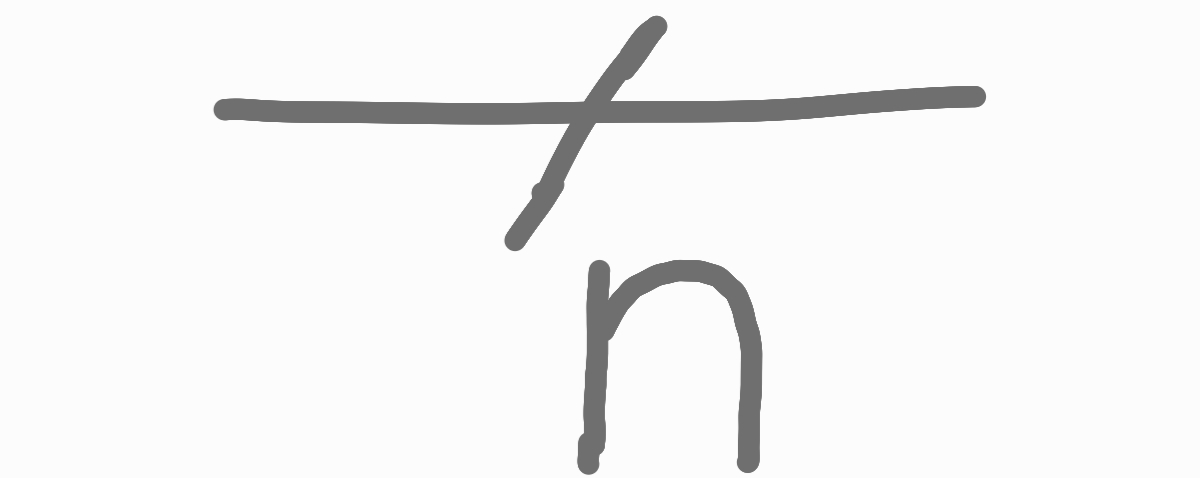} & wire with n qubits \\
        C-U & \includegraphics[width=0.08\textwidth]{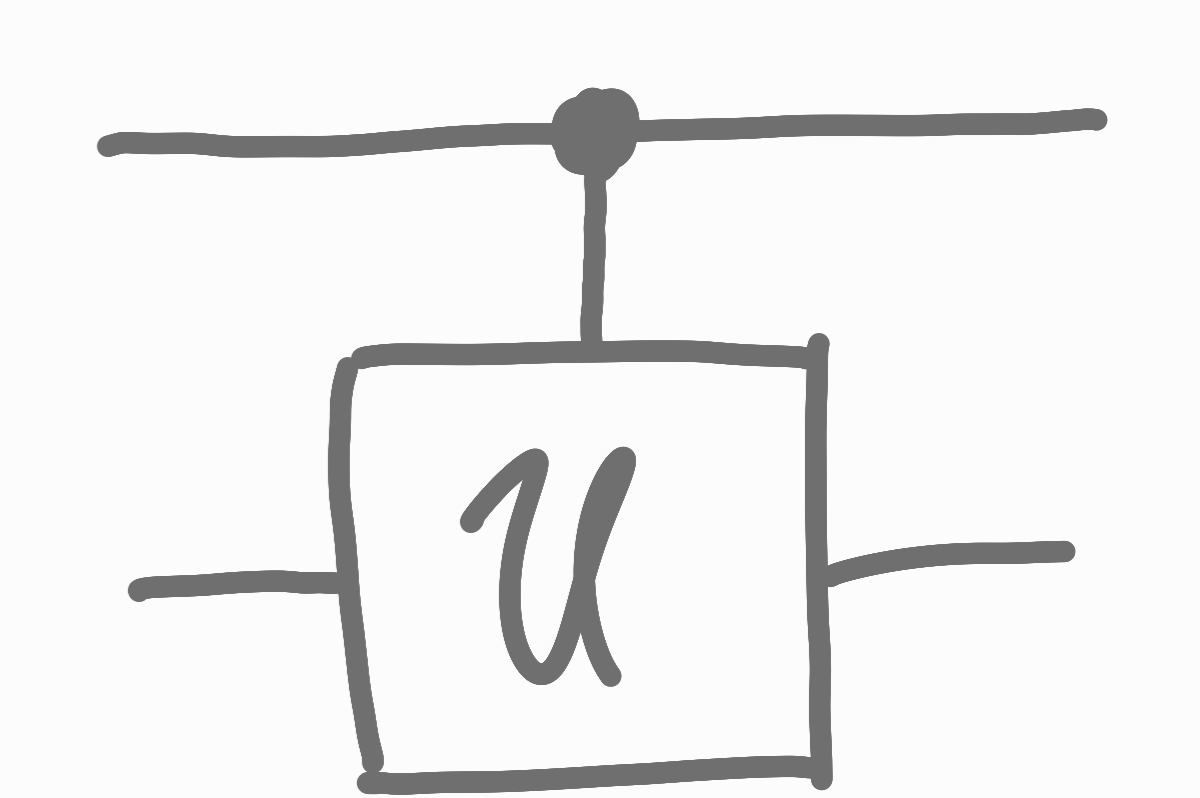} & one qubit control of one qubit unitary\\
         measurement & \includegraphics[width=0.08\textwidth]{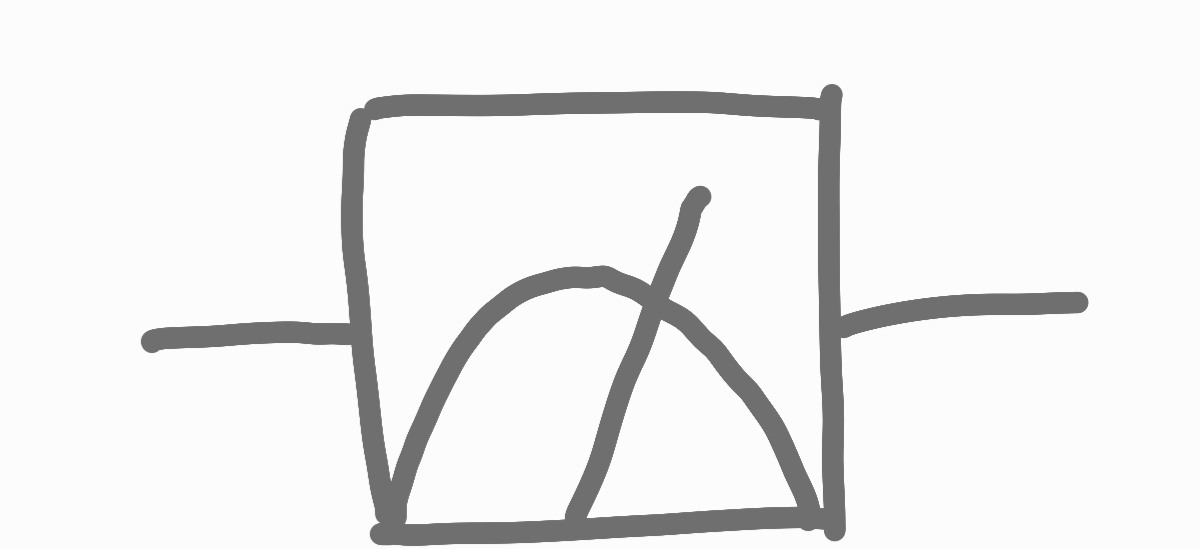} &  projection on one of the states $\ket{0}, \ket{1}$
    \end{tabular}
    \caption{Other common operations.}
    \label{tab: common gates}
\end{table}
\begin{shaded}
\begin{example}
\textbf{No cloning theorem} While copying classical bits is not a problem, we can not copy quantum states. More precisely, we can rule out the existence of a unitary transformation $U$ which would produce a state $\ket{\psi}\ket{\psi}$ when acting on a state $\ket{\psi}\ket{0}$ for any $\ket{\psi}$. This is easy to show by contradiction. Imagine we have such a unitary. Then, we can write 
\begin{align}
U(\ket{\psi,0})=\ket{\psi,\psi},\quad U(\ket{\phi,0})=\ket{\phi,\phi}.     
\end{align}
It follows that 
\begin{align}
    \braket{\psi, 0}{\phi, 0} &= \braket{\psi}{\phi}\\
    \braket{\psi, 0}{\phi, 0} &= \bra{\psi,0}U^\dag U\ket{\phi,0}=\braket{\psi,\psi}{\phi,\phi}=(\braket{\psi}{\phi})^2 
\end{align}
We find the equation $x^2=x$, which has only two solutions, namely $x=0$ and $x=1$. This means that we can copy only one set of orthogonal basis states. We will not exactly copy any other state.
\end{example}
\end{shaded}

\begin{figure}
    \centering
    \includegraphics[width=0.7\textwidth]{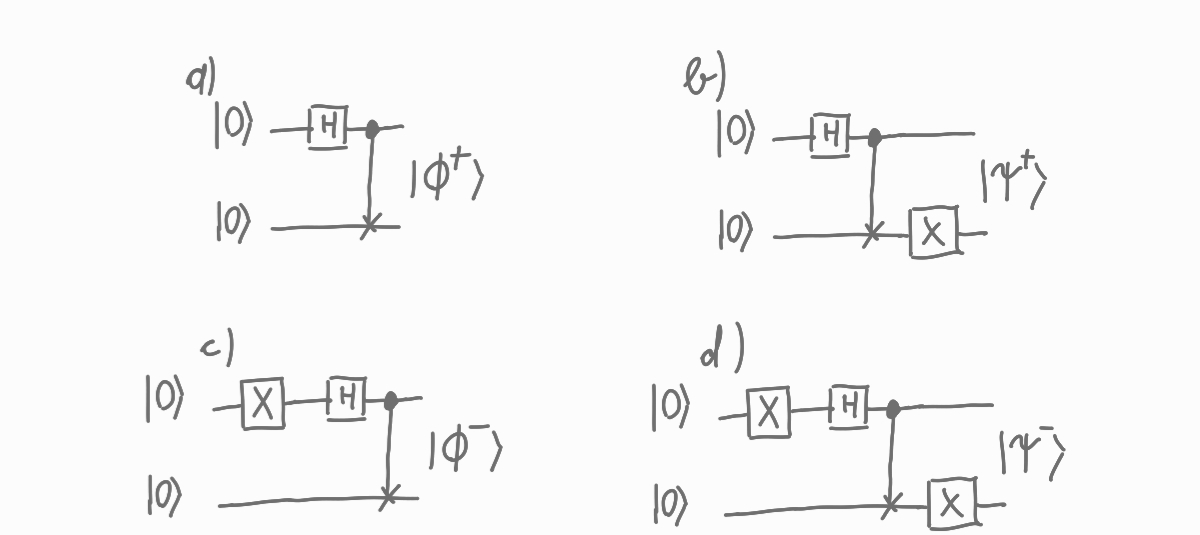}
    \caption{Quantum protocols for preparation of the single qubit Bell basis.}
    \label{fig: phi+}
\end{figure}

\paragraph{Bell states}
We always initialise the circuit-based quantum computers in a state with all qubits in the state $\ket{0}$. As our first protocol, we will discuss in detail how we can transform the two-qubit, product state $\psi_0=\ket{0,0}$ into entangled Bell states
\begin{align}
    \ket{\phi^\pm}=\frac{1}{\sqrt{2}}(\ket{0,0}\pm\ket{1,1}),\quad \ket{\psi^\pm}=\frac{1}{\sqrt{2}}(\ket{0,1}\pm\ket{1,0}).
\end{align}
We describe the circuit constructing the state $\ket{\phi^+}$ shown in \fref{fig: phi+} a. We first transform the initial state by a Hadamard gate acting on the first qubit, which transforms the state of the first qubit as shown in table \ref{tab: qubit gates}, while the second qubit is unchanged. We get the state $\psi_1$.
\begin{align}
    \ket{\psi_1}&=\frac{1}{\sqrt{2}}(\ket{0}+\ket{1})\ket{0}=\frac{1}{\sqrt{2}}(\ket{0,0}+\ket{1,0}).
\end{align}
Then, we transform $\psi_1$ by a controlled-NOT operation with the first qubit as a control. This means that all states where the first qubit is in a state $\ket{0}$ remain unchanged. All states where the first qubit is in a state $\ket{1}$ will have the second qubit flipped after the transformation. By applying this rule to the state $\psi_1$, we get the final result
\begin{align}
    \ket{\psi_2} = \frac{1}{\sqrt{2}}(\ket{0,0}+\ket{1,1}),
\end{align}
To get the state $\psi^+$, we have to flip one of the qubits. We do this with the $X$ gate, which we should apply after the ${\rm CNOT}$ gate (see \fref{fig: phi+} b). We must add a phase to the states with the first qubit in the state $\ket{1}$ to get the remaining states. We can achieve this by a controlled $Z$ gate at the end of the circuit or a bit flip on the first qubit applied before the Hadamard transformation (see \fref{fig: phi+} c) and \fref{fig: phi+} d)).  

\begin{figure}
    \centering
    \includegraphics[width=0.8\textwidth]{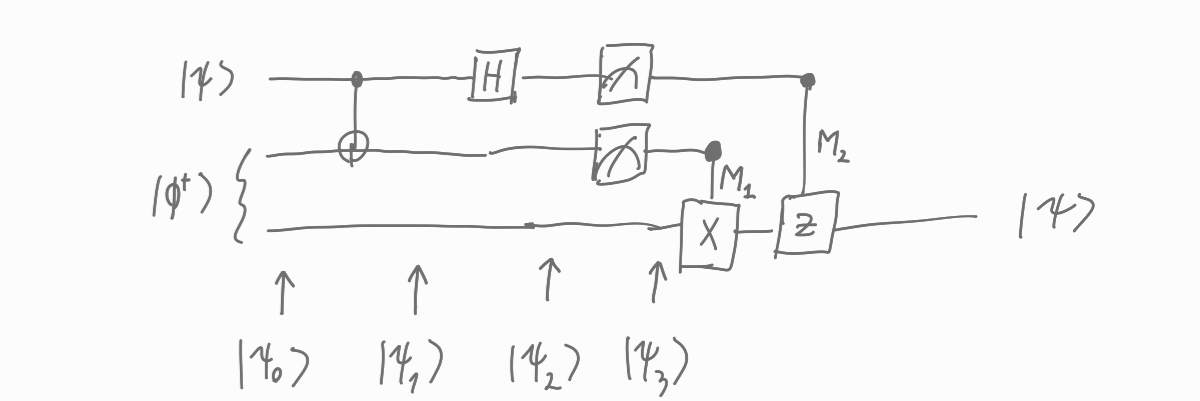}
    \caption{Single qubit teleportation protocol}
    \label{fig: teleportation}
\end{figure}

\paragraph{Teleportation}
One of the most interesting quantum experiments is teleportation. Here, we will consider a qubit teleportation. In this experiment, we have an entangled qubit in a state $\ket{\psi^+}=\frac{1}{\sqrt{2}}(\ket{0,0}+\ket{1,1})$ and an additional unknown qubit in a state $\ket{\psi}=a\ket{0} + b\ket{1}$. By performing a series of transformations, we can transfer the state of the additional qubit to one of the entangled qubits. This interesting point is that the entangled qubit holding the state can be far away from the additional qubit. We show the circuit describing the gates performing this operation in \fref{fig: teleportation}. The system's initial state is $\ket{\psi_0}$. It gets transformed during the protocol as follows
\begin{align}
    \ket{\psi_0}&= \ket{\psi,\phi}=(a\ket{0}+b\ket{1})\frac{1}{\sqrt{2}}(\ket{0,0}+\ket{1,1})\\
    \ket{\psi_1}&=\frac{1}{\sqrt{2}}(a\ket{0}(\ket{0,0}+\ket{1,1})+b\ket{1}(\ket{1,0}+\ket{0,1})) \\
    \ket{\psi_2}&= \frac{1}{2}\left(\ket{0,0}(a\ket{0}+b\ket{1})+\ket{0,1}(b\ket{0}+a\ket{1})\right)+\ket{1,0}(a\ket{0}-b\ket{1})+\ket{1,1}(b\ket{1}-a\ket{0})\\
    \ket{\psi_3}&= \begin{cases}
a\ket{0}+b\ket{1} & (M_1 = 0,\quad M_2 = 0) \\
b\ket{0}+a\ket{1} & (M_1 = 0,\quad M_2 = 1) \\
a\ket{0}-b\ket{1} & (M_1 = 1,\quad M_2 = 0) \\
b\ket{0}-a\ket{1} & (M_1 = 1,\quad M_2 = 1) \\
\end{cases}\\
\end{align}
Based on the measurements $M_1$ and $M_2$, we can transform the state of the last qubit into the state $\psi$ by applying controlled bit flip (gate $X$) and phase flip (gate $Z$). 

The quantum teleportation protocol is important for quantum computation since it provides a way to transform different types of resources: 1 qubit = 2 bits + 1 entangled pair. It is also used as a part of error-correcting protocols and can be interpreted (with few modifications) as a simple example of a measurement-based quantum computation. 

\paragraph{Quantum parallelism}
We will now discuss a quantum circuit that evaluates the value of a function $f(x)$. We will illustrate another essential feature of quantum computation: quantum parallelism. 

We consider a binary function
\begin{align}
    f(x):\{0,1\}\rightarrow\{0,1\}.
\end{align}
There are four such functions: identity, bit flip, constant 0, and constant 1. The idea is to construct a unitary $U_{f}$ to implement these functions. It can not be a one-qubit unitary since the constant functions can not be uniquely inverted. Hence we define the unitary $U_f$ as
\begin{align}
    (x,y)\xrightarrow{U_f}(x,y\oplus f(x)). 
\end{align}
The $\oplus $ denotes $\mod 2$ addition, namely $0\oplus 0 = 0$, $0\oplus 1 = 1$, $1\oplus 0 = 1$, $1\oplus 1 = 0$. In the dirac notation we write $U_f\ket{x,y}=\ket{x,y\oplus f(x)}$. That this is a unitary transformation, it is not difficult to see since 
\begin{align}
    U_f^2\ket{x,y}=U_f\ket{x,y\oplus f(x)}=\ket{x,(y\oplus f(x)\oplus f(x))}=\ket{x,y}.
\end{align}
Applying the operation on a state $\ket{x,0}$ we get $U_f\ket{x,0}=\ket{x,f(x)}$. If we now prepare the first qubit in the state $\frac{1}{\sqrt{2}}(\ket{0}+\ket{1})$ and apply the unitary transformation we get
\begin{align}
    U_f \frac{1}{\sqrt{2}}(\ket{0}+\ket{1})\ket{0} = \frac{1}{\sqrt{2}}(U_f\ket{0,0}+U_f\ket{1,0})=\frac{1}{\sqrt{2}}(\ket{0,f(0)}+\ket{1,f(1)}).
    \label{eq: parallelism}
\end{align}
One application of $U_f$ produced the results of $f(0)$ and $f(1)$ in two orthogonal states. However, it is important to note that we can not access this information immediately since the results of measurements are probabilistic, and we can not choose which state of the first qubit to observe. 

\begin{figure}[!htb]
    \centering
    \includegraphics[width=0.6\textwidth]{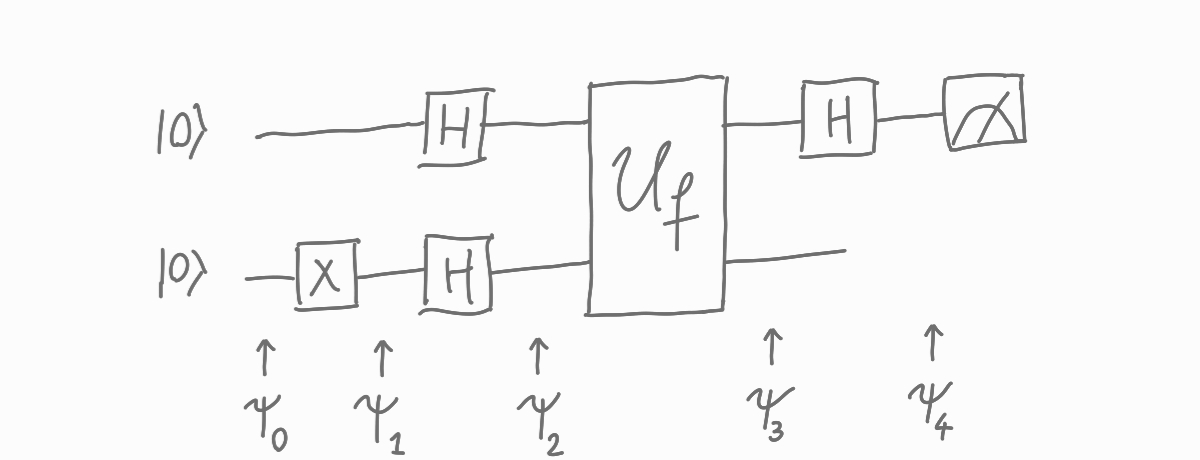}
    \caption{Diagramatic representation of the Deutsch algorithm to determine if the function is balanced or constant.}
    \label{fig: deutsch}
\end{figure}

\paragraph{Deutsch algorithm}
This section will show how we can extract useful information from a quantum state of the form \eref{eq: parallelism}. We consider a problem to determine if the function $f$ is constant or balanced. A constant function has $f(0)=f(1)$, and a balanced has $f(0)\neq f(1)$. We need two applications of $f$ to determine if $f$ is constant or balanced. A quantum algorithm (called Deutsch algorithm) shown in \fref{fig: deutsch} can do that with only one application of $U_f$, namely
\begin{align}
    (H\otimes \mathds{1})U_f(H\otimes H) (\mathds{1}\otimes X)\ket{0,0}.
\end{align}

The transformations of the Deutsch algorithm are as follows (see \fref{fig: deutsch})
\begin{align}
\ket{\psi_0}& = \ket{0,0} \\
\ket{\psi_1}& = (\mathds{1}\otimes X)\ket{\psi_0} &=& \ket{0,1} \\
\ket{\psi_2}& = (H\otimes H)\ket{\psi_1} &=& \frac{1}{2}(\ket{0}+\ket{1})(\ket{0}-\ket{1}) =\frac{1}{2}\left(\sum_{x=0,1}\ket{x}\right)\left(\ket{0}-\ket{1}\right)\\
\ket{\psi_3}& = U_f\ket{\psi_2} &=& \frac{1}{2}\left(\sum_{x=0,1}(-1)^{f(x)}\ket{x}\right)(\ket{0}-\ket{1}) \\
\ket{\psi_4}& = (H\otimes \mathds{1})\ket{\psi_3} &=&  \frac{1}{2}\left(\left((-1)^{f(0)}+(-1)^{f(1)}\right)\ket{0}+\left((-1)^{f(0)}-(-1)^{f(1)}\right)\ket{1}\right)(\ket{0}-\ket{1})
\end{align}
If we now measure the first qubit and observe the state $\ket{0}$, we know that $f(0)=f(1)$, and if we observe the state $\ket{1}$, we know that $f(0)\neq f(1)$. Hence, we evaluated a global property of a function without explicitly calculating its values. 

\begin{figure}[!htb]
    \centering
    \includegraphics[width=0.6\textwidth]{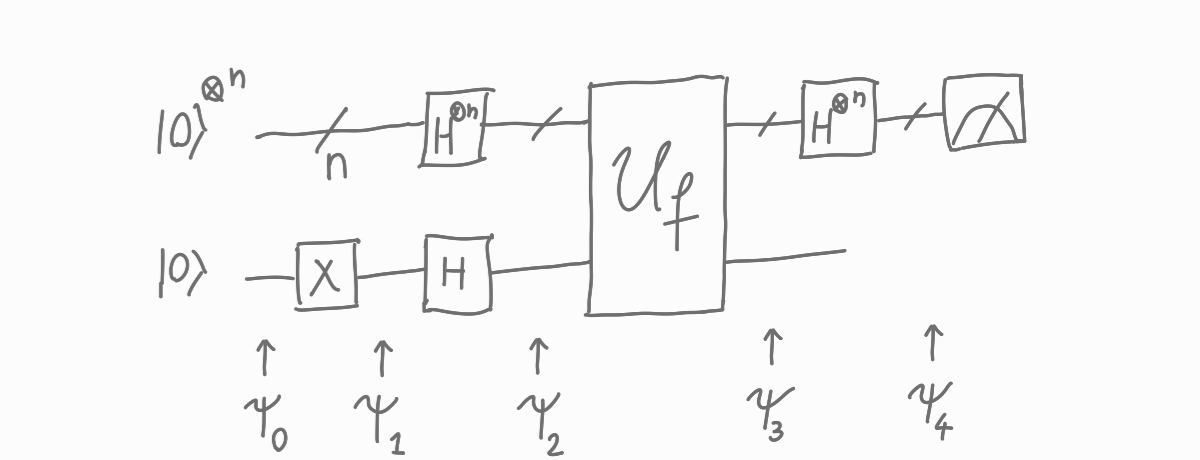}
    \caption{Diagramatic representation of the Deutsch-Jozsa algorithm to determine if a function $f$ is balanced or constant.}
    \label{fig: deutsch-josza}
\end{figure}

\paragraph{Deutsch-Josza algorithm}
We can generalise the Deutsch algorithm to $n$-qubits. In this case, we consider a function 
\begin{align}
    f(x): \{0,1\}^n\rightarrow {0,1}.
\end{align}
The function $f$ is constant if all $2^n$ inputs have the same result, and it is balanced if half of the inputs result in 1 and the other half in 0. To evaluate whether the function is balanced or constant classically, we need to evaluate on average $\mathcal{O}(2^N)$ different inputs. The quantum Deutsch-Josza algorithm shown in \fref{fig: deutsch-josza} does this with one application of $U_f$ defined as
\begin{align}
    U_f\ket{x,z}=\ket{x,z\oplus f(x)},\quad x=(x_1,x_2\ldots x_n), \quad z\in\{0,1\}.
\end{align}

Also, the algorithm is a straightforward generalisation of the Deutsch algorithm to $n$ qubits (plus one auxiliary qubit).
\begin{align}
    (\mathds{H}^{\otimes n}\otimes\mathds{1}_2)U_f(\mathds{H}^{\otimes n}\otimes H) (\mathds{1}_{2^n}\otimes X)\ket{0}^{\otimes n}\ket{0}.
\end{align}
Similarly, as in the one-qubit case, we write the state after each transformation
\begin{align}
\ket{\psi_0}& = \ket{0^{\otimes n},0} \\
\ket{\psi_1}& = (\mathds{1}_{2^n}\otimes X)\ket{\psi_0} &=& \ket{0^{\otimes n},1} \\
\ket{\psi_2}& = (H^{\otimes n}\otimes H)\ket{\psi_1} &=& \frac{1}{\sqrt{2^{n+1}}}\left(\sum_{s_1=0,1}\ket{s_1}\right)\left(\sum_{s_2=0,1}\ket{s_2}\right)\ldots \left(\sum_{s_n=0,1}\ket{s_n}\right)(\ket{0}-\ket{1})\\
&&=&\frac{1}{\sqrt{2^{n+1}}}\left(\sum_{x=0}^{2^n-1}\ket{x}\right)(\ket{0}-\ket{1})\\
\ket{\psi_3}& = U_f\ket{\psi_2} &=& \frac{1}{\sqrt{2^{n+1}}}\left(\sum_{x=0}^{2^n-1}(-1)^{f(x)}\ket{x}\right)(\ket{0}-\ket{1}) \\
\ket{\psi_4}& = (H^{\otimes n}\otimes \mathds{1})\ket{\psi_3} &=&  \frac{1}{\sqrt{2^{n+1}}}\left(\sum_{x,z=0}^{2^n-1}(-1)^{x\cdot z + f(x)}\ket{z}\right)(\ket{0}-\ket{1}).
\end{align}
To find the last equality, we used the identity
\begin{align}
    H^{\otimes n}\ket{x_1,x_2,\ldots x_n}=\frac{1}{\sqrt{2^n}}\sum_{z=0}^{2^n-1}(-1)^{z_1x_1+z_2x_2+\ldots z_nx_n}\ket{z_1,z_2,\ldots z_n}.
\end{align}
The probability of observing a state corresponding to a constant function is 
\begin{align}
    p(0,0,\ldots 0)=\left|\frac{1}{2^n}\sum_{x=0}^{2^n-1} (-1)^{f(x)} \right|^2.
\end{align}
If the function is constant, the probability is one, and if the function is balanced, the probability is 0. The presented Deutsch-Jozsa algorithm is the simplest case where we can show exponential speedup with respect to the best-known classical algorithm. 

\paragraph{Categories of quantum algorithms}
The Deutsch-Jozsa solves a problem which is not practically relevant. However, similar problems are practically important and can be solved efficiently on a quantum computer but not on a classical computer. This is a class of problems related to the quantum Fourier transform, which we will discuss in the next section.

An entirely different class of algorithms are the quantum search algorithms. A prime example is Grover's algorithm. These general algorithms solve unstructured problems; we can not expect exponential speedup as in the Fourier transformation class. For example, Grover's search is quadratically better than the best classical version. However, we can make stronger theoretical claims. While the Fourier algorithms are compared to the best-known classical algorithms, we can prove that Grover's search is the best possible quantum algorithm, quadratically better than the best possible classical algorithm.

The last class of algorithms for fault-tolerant quantum devices are the quantum simulation algorithms. They also achieve exponential speedups with respect to the best-known classical algorithms but do not rely on the Fourier transformation. They are simply simulating a many-body quantum system. The exponential speedup comes from the fact that quantum simulation is classically challenging.

\begin{referencesbox}
\begin{itemize}
  \item Nielsen, M. A., \& Chuang, I. L. (2010). Quantum computation and quantum information. Cambridge University Press. (Chapter 1.3: Quantum computation, Chapter 1.4: Quantum algorithms, Chapter 4: Quantum circuits)
\end{itemize}
\end{referencesbox}

\subsection{Quantum Fourier transformation}
In this section, we will look at the quantum Fourier transformation (QFT) and then discuss a set of routines that can be used to speed up many classical machine learning algorithms.

\begin{figure}[!htb]
    \centering
    \includegraphics[width=\textwidth]{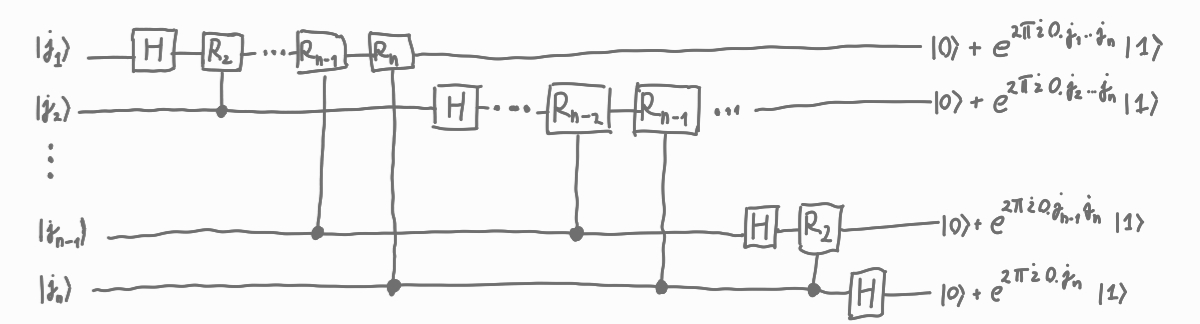}
    \caption{The main part of the QFT algorithm.}
    \label{fig: qft}
\end{figure}

\paragraph{Algorithm}
The QFT algorithm shown in \fref{fig: qft} is a quantum version of the classical discrete Fourier transformation. The classical discrete Fourier transformation maps a vector of complex numbers $x_1,x_2\ldots x_{2^n}$ to a new vector $y_1,y_2\ldots y_{2^n}$ defined as
\begin{align}
    y_k=\frac{1}{\sqrt{2^n}}\sum_{j=0}^{2^n-1}\ee^{\frac{2\pi\ii jk}{2^n}}x_j.
    \label{eq: discrete FT}
\end{align}
The quantum version is then defined by analogy. The basis state $\ket{j}$ is transformed as 
\begin{align}
    \ket{j}\xrightarrow{\rm QFT}\frac{1}{\sqrt{2^n}}\sum_{k=0}^{2^n-1}\ee^{\frac{2\pi\ii jk}{2^n}}\ket{k}.
    \label{eq: qft}
\end{align}
The action on an arbitrary state is then
\begin{align}
    \sum_{j=0}^{2^n-1}x_j\ket{j}\xrightarrow{\rm QFT}\sum_{k=0}^{2^n-1}y_k\ket{k},
    \label{eq: qftx}
\end{align}
where $y_k$ are the discrete Fourier transformed amplitudes defined in \eref{eq: discrete FT}. The presented quantum transformation is a unitary transformation, which we will show by explicitly constructing the circuit for its implementation. 

First, we define some useful conventions/notation, which we will use throughout the lecture notes. A natural number $j$ can be written in a binary representation $j=j_1j_2,\ldots j_n$, which more formally means $j=j_1 2^{n-1}+j_2 2^{n-2}+\ldots j_n 2^{0}$. Similarly, we can represent a binary fraction as $0.j_lj_{l+1}\ldots j_m=j_l/2+j_{l+1}/2^2+\ldots j_m/2^{m-l+1}$.

To obtain a useful representation for a QFT, we rewrite the definition \eref{eq: qft} as follows
\begin{align}
    \ket{j}\xrightarrow{\rm QFT}&\frac{1}{\sqrt{2^n}}\sum_{k=0}^{2^n-1}\ee^{\frac{2\pi\ii jk}{2^n}}\ket{k}\\
    =&\frac{1}{\sqrt{2^n}}\sum_{k_1=0}^1\ldots \sum_{k_n=0}^1 \ee^{2\pi\ii j\left(\sum_{l=1}^nk_l2^{-l}\right)}\ket{k_1,k_2,\ldots,k_n}\\
    =&\frac{1}{\sqrt{2^n}}\sum_{k_1=0}^1\ldots \sum_{k_n=0}^1 \prod_{l=1}^n\ee^{2\pi\ii j k_l2^{-l}}\ket{k_1,k_2,\ldots,k_n} \\
    =&\frac{1}{\sqrt{2^n}}\bigotimes_{l=1}^n\left[\sum_{k_l=0}^1 \ee^{2\pi\ii j k_l2^{-l}}\ket{k_l}\right] \\
    =&\frac{1}{\sqrt{2^n}}\bigotimes_{l=1}^n\left[\ket{0}+ \ee^{2\pi\ii j 2^{-l}}\ket{1}\right] \\
    =&\frac{1}{\sqrt{2^n}}(\ket{0}+\ee^{2\pi\ii 0.j_n}\ket{1})(\ket{0}+\ee^{2\pi\ii 0.j_{n-1}j_n}\ket{1})\ldots (\ket{0}+\ee^{2\pi\ii 0.j_1j_2\ldots j_n}\ket{1})
    \label{eq: product qft}
\end{align}
The final representation of the QFT \eref{eq: product qft} is convenient to derive an efficient circuit for its implementation presented in \fref{fig: qft}. The gate $R_k$ denotes a single qubit phase shift
\begin{align}
    R_{k} = \begin{pmatrix}1&0\\ 0&\ee^{2\pi\ii/2^k}\end{pmatrix}.
\end{align}

To show that the algorithm \fref{fig: qft} represents the QFT, we focus on the transformation of the first qubit of the state $\ket{j}=\ket{j_1,j_2,\ldots j_n}$ (top wire in \fref{fig: qft}). Application of the Hadamard on the first qubit produces the state
\begin{align}
    \frac{1}{\sqrt{2}}\left(\ket{0}+\ee^{2\pi\ii 0.j_1}\ket{1}\right)\ket{j_2\ldots j_n}.
\end{align}
Above follows from the fact that $\ee^{2\pi\ii0.j_1}=-1$ if $j_1=1$ and $\ee^{2\pi\ii0.j_1}=1$ if $j_1=0$. Then, we apply the $R_2$ gate and again shift the phase of the state $\ket{1}$ as
\begin{align}
    \frac{1}{\sqrt{2}}\left(\ket{0}+\ee^{2\pi\ii 0.j_1j_2}\ket{1}\right)\ket{j_2\ldots j_n}.
\end{align}
This follows directly from the definition of the $R_k$ gate. Then we apply the controlled rotations $R_3$,\ldots $R_{n}$ and get the state
\begin{align}
    \frac{1}{\sqrt{2}}\left(\ket{0}+\ee^{2\pi\ii 0.j_1j_2\ldots j_n}\ket{1}\right)\ket{j_2\ldots j_n}.
\end{align}
Transformations of other qubits follow the same pattern with a decreasing number of conditional rotations, resulting in fewer digits in the phase shift. Comparing the representation \eref{eq: product qft} of the QFT and the output of the circuit \fref{fig: qft}, we see that the outputs are precisely reversed. Therefore, we have to reverse the order of the states obtained by the presented algorithm by applying $n-1$ swap gates. 

The presented algorithm is efficient in the number of qubits since it requires $\mathcal{O}(n^2)$ simple two-qubit gates. In contrast, the classical algorithm needs to apply $\mathcal{O}(n2^n)$ elementary operations. The main caveat is that there is no known efficient way of extracting all the information stored in the amplitudes. Therefore, we use QFT as a part of algorithms to extract some global properties of functions, which would otherwise require the computation of their values for all possible inputs. A prime example is the Deutsch-Jozsa algorithm discussed in the previous section.

\begin{figure}[!htb]
    \centering
    \includegraphics[width=\textwidth]{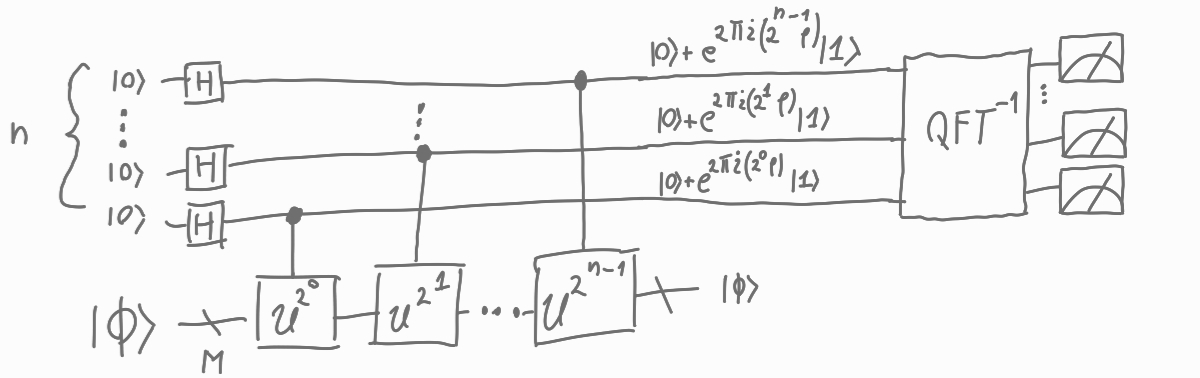}
    \caption{The circuit implementing the quantum phase estimation protocol}
    \label{fig: qpa}
\end{figure}

\paragraph{Quantum phase estimation}
One of the most used quantum procedures based on the QFT is the phase estimation protocol, which encodes the phase $\varphi$ of an amplitude $\alpha=|\alpha|\ee^{\ii\varphi}$ into a state of an auxiliary quantum system. Formally, we define the problem as follows. Given a unitary $U$ and an eigenstate $\ket{\phi}$ such that 
\begin{align}
    U\ket{\phi}=\ee^{\ii\varphi} \ket{\phi}
    \label{eq:qpa eigenstate}
\end{align} we want to estimate the phase $\varphi$. We show the protocol, which efficiently estimates the phase $\varphi$, in \fref{fig: qpa} and uses the inverse quantum Fourier transformation. 

Here, we will focus only on the first part, which consists of $n$ Hadamard transformations and $n$ controlled unitary transformations with exponentially increasing duration (or power). The first controlled unitary after the Hadamard transformation transforms the state to 
\begin{align}
    \frac{1}{\sqrt{2^n}}(\ket{0}+\ket{1})^{\otimes n-1} (\ket{0}+\ee^{2\pi\ii \varphi}\ket{1}).
\end{align}
We used that $\ket{\phi}$ is an eigenstate of $U$, i.e. \eref{eq:qpa eigenstate}. After the second controlled unitary, we have the state
\begin{align}
    \frac{1}{\sqrt{2^n}}(\ket{0}+\ket{1})^{\otimes n-2} (\ket{0}+\ee^{2\pi\ii 2^1 \varphi}\ket{1})(\ket{0}+\ee^{2\pi\ii \varphi}\ket{1}).
\end{align}
The rest of the controlled unitary transformations add the phases in an exponentially increasing fashion, such that the state before the application of the inverse QFT is 
\begin{align}
    \frac{1}{\sqrt{2^n}}(\ket{0}+\ee^{2\pi\ii 2^{n-1} \varphi}\ket{1})(\ket{0}+\ee^{2\pi\ii 2^{n-2} \varphi}\ket{1})\ldots(\ket{0}+\ee^{2\pi\ii \varphi}\ket{1})=\frac{1}{\sqrt{2^n}}\sum_{k=0}^{2^n-1}\ee^{2\pi\ii \varphi k}\ket{k}
    \label{eq: qpt middle}
\end{align}
Comparing \eref{eq: qftx} with the result of the Fourier transformation \eref{eq: qftx}, we observe that the obtained state \eref{eq: qftx} encodes a Fourier transformation of the phase $\varphi$. To obtain an approximation of the phase, we must apply an inverse Fourier transformation and read out the bits in the auxiliary system. If $\varphi=b=0.b_1\ldots b_n$ our final state will be $\ket{b}=\ket{b_1b_2\ldots b_n}$ with probability 1. 

To calculate the efficiency of the protocol in general, we have to estimate the probability of observing a state with an error larger than $e$. The state of the ancillary register after the inverse QFT is
\begin{align}
    \frac{1}{2^n}\sum_{k,l=0}^{2^n-1}\ee^{-2\pi\ii kl/2^n}\ee^{2\pi\ii \varphi k}\ket{l}.
\end{align}
Now consider the amplitude $\alpha_l$ of the state $\ket{(b+l)\mod 2^n}$
\begin{align}
    \alpha_l = \frac{1}{2^n}\sum_{k=0}^{2^n-1}\left(\ee^{2\pi\ii(\varphi-(b+l)/(2^n))}\right)^k.
    \label{eq:alphal}
\end{align}
The sum in \eref{eq:alphal} is a geometric series, hence 
\begin{align}
    \alpha_l= \frac{1}{2^n}\frac{1-\ee^{2\pi\ii (2^n\varphi -(b+l))}}{1-\ee^{2\pi\ii (\varphi-(b+l)/2^n)}} = \frac{1}{2^n}\frac{1-\ee^{2\pi\ii (2^n\delta - l)}}{1-\ee^{2\pi\ii (\delta-l/2^n)}},
\end{align}
where $\delta = \varphi-0.b_1b_2\ldots b_n$. We will estimate the probability of obtaining the value $m$ such that $|m-b|>e$, where $e$ is a positive integer characterising our desired tolerance to error. The probability of observing such an $m$ is given by 
\begin{align}
    p(|m-b|>e)=\sum_{-2^{n-1}<l\leq -(e+1)}|\alpha_l|^2+ \sum_{e+1\leq l\leq 2^{n-1}}|\alpha_l|^2
\end{align}
For any real $\theta$, we have $|1-\exp\ii\theta|\leq2$ so 
\begin{align}
    |\alpha_l|\leq\frac{1}{2^n}\frac{2}{|1-\ee^{2\pi\ii(\delta-l/2^n)}|}\leq\frac{1}{2^{n+1}|\delta -l/2^n|}. 
\end{align}
We obtain the last inequality using $|1-\exp(\ii \theta)|\geq 2\theta/\pi$. Combining the last two expressions, we get the bound 
\begin{align}
    p(|m-b|>e)\leq\frac{1}{4}\left[\sum_{l=-2^{n-1}+1}^{ -(e+1)}\frac{1}{(2^n\delta -l)^2}+ \sum_{l=e+1}^{2^{n-1}}\frac{1}{(2^n\delta -l)^2}\right]
\end{align}
Since $0<2^n\delta<1$ we have
\begin{align}
    p(|m-b|>e)&\leq \frac{1}{4}\left[\sum_{l=-2^{n-1}+1}^{ -(e+1)}\frac{1}{l^2}+ \sum_{l=e+1}^{2^{n-1}+}\frac{1}{(l-1)^2}\right]\\
    &\leq \frac{1}{2}\sum_{l=e}^{2^{n-1}-1}\frac{1}{l^2}\\ 
    &\leq \frac{1}{2}\int_{e-1}^{2^{n-1}-1}\dd l \frac{1}{l^2}\\
    &=\frac{1}{2(e-1)}.
\end{align}
Suppose we want to estimate $\varphi$ up to $2^{-t}$, which translates to $e=2^{n-t}-1$. The probability of getting such a state is at least $1-\frac{1}{2(2^{n-t}-2)}$. Thus, to successfully obtain $\varphi$ to $t$ bits with a probability of success at least $1-\epsilon$, we need to use $n=t+\left\lfloor \log(2+1/2\epsilon)\right\rfloor$ qubits. 

As noted above, many exponentially fast algorithms use the quantum phase estimation protocol. The most famous are order finding and factoring. The second is essential since most of the current internet security is based on the 2048 RSA key encryption, which can be decoded efficiently using the factoring algorithm. However, classical security protocols exist which can not be efficiently decrypted even with a quantum algorithm. A field studying these types of problems is post-quantum cryptography.

Besides the mentioned algorithms, the QPE can perform many linear algebra transformations efficiently. These can then accelerate many classical machine learning algorithms. 

\begin{figure}
    \centering
    \includegraphics[width=0.7\textwidth]{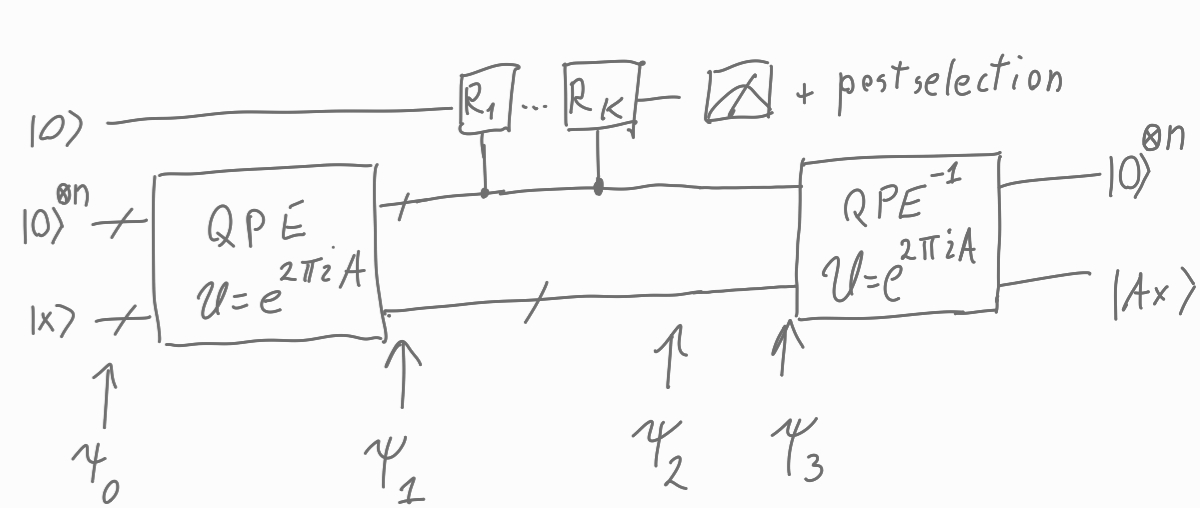}
    \caption{Matrix multiplication and matrix inversion protocols. The difference between the protocols is the choice of the rotations $R_1$\ldots $R_K$.}
    \label{fig: matrix multiplication}
\end{figure}

\paragraph{Matrix multiplication}
The first linear algebra routine we can speed up using the QPA protocol is vector-matrix multiplication. This protocol applies to Hermitian matrices $A$ with eigenvalues $\lambda_r$ smaller than one. Classically, we can write
\begin{align}
    Av_r=\lambda_rv_r,\quad x=\sum_{r=1}^R(v_r^{\rm T} x)v_r \Rightarrow Ax=\sum_{r=1}^R \lambda_r(v_r^{\rm T}x)v_r.
\end{align}
The quantum formulation of the problem is as follows. Given a quantum state $\ket{x}$, we want to calculate the state $\ket{Ax}$. More formally, we want to find the transformation
\begin{align}
    \ket{x}=\sum_{r=1}^R\braket{v_r}{x}\ket{v_r}\rightarrow \sum_{r=1}^{R}\lambda_r\braket{v_r}{x}\ket{v_r}.
\end{align}
The corresponding quantum matrix multiplication protocol is shown in \fref{fig: matrix multiplication}.
We start the protocol in a state $\ket{\psi_0}=\ket{0}\ket{0}^n\ket{x}$ and transform it with a quantum phase transformation protocol related to the unitary transformation $U=\ee^{2\pi\ii A}$. To determine the state $\ket{\psi_1}$ we recall the action of the QPA protocol on an eigenstate ${\rm QPA} \left(U=\ee^{2\pi\ii A}\right)\ket{0}^{\otimes n}\ket{v_r}  \sim \ket{\lambda_r}\ket{v_r}$. Therefore we have 
\begin{align}
    \ket{\psi_1} \sim \sum_{r=1}^R\ket{0}\ket{\lambda_r}\braket{v_r}{x}\ket{v_r}. 
\end{align}
In the next step, we use conditional rotations to rotate the first auxiliary qubit to a state $\sqrt{1-\lambda_r^2}\ket{0}+\lambda_r\ket{1}$. This can be done by applying $K=n$ conditional rotations, which successively move the state $\ket{0}$ to the state $\ket{1}$. For example, the $k$-th rotation would move the current state of the auxiliary qubit by $1/2^k$ if $q_k=1$, where $\lambda_r=0.q_1q_2,\ldots q_n$. After the rotations, we get the state
\begin{align}
    \ket{\psi_2} \sim \sum_{r=1}^R(\sqrt{1-\lambda_r^2}\ket{0}+\lambda_r\ket{1})\ket{\lambda_r}\braket{v_r}{x}\ket{v_r}. 
    \label{eq: mm rotations}
\end{align}
Next, we measure the first auxiliary qubit. If we observe the state $\ket{1}$, we proceed with the calculation; otherwise, we start from the beginning. This process is called post-selection. The successful state after the post-selection is
\begin{align}
    \ket{\psi_3} \sim \sum_{r=1}^R\lambda_r\ket{1}\ket{\lambda_r}\braket{v_r}{x}\ket{v_r}
\end{align}
This state is almost what we need to get. The problem is that the auxiliary qubits are entangled with the main qubits. We can disentangle them by using the inverse QPA algorithm. Finally, we get
\begin{align}
    \ket{1}\ket{0}^{\otimes n} \sum_{r=1}^R\lambda_r\braket{v_r}{x}\ket{v_r},
\end{align}
which is exactly the desired result.

\paragraph{Matrix inverse}
The matrix inverse differs from the matrix multiplication only in the rotation part. Instead of the transformation leading to \eref{eq: mm rotations}, we use the rotations that implement the "inverse" transformation
\begin{align}
    \ket{0}\ket{\lambda}\rightarrow (\sqrt{1-C/\lambda_r}\ket{0}+C/\lambda_r\ket{1}).
\end{align}
Since $|\lambda_r|<1$, we have to use a constant essentially determined by the smallest eigenvalue of $A$. 

The postselection procedure is a non-unitary operation which requires repeating the experiment $\mathcal{O}(1/p_{\rm acc})$ times. For matrix multiplication the acceptance probability is $p_{\rm acc} = \sum_{r}\lambda_r^2|\braket{v_r}{x}|^2$. For the matrix inversion the acceptance probability is $p_{\rm acc} = \sum_{r}\frac{C^2}{\lambda_r^2}|\braket{v_r}{x}|^2\leq \kappa^{-2}$, where $\kappa$ is the conditional number of the matrix. The quantum algorithm will use many tries to succeed for a badly conditioned matrix. We can circumvent this by using preconditioning techniques.

\paragraph{qBlas}
Besides matrix multiplication and matrix inversion, several important linear algebra routines have been translated to the quantum domain. Most important are the singular value decomposition and density matrix exponentiation. These routines have logarithmic requirements regarding data size and number of examples. However, they can typically be used only with subtle restrictions. We will put the limitations aside and focus on the machine learning implications. We can not immediately transform many machine learning algorithms into their quantum versions. We often have to reformulate them to a less efficient version on which we can then apply quantum linear algebra routines, such that the final quantum protocol still represents a significant speedup with respect to the original algorithm. Prominent machine learning algorithms that achieve a possible exponential speed-up are principal component analysis, support vector machines, cluster finding, topological data analysis, and Gaussian processes.

\begin{referencesbox}
\begin{itemize}
  \item Schuld, M., \& Petruccione, F. (2021). Machine learning with quantum computers (Vol. 676, pp. 163-169). Berlin: Springer. (Chapter: Fault-Tolerant Quantum Machine Learning )
  \item Nielsen, M. A., \& Chuang, I. L. (2010). Quantum computation and quantum information. Cambridge University Press. (Chapter 5.1: The quantum Fourier transform, Chapter 5.2: Phase estimation)
\end{itemize}
\end{referencesbox}

\subsection{Quantum search}
The second branch of algorithms, which achieves less dramatic but more rigorous/strict speedups, is the quantum search algorithm. Finding one or more examples in an unstructured database with $N$ samples can classically be solved in time $\mathcal{O}(N)$. Remarkably, a quantum algorithm (Grover's search) can do that in time $\mathcal{O}(\sqrt{N})$. 

\begin{figure}[!htb]
    \centering
    \includegraphics[width=0.7\textwidth]{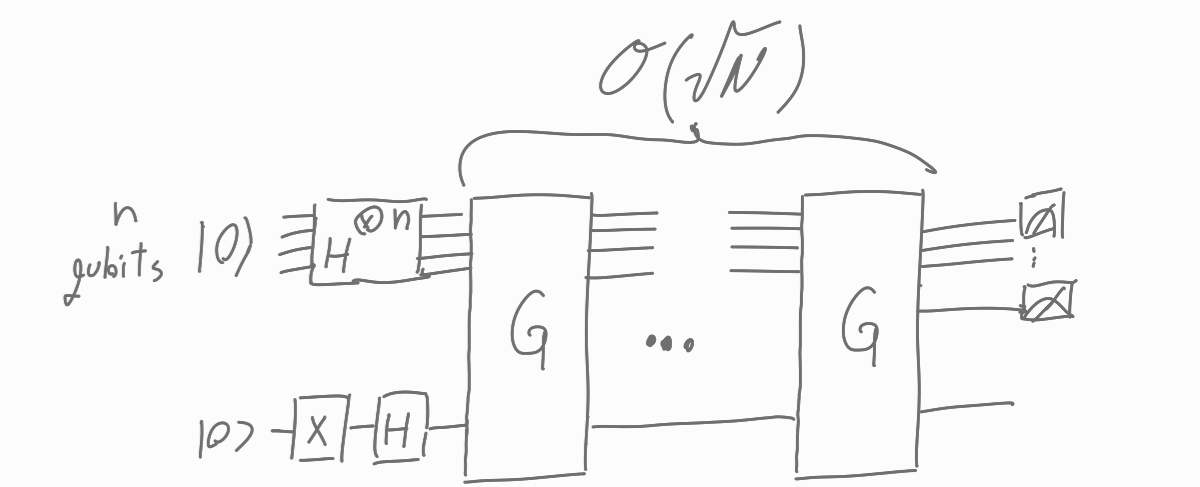}
    \caption{Circuit implementing the Grover search algorithm.}
    \label{fig: Grover}
\end{figure}

\begin{figure}[!htb]
    \centering
    \includegraphics[width=0.6\textwidth]{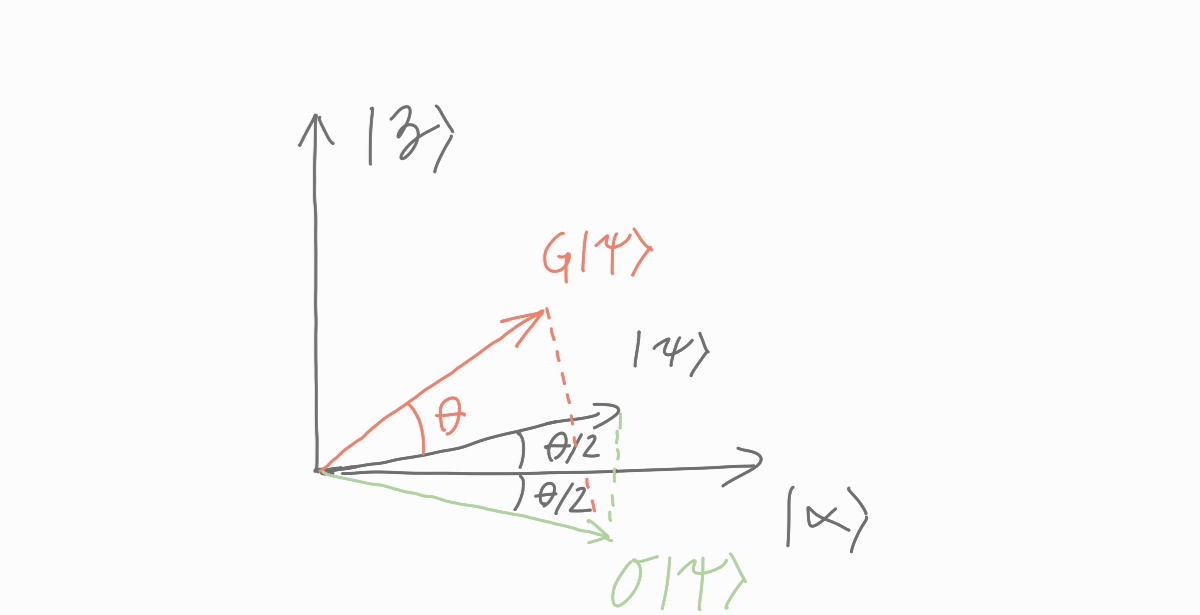}
    \caption{Geometric view of the Grover rotation.}
    \label{fig: Grover geometric}
\end{figure}

Suppose you have a quantum unitary/oracle $O$ implementing the classical function $f(x): \{0,1\}^{\otimes n}\rightarrow \{0,1\}$
\begin{align}
    \ket{x}\ket{y}\xrightarrow{O}\ket{x}\ket{y\oplus f(x)}.
    \label{eq: Grover oracle}
\end{align}
The task is to find an example $x$ for which $f(x)=1$. We can achieve this in $\mathcal{O}(\sqrt{N/M})$ steps by performing the algorithm presented in \fref{fig: Grover}.
The initial state of the algorithm is $\ket{\phi_0}=\ket{0}^n\ket{0}$. Like many quantum protocols, the algorithm starts with applying the $n$-fold tensor product of Hadamard gates on the first $n$ qubits and a bit flip, followed by a Hadamard gate on the last qubit. These transformations result in the state
\begin{align}
    \ket{\psi_1}=\frac{1}{\sqrt{2^n}}\sum_{x=0}^{2^n-1}\ket{x}\frac{1}{\sqrt{2}}(\ket{0}-\ket{1}).
\end{align}
Then we apply several Grover rotations $G=((2\ketbra{\psi}{\psi}-\mathds{1}_{2^n})\otimes{\mathds{1}_2})O$, where $\ket{\psi}=\frac{1}{\sqrt{2^{n+1}}}\sum_{x=0}^{2^n-1}\ket{x}$ and $O$ is the unitary defined in \eref{eq: Grover oracle}. We can understand the Grover rotation $G$ geometrically. First, we write the state $\ket{\psi}$ as 
\begin{align}
    \ket{\psi}&=\sqrt{\frac{N-M}{N}}\ket{\alpha}+\sqrt{\frac{M}{N}}
    \ket{\beta}\\ 
    &=\cos\frac{\theta}{2}\ket{\alpha}+\sin\frac{\theta}{2}\ket{\beta}.
\end{align}
The state $\ket{\alpha}$ is given by a superposition of $N-M$ states for which $f(x)=0$. Similarly, the state $\ket{\beta}$ represents a superposition of $M$ states for which $f(x)=1$. We now consider the action of the Grover rotation in the basis spanned by the vectors $\ket{\alpha}, \ket{\beta}$. The application of the oracle unitary $O$ results in a reflection around the $\alpha$ axis (see \fref{fig: Grover geometric}), namely \begin{align}
    O(a\ket{\alpha}+b\ket{\beta})\frac{1}{\sqrt{2}}(\ket{0}-\ket{1})=a\ket{\alpha}\frac{1}{\sqrt{2}}(\ket{0}-\ket{1})+b\ket{\beta}\frac{1}{\sqrt{2}}(\ket{1}-\ket{0}) = (a\ket{\alpha}-b\ket{\beta})\frac{1}{\sqrt{2}}\left(\ket{0}-\ket{1}\right).
\end{align}
Similarly the application of $2\ketbra{\psi}{\psi}-\mathds{1}_{2^n}$ results in a reflection around the state $\ket{\psi}$. The two actions do not leave the manifold spanned by the states $\ket{\alpha}$ and $\ket{\beta}$ and constitute a rotation. From \fref{fig: Grover geometric} it is clear that their combination implements a rotation
\begin{align}
    G\ket{\psi} = \cos\frac{3\theta}{2}\alpha +\sin\frac{3\theta}{2}\ket{\beta}. 
\end{align}
The state after $K$ application of $G$ is 
\begin{align}
    G^K\ket{\psi} = \cos\frac{2K+1}{2}\theta\ket{\alpha} + \sin\frac{2K+1}{2}\theta\ket{\beta}.
\end{align}
Finally, we measure the first $n$ qubits. If the angle $\frac{2K+1}{2}\theta$ is equal to $\pi/2$, we will likely observe one of the states in $\ket{\beta}$, which is the desired result. It follows that we have to perform $R\approx\left\lfloor \frac{\pi}{4}\sqrt{\frac{N}{M}}\right\rfloor$ operations to get the desired states with a high probability.

\begin{referencesbox}
\begin{itemize}
  \item Schuld, M., \& Petruccione, F. (2021). Machine learning with quantum computers (Vol. 676, pp. 163-169). Berlin: Springer. (Chapter: Quantum computing)
  \item Nielsen, M. A., \& Chuang, I. L. (2010). Quantum computation and quantum information. Cambridge University Press. (Chapter 6.1: The quantum search algorithm)
\end{itemize}
\end{referencesbox}

\subsection{Deterministic Quantum Computation with 1 Qubit} In the previous example, we used a pure quantum state of n qubits to achieve a quantum speedup. This is the case for most quantum algorithms. A notable exception is the deterministic Quantum Computation with One Clean Qubit (DQC1) model, introduced by Knill and Laflamme (1998). The DQC1 model of quantum computation challenges the conventional belief that quantum computation requires highly entangled pure states. 

\begin{figure}[!htb]
    \centering
    \includegraphics[width=0.5\linewidth]{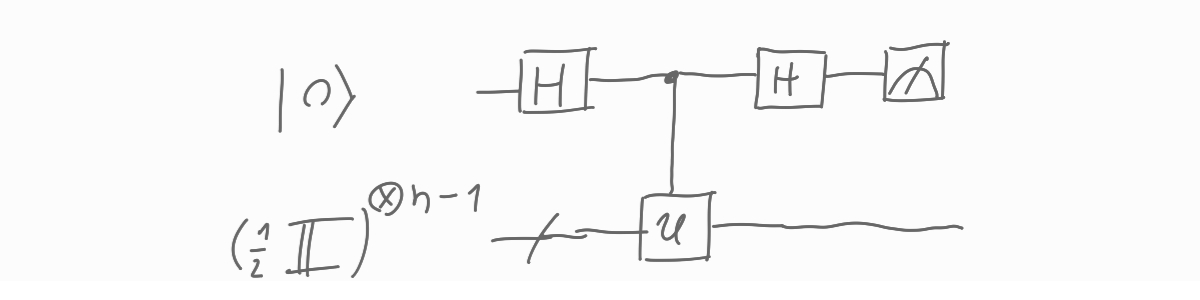}
    \caption{A circuit representation of the DQC1 computational model; a Hadamard test on a mixed state.}
    \label{fig: DQC1}
\end{figure}

The standard setup of the DQC1 model is shown in \fref{fig: DQC1} and consists of one pure qubit in a known quantum state $\ket{0}$ and $n-1$ qubits in a completely mixed state $\rho=\mathds{1}/2^{n-1}$. We then perform a unitary on all $n$ qubits and measure the expectation value of the observable $\sigma^{\rm z}$ on the first clean qubit. In the most basic example, the unitary implements the Hadamard test, which is given by Hadamard on the first qubit, followed by a controlled unitary (the first qubit is the control) and another Hadamard on the first qubit, namely
\begin{align}
    U=(H\otimes\mathds{1}_{n-1} )CU (H\otimes\mathds{1}_{n-1}),
\end{align}
where $CU$ is a controlled unitary, the first qubit is the control, and the unitary is applied to the rest of the $n-1$ qubits. 
To calculate the action of the unitary $U$ on a mixed state, we first calculate its action on a pure basis state and then invoke the linearity of the map. The action of the unitary $U$, also known as the Hadamard test, on a pure state $\ket{0}\ket{\psi}$ is 
\begin{align}
    U\ket{0}\ket{\psi} &= \frac{1}{\sqrt{2}}(H\otimes\mathds{1}_{n-1} )CU (\ket{0}+\ket{1})\ket{\psi}\\ \nonumber
    &=\frac{1}{\sqrt{2}}\left( (H\otimes\mathds{1}_{n-1})(\ket{0}\ket{\psi}+\ket{1}U\ket{\psi}) \right) \\ \nonumber
    &=\frac{1}{2}\left((\ket{0}+\ket{1})\ket{\psi}+(\ket{0}-\ket{1})U\ket{\psi} \right) \\ \nonumber   
    &=\frac{1}{2}\left(\ket{0}(1+U)\ket{\psi}+\ket{1}(1-U)\ket{\psi} \right) \\ \nonumber    
\end{align}
The probabilities $P_j$ of observing the first qubit in state $\ket{j}$ for $j=0,1$ are 
\begin{align}
    P_0 &= \frac{1}{2}(1+{\rm Re}(\bra{\psi}U\ket{\psi})), \\ \nonumber
    P_1 &= \frac{1}{2}(1-{\rm Re}(\bra{\psi}U\ket{\psi})).
\end{align}
The expectation value $\average{\sigma^{\rm z}}$ in the Hadamard test with a pure initial state is then 
\begin{align}
    \average{\sigma^{\rm z}} = {\rm Re}(\bra{\psi}U\ket{\psi})
\end{align}
By linearity, we have in the case of the DQC1 circuit (a Hadamard test of a completely mixed initial state)
\begin{align}
    \average{\sigma^{\rm z}} = \frac{1}{2^{n-1}}\sum_{k=0}^{2^{n-1}-1}{\rm Re}(\bra{k}U\ket{k}) = \frac{1}{2^{n-1}} {\rm Re}(\tr{U}).
\end{align}
Since we can calculate the trace of an exponentially large matrix with only a few gates, we have demonstrated that a quantum computer with only one "clean" qubit (fully quantum-coherent) and all others in maximally mixed states can still efficiently solve problems believed to be classically intractable. DQC1 thus serves as a key model for studying the boundary between classical and quantum computation, offering insights into how minimal quantum resources can still yield a computational advantage.

\begin{referencesbox}
\begin{itemize}
  \item Knill, E., \& Laflamme, R. (1998). Power of one bit of quantum information. Physical Review Letters, 81(25), 5672.
\end{itemize}
\end{referencesbox}


\subsection{Adiabatic quantum computation}
Adiabatic Quantum Computation (AQC) is a distinct quantum computing model that differs from the circuit model that we have discussed so far by evolving a quantum system from an initial Hamiltonian with an easily prepared ground state to a final Hamiltonian whose ground state encodes the solution. The adiabatic theorem ensures the system remains in its ground state if the evolution is slow enough. AQC has strong ties to condensed matter physics, computational complexity, and heuristic algorithms.  

The concept originated in the late 1980s as quantum stochastic optimisation, later renamed quantum annealing, which was inspired by simulated annealing. Early research suggested quantum annealing could outperform classical methods, leading to experimental implementations and the development of the quantum adiabatic algorithm. While optimisation problems typically use stoquastic Hamiltonians, non-stoquastic AQC is as powerful as the standard circuit-based quantum model, making AQC a general framework beyond optimisation.

\paragraph{Quantum Adiabatic Theorem}
In this section, we present an elementary argument for the quantum adiabatic theorem, which serves as the cornerstone of AQC.

Consider a time-dependent Hamiltonian $H(t)$ with an initial state $\ket{\psi(0)} = \ket{\phi_k(0)}$, where $\ket{\phi_k(0)}$ is an instantaneous eigenstate of the Hamiltonian at time $t = 0$. The adiabatic theorem states that if the Hamiltonian evolves slowly compared to the energy gaps, the time-evolved state remains approximately proportional to the instantaneous eigenstate. More precisely, the deviation satisfies  
\begin{align}
    \lVert \ket{\psi(t)} - \ket{\phi_k(t)} \rVert_2 = \mathcal{O}\left(\frac{1}{T}\right),\quad 0<t<T,
\end{align}
where $T$ is the annealing time.

To formalise this, we define the instantaneous eigenstates of the time-dependent Hamiltonian:
\begin{align}
    \label{eq:instant_eigenstates}
    H(t) \ket{\psi_l(t)} = E_l(t) \ket{\psi_l(t)}.
\end{align}
Expanding the state $\ket{\psi(t)}$ in terms of these eigenstates, we write
\begin{align}
    \ket{\psi(t)} = \sum_k c_k(t) \ket{\psi_k(t)}.
\end{align}
Substituting this into the Schrödinger equation
\begin{align}
    \ii\partial_t \ket{\psi(t)} = H(t) \ket{\psi(t)},
\end{align}
we obtain
\begin{align}
    \ii \sum_k \dot{c}_k(t) \ket{\psi_k(t)} + c_k(t)\ket{\dot{\psi}_k(t)} = \sum_k c_k(t) E_k(t) \ket{\psi_k(t)}.
\end{align}
Projecting onto $\bra{\psi_l(t)}$, we derive
\begin{align}
    \ii \dot{c_l}(t) &= E_l(t) c_l(t) - i \sum_k \braket{\psi_l(t)}{\dot{\psi}_k(t)} c_k(t) \\ \nonumber
    &= \left[E_l(t) - i \braket{\psi_l(t)}{\dot{\psi}_l(t)}\right] c_l(t) - i \sum_{k \neq l} \braket{\psi_l(t)}{\dot{\psi}_k(t)} c_k(t).
\end{align}
Neglecting the last term under the adiabatic approximation, we arrive at
\begin{align}
    i \dot{c_l}(t) = \left[E_l(t) - i \braket{\psi_l(t)}{\dot{\psi}_l(t)}\right] c_l(t).
\end{align}
This differential equation has the solution
\begin{align}
    \label{eq:adiabatic_solution}
    c_l(t) = e^{-i \int_0^t \left[E_l(u) - i \braket{\psi_l(u)}{\dot{\psi}_l(u)}\right] \dd u} c_l(0).
\end{align}

Now, assume the system starts in the state defined by $c_l(0) = 1$ with all other coefficients initially zero. To justify neglecting the term in \eqref{eq:adiabatic_solution}, we differentiate the eigenvalue equation \eqref{eq:instant_eigenstates} with respect to time:
\begin{align}
    \dot{H} \ket{\psi_k} + H \ket{\dot{\psi}_k} = \dot{E}_k \ket{\psi_k} + E_k \ket{\dot{\psi}_k}.
\end{align}
Taking the inner product with $\bra{\psi_l}$ for $l \neq k$ yields
\begin{align}
    \bra{\psi_l} H \ket{\dot{\psi}_k} + E_k \braket{\psi_l}{\dot{\psi}_k} = E_k \braket{\psi_l}{\dot{\psi}_l}.
\end{align}
From this, we extract
\begin{align}
    \braket{\psi_l}{\dot{\psi}_l} = \frac{\bra{\psi_l} \dot{H} \ket{\psi_k}}{E_k - E_l} = \frac{\dot{H}_{lk}}{E_k - E_l}.
\end{align}
For a time-dependent Hamiltonian of the form
\begin{align}
    H(t) = H_0 + \frac{t}{T} V, \quad 0 < t < T,
\end{align}
where $H_0$ is the initial Hamiltonian, $V$ is an interaction term, and $T$ is the annealing time, we find
\begin{align}
    \label{eq:H_derivative}
    \dot{H}_{lk} = \frac{1}{T} V_{lk}.
\end{align}
Thus, if the change is sufficiently slow (large $T$), the neglected term in \eqref{eq:adiabatic_solution} remains small, ensuring that the time-evolved state remains close to the instantaneous eigenstate.

\begin{figure}
    \centering
    \includegraphics[width=0.8\textwidth]{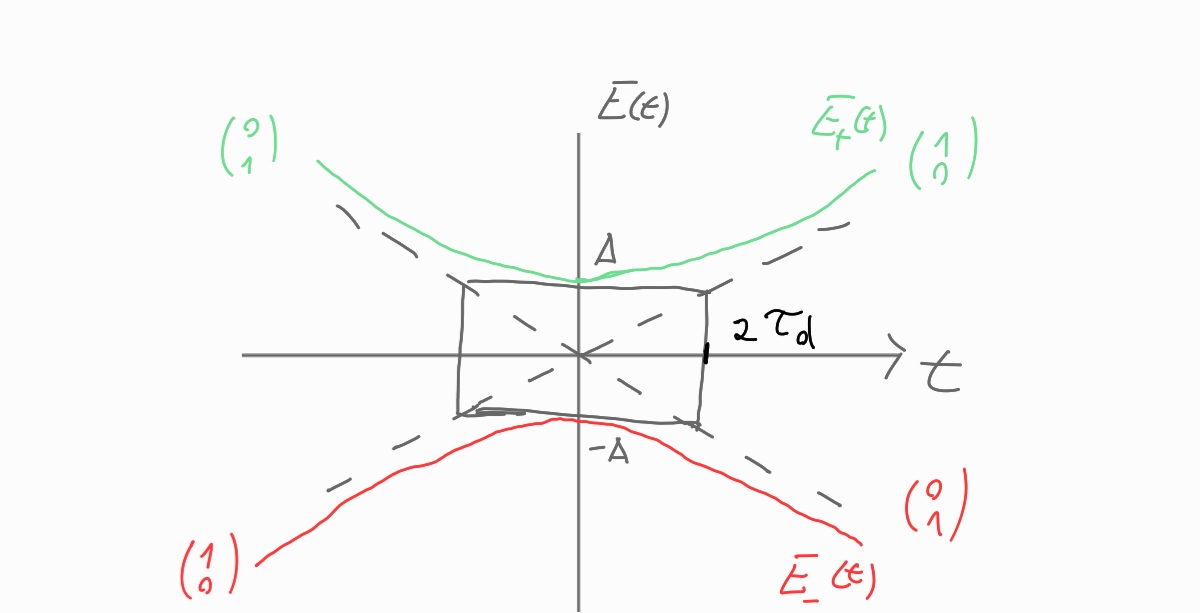}
    \caption{The Landau-Zenner energy diagram. The vectors at the ends of the energy lines $E_\pm$ correspond to asymptotic eigenvectors of the Hamiltonian $H(t)$ at times $t=\pm \infty$.}
    \label{fig: LZ}
\end{figure}

\begin{shaded}
\begin{example}
Consider the time-dependent Hamiltonian:
\begin{align}
H(t) = \begin{bmatrix} \frac{\alpha t}{2} & \Delta \\ \Delta & -\frac{\alpha t}{2} \end{bmatrix},
\end{align}
where $ \alpha $ is the sweep rate and $ \Delta $ is the coupling between the two states. The instantaneous eigenvalues (energy levels) of the Hamiltonian are given by:
\begin{align}
E_{\pm}(t) = \sqrt{\Delta^2 + \frac{\alpha^2 t^2}{4}}.
\end{align}

Figure~\ref{fig: LZ} illustrates these instantaneous energy levels and the relevant time scales where the eigenvector rotation occurs. The characteristic time scale 
\begin{align}
\tau_d = \frac{\Delta}{\alpha}
\end{align}
determines the duration over which the eigenvector transformation takes place. For an adiabatic transition, this time scale must be much larger than the inverse of the energy gap $ \left( \frac{1}{2\Delta} \right) $, leading to the adiabatic condition:
\begin{align}
\frac{\Delta^2}{\alpha} \gg 1.
\end{align}

We now focus on the probability of a \textit{non-adiabatic} transition from the ground state to the excited state. While the full derivation is lengthy, we state the well-known result. The probability of a non-adiabatic transition is given by:
\begin{align}
P_{\text{n.ad.}} = \exp\left(- 2 \pi  \frac{\Delta^2}{\alpha}\right).
\end{align}
This result is consistent with our heuristic adiabatic condition: as the time scale $ \tau_d $ increases while keeping the energy gap $ 2\Delta $ constant, the transition probability decreases exponentially, ensuring adiabatic evolution.
\end{example}
\end{shaded}





\begin{referencesbox}
\begin{itemize}
  \item MITOpenCourseWare: \href{https://ocw.mit.edu/courses/8-06-quantum-physics-iii-spring-2018/95ffb7405e75bec4f75564318e173ce0\_MIT8\_06S18ch6.pdf}{Chapter 6: Adiabatic approximation}
\end{itemize}
\end{referencesbox}

\section{Quantum computing with data}
The previous section covered essential quantum algorithms applicable to universal fault-tolerant quantum devices with at least $10^6$ qubits. The best current quantum devices are noisy and have around a few hundred qubits. Hence, the name Noisy Intermediate Scale Quantum (NISQ) devices. It is also unclear how fast we can scale up quantum devices and make them fault-tolerant. Therefore, most of the current research is focused on algorithms that apply to NISQ devices and offer significant benefits with respect to the best classical algorithms. A new hybrid, variational, quantum-classical algorithm is emerging in this respect. The limitations of NISQ devices also force us to consider the data encoding schemes, i.e., how to encode classical information on a quantum computer. 

This section will first discuss the data encoding strategies and then introduce the variational quantum circuits as the leading candidates for a practical algorithm on NISQ devices. 

\subsection{Data encoding}
\label{sec: embeddings}
We divide the encoding strategies into four groups: \textit{basis encoding}, \textit{amplitude encoding}, \textit{qusample encoding}, and \textit{Hamiltonian encoding}. Each of the strategies applies to certain types of problems. They also differ in storage capacity and computational complexity. Table \ref{tab: encoding complexity} shows an overview of the runtime complexity of the various encoding schemes discussed in the following.
\begin{table}[!htb]
    \centering
    \begin{tabular}{c|c|c|c}
        Encoding & Number of qubits & Runtime of state prep & Input features  \\
         \hline
        Basis & $N$& $\mathcal{O}(MN)$ & Binary\\
        Amplitude & $\log N$& $\mathcal{O}(MN)/\mathcal{O}(\log(MN))$ & Continuous\\
        Qusample & $N$& $\mathcal{O}(2^N)/\mathcal{O}(N)$ & Binary \\
        Hamiltonian & $\log N$& $\mathcal{O}(MN)/\mathcal{O}(\log(MN))$ & Continuous\\
    \end{tabular}
    \caption{Storage capacity and encoding runtime complexity of the encoding strategies for the dataset of $M$ samples with $N$ features. }
    \label{tab: encoding complexity}
\end{table}

\paragraph{Basis encoding}
Basis encoding is the backbone of many algorithms for universal quantum devices. Assume we have a dataset $\mathcal{D}$ with $M$ samples. Each sample is an $N$ dimensional binary vector $x^m=(b^m_1,b^m_2,\ldots,b^m_N)$, where $b^m_j\in\{0,1\}$, $j=1,2\ldots,N$, and $m=1,2,\ldots,M$. We can encode the dataset as a quantum superposition as
\begin{align}
    \ket{\mathcal{D}} = \frac{1}{\sqrt{M}}\sum_{m=1}^M\ket{b_1^m,b_2^m,\ldots, b_N^m}.
    \label{eq: basis encoding}
\end{align}
If the features are not binary, we must first convert them to binary codes and then use them as binary features.
\begin{shaded}
\begin{example}
    Assume we have a dataset $\mathcal{D}=\{x^1,x^2\}$ with vectors $x^1=(2,0)$ and $x^2=(1,3)$. We first convert the features to binary representations and then concatenate the representations of the first and the second feature of the vectors
    \begin{align}
        (2,0)\rightarrow((0,1),(0,0))\rightarrow (0,1,0,0) \\
        (1,3)\rightarrow((1,0),(1,1))\rightarrow (1,0,1,1)
    \end{align}
    We can now use the basis encoding for binary strings defined in \eref{eq: basis encoding}. The quantum state representing the two vectors is then
    \begin{align}
        \ket{\mathcal{D}}=\frac{1}{\sqrt{2}}(\ket{0,1,0,0}+\ket{1,0,1,1}).
    \end{align}
\end{example}
\end{shaded}
The benefit of encoding is that it is a simple preparation of single states since we have to apply only a local bit flip. Efficient preparation routines have also been developed for a superposition of states, i.e. a dataset. The main disadvantage of the basis encoding is that we need many qubits to represent features. For example, for one floating-point number, we need 16 qubits.

\paragraph{Amplitude encoding}
Also, amplitude encoding is important for many quantum algorithms. In particular, the algorithms that apply the qBLAS routines. Assume we have a dataset $\mathcal{D}$ with $M$ samples of $N$ dimensional real vectors $x^m=(x_1^m,x_2^m,\ldots,x_N^m)$, where $x^m_j\in \mathds{R}$, $j=1,2\ldots,N$, and $m=1,2,\ldots,M$. We encode the dataset $\mathcal{D}$ in a quantum state as
\begin{align}
    \mathcal{D}=\frac{1}{\sqrt{M}}\sum_{m=1}^M\sum_{j=1}^Nx^m_j\ket{j}\ket{m}=\frac{1}{\sqrt{M}}\sum_{m=1}^M\ket{\psi_{x^m}}\ket{m}.
    \label{eq: amplitude encoding}
\end{align}
This encoding needs $MN$ orthogonal vectors, which can be achieved with $\log MN$ qubits. Due to the normalisation of the quantum state, the states $\ket{\psi_{x^m}}$ and hence the vectors $x^m$ need to be $L_2$ normalised. We can overcome this restriction by adding a qubit that encodes the norm of the entire vector. Efficient, i.e. linear in $MN$, schemes have been developed to encode the classical dataset in a quantum state of the form \eref{eq: amplitude encoding}. The benefits of amplitude encoding are the efficient use of qubits and the fact that there is no restriction on the precision of the calculations. On the other hand, we can not easily access the information presented in the amplitudes. In general, we need exponentially many measurements (in the number of qubits) to determine the state $\ket{\mathcal{D}}$.

\paragraph{Qsample encoding}
Q-sample encoding is a mixture of basis and amplitude encoding. It encodes a classical probability distribution $(p_1,p_2,\ldots,p_N)$. The square-root probabilities $\sqrt{p_j}$ are encoded in an amplitude encoding of the basis vectors $\ket{j}$ represented in the basis encoding $j\rightarrow \ket{j}=\ket{j_1,j_2\ldots,j_{\lfloor\log N\rfloor}}$. We encountered this encoding in the first section, where we introduced the quantum probabilities as an interesting way to encode classical probability distributions. It has the benefit that the marginal distributions (reduced density matrices) encode the conditional probabilities. Hence, it is possible to reconstruct the joint probability by knowing only the marginal probabilities (see \exref{ex: probability reconstruction}). Further, it is trivial to obtain marginal probabilities and perform rejection sampling. 

\begin{shaded}
\begin{example}
    \textbf{Rejection sampling} is related to branch selection (or post-selection). In a branch selection, we have a state of the form
    \begin{align}
        \sum_{i=1}^N\sqrt{a_i}\ket{i}(\sqrt{1-\sqrt{b_i}}\ket{0}+\sqrt{b_i}\ket{1}).
    \end{align}
    We assume $a_i,b_i$ are real numbers in the unit interval $[0,1]$. By measuring the register $\ket{i}$ we sample from a distribution $(a_1,a_2,\ldots,a_N)$, but we want to sample from a distribution $(a_1b_1,a_2b_2,\ldots,a_Nb_N)$. We can do that by first sampling the auxiliary qubit. If we observe a state $\ket{0}$, we disregard the state, but if we observe $\ket{1}$, we continue with sampling the main register $\ket{i}$. The second sampling is then performed according to the desired distribution. The acceptance rate of the postselection is given by the square amplitude of the selected branch
    \begin{align}
        p_{\rm acc}=\sum_{i=1}^Nb_ia_i.
    \end{align}
    
    We now identify the distribution given by $\{a_j\}$ as the easy-to-sample distribution ($p_j$) and the distribution given by $\{a_jb_j\}$ as the complex distribution $q_j$. The described branch selection is analogous to the rejection sampling of $q_j$, where the samples are drawn from the probability $p_j$. 
\end{example}
\end{shaded}

\paragraph{Hamiltonian encoding}
Hamiltonian encoding determines the state implicitly by specifying the unitaries with which we transform the fixed initial state. Given a Hermitian matrix $H\in\mathcal{R}^{2^n\times2^n}$, we encode it as a unitary transformation determined by $U_H=\ee^{\ii H}$. If the matrix is unitary, we can consider it a quantum transformation. If a matrix $A$ does not have a special form, we can still apply the Hamiltonian encoding on a matrix
\begin{align}
    H=\begin{pmatrix}0&A\\A^\dag&0\end{pmatrix}.
\end{align}
To apply Hamiltonian encoding, we need to simulate the unitaries of the form $\ee^{\ii H}$. The problem of simulating such unitaries is called Hamiltonian simulation and can be done efficiently with local gates (see Solovay-Kitaev theorem in Section \ref{sec: gate model}). However, the best decomposition of a given unitary in terms of local gates can depend on a given implementation of the NISQ device.

\begin{referencesbox}
\begin{itemize}
  \item Schuld, M., \& Petruccione, F. (2021). Machine learning with quantum computers (Vol. 676, pp. 163-169). Berlin: Springer. (Chapter: Representing Data on a Quantum Computer)
\end{itemize}
\end{referencesbox}


\subsection{Block encoding and linear combination of unitaries} Block encoding is a powerful technique that unifies many quantum algorithms. It is based on two ideas:
\begin{itemize}
    \item Any matrix (apart from a normalisation) can be embedded into the top left part of a unitary matrix
    \item We can selectively apply a non-unitary operator to a state by postselecting on the desired outcome of an auxiliary qubit measurement.
\end{itemize}
We can concisely write the two ideas with the following equation
\begin{align}
    U = \begin{bmatrix}
    A/\alpha& \bullet\\ 
    \bullet & \bullet
    \end{bmatrix} \quad \Rightarrow \quad A = \alpha \left(\bra{0}\otimes \mathds{1}\right)U \left(\ket{0}\otimes \mathds{1}\right)
\end{align}
More precisely, we say that the $(s+a)$-qubit unitary $U$ is an $(\alpha,a,\varepsilon)$ block encoding of $A$ if
\begin{align}
    \lVert A- \alpha \left(\bra{0}\otimes \mathds{1}\right)U \left(\ket{0}\otimes \mathds{1}\right)\rVert_2\leq \varepsilon,
\end{align}
where $A$ acts on $s$ qubits and $U$ is implemented using additional $a$ qubits for selecting the correct block of the unitary matrix.

Since there are exponentially many different matrices, we can not efficiently block encode all possible $A$ in the number of qubits $s$. However, we can efficiently encode some interesting operators. 

Here, we will consider the block encoding of a linear combination of unitaries. To streamline the derivation of the block encoding, we note that the space of linear operators is itself a Hilbert space with a Hilbert-Schmidt inner product
\begin{align}
    \label{eq: HS product}
    \dbraket{A}{B} = \frac{1}{2^n}\tr A^\dag B,
\end{align}
where $A,~B$ are $2^n\times 2^n$ matrices, and the factor in front of the trace is chosen for easier normalisation of Pauli operators. If we vectorise the matrices, we find that the Hilbert-Schmidt inner product is the same (apart from the particular normalisation) as the inner product of vectorised matrices. We remark that the Pauli strings, i.e. the $4^n$ operators of the form $\sigma^{j_1}\otimes \sigma^{j_2} \otimes \ldots \otimes^{j_n}$, where $\sigma^0=\mathds{1},~\sigma^1=\sigma^{\rm x},~\sigma^2=\sigma^{\rm y},~\sigma^3=\sigma^{\rm z}$, are the orthonormal bases of operators acting on $\mathcal{C}^{2^n}$ (see \exref{ex: Pauli basis}). Therefore, any operator $A$ can be decomposed as a linear combination of Pauli strings,
\begin{align}
    \label{eq: Pauli decomposition}
    A = \sum_{k_1,k_2,\ldots k_n} \dbraket{\sigma^{k_1}\sigma^{k_2}\ldots \sigma^{k_n}}{A} \sigma^{k_1}\otimes \sigma^{k_2}\otimes\ldots\otimes \sigma^{k_n}
\end{align}
Since Pauli strings are also unitaries, we can express any matrix as a sum of unitaries. Therefore, block encoding of \texttt{linear combinations of unitaries} enables us to block encode any matrix. 

\paragraph{Linear combination of unitaries (LCU):} Assume we have $A=\sum_j \alpha_j U_j$ for $\alpha_j>0$ where the matrices $A, U_j$ act on the $2^s$ dimensional space. To block encode $A$, we need a way to specify the expansion coefficients $\alpha$ and the corresponding unitaries. We can do this by preparing and selecting circuits/oracles. We define the \texttt{Prepare} circuits acting on the $2^a$ dimensional auxiliary space as 
\begin{align}
    Prep\ket{\underline{0}}=\sum_i \frac{\sqrt{\alpha_i}}{\sqrt{\alpha}}\ket{i}, \quad \alpha=\sum_i\alpha_i.
\end{align}
Note that we only defined the action on the zero vector $\ket{\underline{0}}=\ket{0}\otimes\ket{0}\otimes\ldots\otimes\ket{0}$. Only this part will be important when constructing the block encoding. Therefore, we can choose the rest of the oracle/circuit to optimise the actual implementation on a quantum device. Notice that the $Prep$ circuit provides the information about the expansion coefficients. The second circuit, $Select$, provides the complementary information about the corresponding unitaries
\begin{align}
    Select = \sum_i\ket{i}\bra{i}\otimes U_i.
\end{align}
While $Prep$ acts only on the $2^a$--dimensional auxiliary space, the $Select$ circuits act on the full $2^{a+s}$--dimensional space. We can block encode the matrix $A$ with the following circuit and postselection 
\begin{align}
    BlockEncoding(A) & = (\bra{0}\otimes\mathds{1})Prep^\dag~Select~Prep (\ket{0}\otimes\mathds{1}) = \\ \nonumber
    &=\left(\sum_i\frac{\sqrt{\alpha_i}}{\sqrt{\alpha}}\bra{i}\otimes\mathds{1}\right)Select \left(\sum_j\frac{\sqrt{\alpha_j}}{\sqrt{\alpha}}\ket{j}\otimes\mathds{1}\right)\\ \nonumber
    &=\sum_{i,j}\frac{\sqrt{\alpha_i\alpha_j}}{\alpha}\left(\bra{i}\otimes\mathds{1}\right)\left(\sum_k \ket{k}\bra{k}\otimes U_k\right) \left(\ket{j}\otimes\mathds{1}\right)\\ \nonumber
    & = \sum_{i,j,k} \frac{\sqrt{\alpha_i\alpha_j}}{\alpha}\bra{i} \left(\ket{k}\bra{k}\right)\ket{j} \otimes U_k \\ \nonumber
    & = \frac{1}{\alpha}\sum_k \alpha_k U_k = A/\alpha
\end{align}
An explicit example of encoding a sum of unitaries is discussed in \eref{ex: lcu be}.
    
\begin{shaded}
\begin{example}
    \label{ex: Pauli basis} In this example, we show that the Pauli strings are an orthonormal basis. First we note that $\tr \sigma^{\alpha\geq 1}=0$ and $(\sigma^\alpha)^2=\mathds{I}$. In the case of single-qubit Pauli operators, we have 
    \begin{align}
        \dbraket{\sigma^i}{\sigma^j} = \delta_{i,j}.
    \end{align}
    Since the trace of the tensor product is the product of traces over the separate spaces, we have
    \begin{align}
        \dbraket{\sigma^{i_1}\otimes\sigma^{i_2}\otimes\ldots \otimes\sigma^{i_n}}{\sigma^{j_1}\otimes\sigma^{j_2}\otimes\ldots \otimes\sigma^{j_n}}&=\dbraket{\sigma^{i_1}}{\sigma^{j_1}}\dbraket{\sigma^{i_2}}{\sigma^{j_2}}\ldots \dbraket{\sigma^{i_n}}{\sigma^{j_n}} \\\nonumber
        &=\delta_{i_1,j_1}\delta_{i_2,j_2}\ldots \delta_{i_n,j_n}
    \end{align}
    The above equation is the definition of orthonormal operators. At most, $4^n$ linearly independent operators act on $\mathcal{C}^{2^n}$, which is exactly the number of Pauli strings with length $n$. Therefore, Pauli strings form an orthonormal basis according to the Hilbert-Schmidt inner product \eref{eq: HS product} 
\end{example}
\end{shaded}

\begin{shaded}
\begin{example}
    \label{ex: lcu be}
    Consider two unitaries $U, V$ that block encode $A$ and $B$, i.e., 
    \begin{align}
            U = \begin{bmatrix}
    A/\alpha& \bullet\\ 
    \bullet & \bullet
    \end{bmatrix},, \quad  V = \begin{bmatrix}
    B/\beta& \bullet\\ 
    \bullet & \bullet
    \end{bmatrix}.
    \end{align}
    Given the unitaries $U$ and $V$ we can block encode $A/\alpha + B/\beta$ with the following unitary 
    \begin{align}
        U_{A+B} = (H\otimes\mathds{1})(\ketbra{0}{0}\otimes U+\ketbra{1}{1}\otimes V)(H\otimes\mathds{1}) 
    \end{align}
    In the above case, the $Prep$ unitary is a simple Hadamard gate, and the $Select$ unitary is a sequence of two controlled unitary operations. Suppose we initialise the auxiliary qubit in the state $\ket{0}$ and that after applying the unitary $U_{A+B}$, we also observe the state $\ket{0}$. In this case, the top right corner of the applied transformation on the remaining qubits is described by the sum $(A/\alpha+B/\beta)/2$.
\end{example}
\end{shaded}

\begin{referencesbox}
\begin{itemize}
  \item Chakraborty, Shantanav, András Gilyén, and Stacey Jeffery. "The power of block-encoded matrix powers: improved regression techniques via faster Hamiltonian simulation." arXiv preprint arXiv:1804.01973 (2018).
  \item \href{https://lucidalu.github.io/Quantum-algorithms-A-survey-of-applications-and-end-to-end-complexities/quantum-algorithmic-primitives/quantum-linear-algebra/block-encodings/}{Quantum algorithms: block encoding} 
  \item Sünderhauf, C., Campbell, E., \& Camps, J. (2024). Block-encoding structured matrices for data input in quantum computing. Quantum, 8, 1226.
\end{itemize}
\end{referencesbox}

\subsection{Hybird quantum-classical algorithms}
\begin{figure}[!htb]
    \centering
    \includegraphics[width=0.8\textwidth]{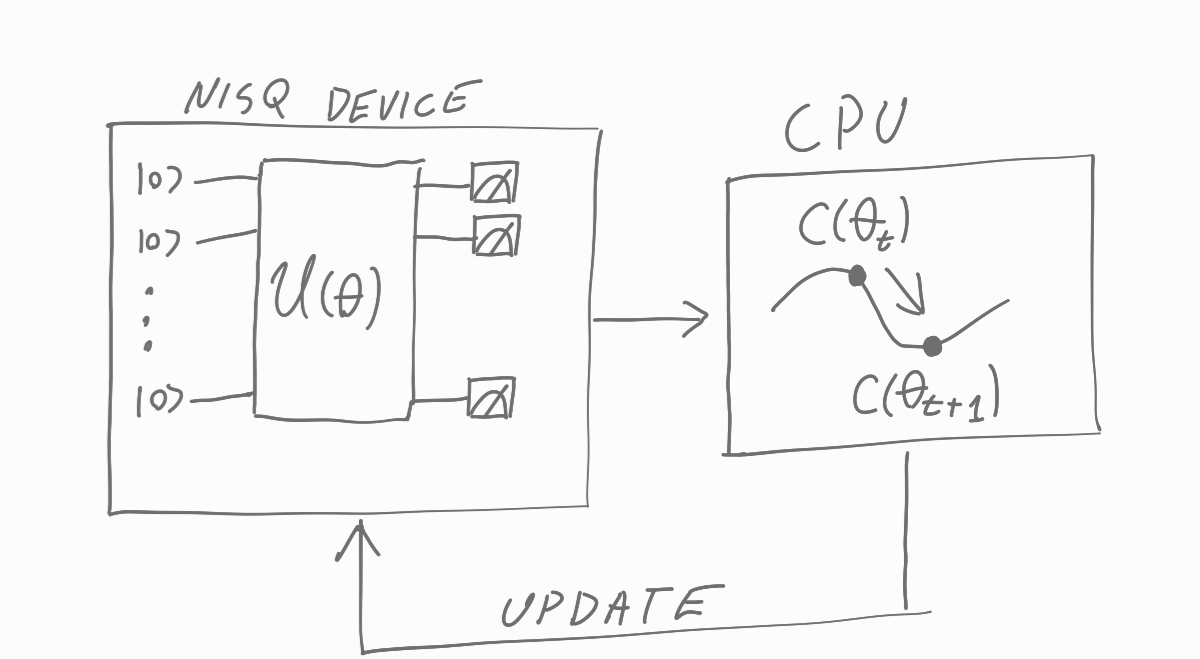}
    \caption{The figure shows a schematic representation of the variational, quantum-classical algorithm.}
    \label{fig: variational circuit}
\end{figure}

The most important algorithms on the available NISQ devices are variational, quantum-classical algorithms. The main idea is to perform only a small but essential part of the entire calculation on the quantum computer. This is done by formulating our problem as a minimisation problem of a certain cost function $C$. The cost function should then depend on the output of the quantum circuit, which we can optimise by modifying some parameters $\theta$. After we obtain the output of the quantum device, we proceed on the classical device by calculating the cost function and the updates of the variational parameters $\theta$. Then, we update the parameters of the quantum circuit and repeat the cycle until our objective converges. We show the described idea in \fref{fig: variational circuit}

The variational quantum-classical computation paradigm already includes several important algorithms, e.g., quantum approximate optimisation algorithm, variational eigensolver and variational classifiers.

\paragraph{Quantum approximate optimisation algorithm}
The quantum approximate optimisation algorithm (QAOA) is inspired by adiabatic quantum computation. Formally, the adiabatic path is determined by a unitary
\begin{align}
    U(t)={\rm T}_{\leftarrow} \ee^{\ii\int_0^t\dd t' H(t')},
    \label{eq: adiabatic solution}
\end{align}
where the notation ${\rm T}_{\leftarrow}$ denotes a time-ordered exponential. The adiabatic Hamiltonian is typically given as
\begin{align}
    H(t)=(1-\lambda(t))H_0+\lambda(t)H_1,
\end{align}
where $\lambda(t)$ determines the adiabatic schedule and is initially 0 and at the end of the calculation 1. The solution to our problem is then encoded in the ground state of the Hamiltonian $H_1$. 

To simplify the formal solution \eref{eq: adiabatic solution}, we use the Baker-Campbell-Hausdorff formula
\begin{align}
    \ee^{t (X+Y) = \ee^{tX}\ee^{tY}\ee^{-\frac{t}{2}[X,Y]}}\ldots=\ee^{tX}\ee^{tY} +\mathcal{O}(t^2).
\end{align}
A consequence of the above formula is the Trotter-Suzuki decomposition
\begin{align}
    \ee^{X+Y}=\lim_{n\rightarrow\infty}\left(\ee^{A/n}\ee^{B/n}\right)^n.
    \label{eq: bch}
\end{align}
By applying the truncated (finite $n$) Trotter-Suzuki formula, we can approximate the formal adiabatic path as
\begin{align}
    U(T)&\approx U(T,T-\Delta t)U(T-\Delta t,T-2\Delta t)\ldots U(\Delta t,0)\\
    &\approx\prod_{j=1}^n\ee^{-\ii H(j\Delta t)\Delta t}
\end{align}
On a NISQ device, the unitary $\ee^{\ii H(t)\Delta t}$ is also difficult to implement. Therefore, we split it into two parts by using the \eref{eq: bch} as $\ee^{\ii H(t)}=\ee^{\ii \lambda(t)H_1\Delta t}\ee^{\ii(1-\lambda(t))H_0\Delta t}$. Finally, we obtain an approximation of the adiabatic path
\begin{align}
    U(T)\approx \prod_{j=1}^n\ee^{\ii \lambda(j\Delta t)H_1\Delta t}\ee^{\ii(1-\lambda(j\Delta t))H_0\Delta t}
\end{align}
Due to the discussed practical limitation, we can not use a large $n$. Therefore, we use the following ansatz as an approximation 
\begin{align}
    \prod_{j=1}^n\ee^{\ii \lambda(t_j)H_1\Delta t_j}\ee^{\ii(1-\lambda(t_j))H_0\Delta t_j},
    \label{eq: qaoa}
\end{align}
where the times $t_j$ become now variational parameters. The parameters $t_j$ are then modified at each step of the variational optimisation, as discussed in the introduction of the section. 

The described quantum approximate optimisation algorithm can be applied on NISQ devices to any optimisation problem for which we can efficiently implement the unitaries in the decomposition \eref{eq: qaoa}. The QAOA can also be applied to non-optimisation problems, e.g. Gibbs sampling (see \exref{ex: Gibbs sampling}).

\begin{shaded}
\begin{example}
\label{ex: Gibbs sampling} In quantum computation, we want to avoid environmental interaction. However, there are situations where this interaction can be useful, e.g. in sampling from a Gibbs distribution $p(i)=\ee^{-E_i/T}/Z$, where $Z=\sum_i\ee^{-E_i/T}$ and $E_i$ are typically Ising energies. As we have seen, the Ising model has a quantum correspondence
\begin{align}
    H_1=-\sum_{i,j=1}^NJ_{i,j}\sigma^{\rm z}_i\sigma^{\rm z}_j-\sum_{i=1}^Nh_i\sigma^{\rm z}_i.
\end{align}
The quantum formulation of the Gibbs distribution is then $\rho_{T}=\ee^{-H_1/T}/Z$, where $Z=\tr{\ee^{-H_1/T}}$. 

We can prepare the state $\rho_T$ using a coherent algorithm. First, we add $N$ auxiliary qubits and pair them with the main qubits. Then, we rotate each of the pairs in an entangled state
\begin{align}
    \ket{\psi_{\rm pair}}\frac{1}{\sqrt{Z_{\rm pair}}}\left(\ee^{-\frac{1}{2T}}\ket{+}\ket{+} + \ee^{\frac{1}{2T}}\ket{-}\ket{-}\right),
\end{align}
where the states $\ket{\pm}$ represent the eigenstates of the $\sigma^{\rm x}$ Hamiltonian. Next, we trace out the auxiliary qubits. The Gibbs state then describes the state of the main qubits
\begin{align}
    \rho_0=\frac{1}{Z_0}\ee^{-H_0/T},
\end{align}
where $H_0=\sum_i\sigma^{\rm x}_i$. Finally, the desired Gibbs state can be obtained by using the QAOA. 

Alternatively, we can implement the interactions $H_1$ and wait for the system to equilibrate with the environment with temperature $T_{\rm environment}$. This will then automatically prepare the correct state. The caveat of the second procedure is that due to interference with the measurement devices, the environment temperature does not precisely match the desired $T$. Therefore, we must find efficient methods to estimate the difference between the desired and environmental temperatures.
\end{example}
\end{shaded}

\begin{shaded}
\begin{example} An NP-hard optimisation problem that the QAOA algorithm can solve is the max cut problem. Assume we have a graph defined by a list of edges $(i,j)\in\mathcal{E}$. We aim to separate it into two pieces by cutting as many edges as possible. This is the same as minimising the objective function 
\begin{align}
    C(z)=\sum_{(i,j)\in\mathcal{E}}(z_iz_j-1),
\end{align}
where $z_j\in{-1,1}$. The objective function can be exactly mapped to the Ising model. Therefore, we can immediately apply the QAOA.
\end{example}
\end{shaded}

\paragraph{Variational eigensolver}
The variational eigensolver is an algorithm that is mostly applied in quantum chemistry. The idea is to variationally find the lowest energy states of a given Hamiltonian $H$. This can be formulated as an optimisation problem, where the cost function is simply the energy of the Hamiltonian
\begin{align}
    C(\theta)=\frac{\bra{\psi(\theta)}H\ket{\psi(\theta)}}{\braket{\psi(\theta)}{\psi(\theta)}}.
    \label{eq: eigensolver objective}
\end{align}
The parameter values $\theta$ that minimise the objective function \eref{eq: eigensolver objective} determine the approximate eigenstate, and the value of the objective function is the approximate energy. We have to perform exponentially many measurements to calculate the expectation values of the Hamiltonian $H$. However, we can do that efficiently for a restricted but important class of local Hamiltonians. These Hamiltonians can be expressed as a sum of terms that act only on a few (typically 2) qubits, e.g. $H_{\rm local}=\sum_{\langle i,j\rangle} H_{i,j}$, where $H_{i,j}$ acts nontrivially only on qubits $i$ and $j$. 

\paragraph{Variational classifiers}
A generalisation of the variational eigensolver is the variational classifier. The idea is to interpret the expectation value of a variational circuit as the output of a classifier
\begin{align}
    f(x;\theta) = \bra{\psi(x;\theta)}O\ket{\psi(x;\theta)}.
\end{align}
For $M$ classes, the observable $O$ needs $M$ different eigenvalues determined by the projectors $\Lambda_j$. The logits of the variational classifier are then obtained as 
\begin{align}
    l_j=\bra{\psi(x;\theta)}\Lambda_j\ket{\psi(x;\theta)},\quad j=1,2,\ldots,M.
\end{align}
Once we have the logits $l_j$, we can utilise standard cost functions, e.g., the cross-entropy loss, as our optimisation objective.

\paragraph{ML with DQC1 computational model}
The hybrid quantum-classical machine learning model is not restricted to any specific type of quantum circuit. Therefore, we can also apply the same idea in the DQC1 computational paradigm, where the resource requirements are less stringent. The main idea is to represent our model as a trace of the unitary $f(x) = \frac{1}{2^n}\tr U(\boldsymbol{x},\boldsymbol{\theta})$, where $\boldsymbol{\theta}$ denotes a vector of trainable parameters and $\boldsymbol{x}$ a vector input. 

The circuit prepares the state 
\begin{align}
    \rho = \frac{1}{2^{n + 1}} \left[ I_{n+1} + \alpha \left( \ket{0}\bra{1} \otimes U^\dagger + \ket{1}\bra{0} \otimes U \right) \right]
\end{align}
By measuring the expectation values of $\sigma^{\rm x,y}$, we find the real and the imaginary part of the trace
\begin{align}
    \average{\sigma^{\rm x}} = \frac{\alpha}{2^n}{\mathrm Re}\tr U(\boldsymbol{x},\boldsymbol{\theta}),\quad \average{\sigma^{\rm y}} = \frac{\alpha}{2^n}{\mathrm Im}\tr U(\boldsymbol{x},\boldsymbol{\theta}).
\end{align}
Let us consider the unitary operator in the form 
\begin{equation}\label{eq:U}
U(\boldsymbol{x},\boldsymbol{\theta})=\prod_{l=1}^{L}W_l(\boldsymbol{\theta}_l)V_l(\boldsymbol{x}_{l}),
\end{equation}
where $\boldsymbol{\theta}_l$ and $\boldsymbol{x}_l$ represent subsets of the parameter vector $\boldsymbol{\theta}$ and the data vector $\boldsymbol{x}$, respectively, which are loaded at the $l$-th step. We write the trainable unitary $W_l(\boldsymbol{\theta}_l)$ as 
\begin{equation}
W_l(\boldsymbol{\theta}_l)=\prod_{k=1}^{k'}\exp(-i (\boldsymbol{\theta}_l)_k H_{lk})T_{lk},
\end{equation}
where $(\boldsymbol{\theta}_l)_k$ represents the $k$-th element of $\boldsymbol{\theta}_l$, $H_{lk}$ is an $2^n\times 2^n$ Hermitian operator and $T_{lk}$ is an unparametrized unitary. 

We further simplify the calculation and consider a special case where the trainable unitary has the simple form
\begin{equation}
    W_l(\boldsymbol{\theta}_l)=\exp(-i \sum_{k}(\boldsymbol{\theta}_l)_k \sigma^{\rm k}),
\end{equation}
and all $\sigma^k$ operators in the summation commute with each other. In this case, the partial derivative with respect to the variational parameter reduces to
\begin{equation}
    \frac{\partial f(\boldsymbol{x},\boldsymbol{\theta})}{\partial (\boldsymbol{\theta}_l)_k} = \frac{-i}{2^n}\mathrm{tr}\left(\sigma^k\prod_{j=l}^{L}W_{j}(\boldsymbol{\theta}_{j})V_{j}(\boldsymbol{x})\prod_{j=1}^{l-1} W_{j}(\boldsymbol{\theta}_{j})V_{j}(\boldsymbol{x}) \right).
\end{equation}
This enables us to calculate all derivatives with the same DQC1 computational paradigm and train the model via standard gradient descent training methods. 

The DQC1 computational model is not universal, but it still enables exponential speedup with respect to the best-known classical algorithms. Similarly, as we will see in the section about Fourier analysis of hybrid quantum-classical machine learning models, the DQC1 ML paradigm offers comparable expressivity to the standard circuit-based computational model.

\begin{referencesbox}
\begin{itemize}
  \item Schuld, M., \& Petruccione, F. (2021). Machine learning with quantum computers (Vol. 676, pp. 163-169). Berlin: Springer. (Variational Circuits as Machine Learning Models)
  \item \href{https://lucidalu.github.io/Quantum-algorithms-A-survey-of-applications-and-end-to-end-complexities/quantum-algorithmic-primitives/variational-quantum-algorithms/#further-reading}{Quantum algorithms: variational quantum algorithms} 
\end{itemize}
\end{referencesbox}

\subsection{Updating parametrised quantum circuits}
In the previous section, we discussed calculating the gradients in a restricted DQC1 computational setting. Here, we will show how we can measure gradients more broadly. 

The parameter update step is the most crucial in all variational quantum-classical algorithms. Broadly, we can split the parameter update methods into derivative-based and derivative-free methods. The latter can be further divided into iterative methods and one-shot methods. The most well-known iterative algorithms are the Nelder-Mead method and genetic algorithms. The one-shot methods rely on heuristic algorithms that solve the optimisation problem of determining the best parameters with a classical computer. 

The gradient-based methods can also be further divided into numeric and analytic variants. Numerically, we can determine the gradient by finite difference methods as 
\begin{align}
    \frac{\partial C(\theta)}{\partial \theta_j} = \frac{C(\theta_1,\ldots, \theta_j,\ldots,\theta_N)-C(\theta_1,\ldots, \theta_j+\Delta\theta_j,\ldots,\theta_N)}{\Delta\theta_j} +\mathcal{O}(\Delta\theta_j^2)+\mathcal{O}\left(\frac{\epsilon}{\Delta\theta_j}\right),
\end{align}
where $\epsilon$ denotes the probability of error in the circuit. The finite difference methods have limited precision due to the environmental noise.

Fortunately, we can calculate the derivative of the objective functions analytically with the same complexity as the objective function. The stochastic parameter shift rule is the general rule for calculating the analytic gradients.

\paragraph{Stochastic parameter shift rule} We will now introduce the procedure for the analytic calculation of the parametric derivatives of the objective function
\begin{align}
    C(\theta) = \bra{\psi_0}U^\dag(\theta)OU(\theta)\ket{\psi_0}.
\end{align}
We implement the unitary with a variational circuit as 
\begin{align}
U(\theta)&=\prod_{t=1}^TU_t(\theta)=U_T(\theta)U_{T-1}(\theta)\ldots U_1(\theta)
\end{align}
where 
\begin{align}
U_t=\ee^{\ii X_t(\theta)}.    
\end{align}
The Hamiltonians $X_t$ can be expanded in the Pauli basis $\sigma_{\nu}=\sigma^{\nu_1}\otimes\sigma^{\nu_2}\otimes\ldots\otimes\sigma^{\nu_N}$,
\begin{align}
    X_t=\sum_{\nu}x_{t,\nu}\sigma_\nu,
    \label{eq:x Pauli expansion}
\end{align}
where $\sigma^{\nu_j}\in \{\sigma^{\rm x}, \sigma^{\rm y}, \sigma^{\rm z}\}$. Given an operator $X_t$ we can calculate the coefficients $x_{t,\nu}$ by using the fundamental Pauli identities
\begin{align}
    \tr\sigma^{\nu_j}=0,\quad (\sigma^{\nu_j})^2=\mathds{1}_2, \quad \tr{\mathds{1}_2}=2.
\end{align}
We find that 
\begin{align}
    x_{t,\nu}(\theta)=\tr{X_t(\theta)\sigma_\nu}/2^N.
\end{align}
The immediate calculation of the cost function's parameter derivatives is impractical. Therefore, we use the chain rule to calculate and write 
\begin{align}
    \frac{\partial C(\theta)}{\partial \theta_j}= \sum_t\sum_\nu\frac{\partial C(\theta)}{\partial x_{t,\nu}(\theta)} \frac{\partial x_{t,\nu}(\theta)}{\partial \theta_j}.
    \label{eq: chain rule}
\end{align}
The strategy now is to calculate the derivatives of the cost function with respect to the expansion coefficients $x_{t,\nu}(\theta)$. The parameter derivatives can then be obtained via the chain rule \eref{eq: chain rule}. 

From now on we will fix $t$ and $\nu=(\nu_1,\nu_2,\ldots,\nu_N)$ and write
\begin{align}
    x&=x_{t,\nu}, 
    &V=&\sigma_\nu,
    &H=&\sum_{\mu\neq\nu}x_{t,\mu}\sigma_\mu,\\ \nonumber \ket{\phi}&=\prod_{s=1}^{t-1}U_s\ket{\psi_0},
    &A =& U^\dag_{t^+}OU_{t^+}, &U_{t^+}=&\prod_{s=t+1}^TU_s=U_T(\theta)U_{T-1}(\theta)\ldots U_{t+1}.
\end{align}
The objective can now be written as a function of $x$ 
\begin{align}
    C(x)=\bra{\phi}U_t^\dag(\theta)AU_t\ket{\phi} = \bra{\phi}\ee^{-\ii(H+xV)}A\ee^{\ii(H+xV)}\ket{\phi}
\end{align}
By using the cyclic property of the trace, we find that
\begin{align}
    C(x)=\tr A \ee^{\ii(H+xV)}\rho \ee^{-\ii(H+xV)},\quad \rho=\ketbra{\phi}{\phi}.
    \label{eq: C(x)}
\end{align}
We now introduce a commutator map that acts on the state $\rho$ as follows
\begin{align}
    Z[\rho] = \ii \left((H+xV)\rho -\rho(H+xV)\right)
    \label{eq: Z commutator}
\end{align}
Since $Z$ is a linear map, it can be considered a matrix acting on a vectorised $\rho$. We will also need the Baker-Campbell-Hausdorff formula
\begin{align}
    \ee^{[X,\bullet]}Y=\ee^XY\ee^{-X}
    \label{eq:bch1}
\end{align}
where $[X,\bullet]$ represents a commutator with $X$. We now combine the equations \eref{eq: C(x)}, \eref{eq: Z commutator}, and \eref{eq:bch1} and compactly write the objective as
\begin{align}
    C(x)=\tr A \ee^Z[\rho].
    \label{eq: C compact}
\end{align}
The above formula \eref{eq: C compact} is just a compact reformulation of the original expression \eref{eq: C(x)}, which will enable us to derive the stochastic parameter shift rule. 

Next, we introduce the commutator map
\begin{align}
    \frac{\partial Z}{\partial x}=\mathcal{V}=\ii[V,\bullet].
\end{align}
By using the property $V^2=\mathds{1}$ we find
\begin{align}
    \ee^{\lambda \mathcal{V}}=\rho+\sin^2(\lambda)(V\rho V-\rho) +\frac{\ii}{2}\sin(2\lambda)[V,\rho].
    \label{eq: lambda eV}
\end{align}
We get the expression in \eref{eq: lambda eV} by using the Taylor expansion of the matrix exponential. By taking the derivative of the \eref{eq: lambda eV} on both sides, we get
\begin{align}
    \mathcal{V}\ee^{\lambda\mathcal{V}}=e^{(\lambda+\frac{\pi}{4})\mathcal{V}}[\rho] - e^{(\lambda-\frac{\pi}{4})\mathcal{V}}[\rho]
    \label{eq:dZdlambda}
\end{align}
Using the expression in \eref{eq:dZdlambda} at $\lambda=0$ we get 
\begin{align}
    \frac{\partial Z}{\partial x}[\rho]=e^{\frac{\pi}{4}\mathcal{V}}[\rho] - e^{-\frac{\pi}{4}\mathcal{V}}[\rho] = \ee^{\ii \frac{\pi}{4}V}\rho \ee^{-\ii \frac{\pi}{4}V} - \ee^{-\ii \frac{\pi}{4}V}\rho \ee^{\ii \frac{\pi}{4}V}.
    \label{eq:dZdx}
\end{align}

To get the final expression, we have to introduce one more identity, namely
\begin{align}
    \frac{\partial \ee^Z}{\partial x} = \int_{0}^1\dd s\ee^{sZ}\frac{\partial Z}{\partial x} \ee^{(1-s)Z}.
    \label{eq:leibnitz}
\end{align}
By combining the expressions \eref{eq:dZdx} and \eref{eq:leibnitz}, we can now calculate the derivative of the cost function with respect to $x$
\begin{align}
    \frac{\partial C(x)}{\partial x}=&\tr{A\partial_{x}\ee^{Z}[\rho]}\\ \nonumber 
    =&\int_0^1\dd s\left[\tr\left( A \ee^{-\ii s(H+xV)}\ee^{\ii\frac{\pi}{4}V}\ee^{\ii(1-s)(H+xV)} \rho \ee^{-\ii s(H+xV)}\ee^{-\ii\frac{\pi}{4}V}\ee^{-\ii(1-s)(H+xV)}\right) -\right. \\ \nonumber
    &\left.-\tr\left( A \ee^{-\ii s(H+xV)}\ee^{-\ii\frac{\pi}{4}V}\ee^{\ii(1-s)(H+xV)} \rho \ee^{-\ii s(H+xV)}\ee^{\ii\frac{\pi}{4}V}\ee^{-\ii(1-s)(H+xV)}\right) \right]
\end{align}
We can compactly write the above expression by introducing the shifted objectives
\begin{align}
    C_{\pm}(x,s)=&\bra{\phi}U^\dag_{\pm}(x,s)AU_\pm(x,s)\ket{\phi},\quad \mbox {where}\\ 
    U_{\pm}(x,s)=&\ee^{-\ii s(H+xV)}\ee^{\pm\ii\frac{\pi}{4}V}\ee^{\ii(1-s)(H+xV)}.
\end{align}
We then have
\begin{align}
    \partial_xC(x)=\int_0^1(C_+(x,s)-C_-(x,s))\dd s.
    \label{eq:spsr}
\end{align}

To calculate the derivative of the objective function, we have to implement two more local gates for each term in the expansion \eref{eq:x Pauli expansion}. Then we need to perform the integral over $s$, which can be done using Monte Carlo and does not significantly increase the number of measurements since we need to calculate quantum expectations using many measurements. 

Below, we summarise the stochastic parameter shift rule algorithm based on the equation \eref{eq:spsr}.

\begin{enumerate}
    \item Sample $s$ from the uniform distribution in [0,1];
    \item Initialize the computer in the state $\ket{\phi}$
    \item Apply the gate $\ee^{\ii(1-s)(H+xV)}$
    \item Apply the gate $\ee^{\ii\frac{\pi}{4}V}$
    \item Apply the gate $\ee^{\ii s(H+xV)}$
    \item Measure the observable $A$ and call the result $r_+$
    \item Repeat steps 2 to 5, but on point 4, apply $\ee^{-\ii\frac{\pi}{4}V}$
    \item Measure A and call the result $r_-$
    the sample $g_{t,\nu}=r_+-r_-$ is such that $\partial C/\partial x_{t,\nu}=\mathds{E}[g_{t,\nu}]$.
\end{enumerate}

\begin{referencesbox}
\begin{itemize}
  \item Banchi, L., \& Crooks, G. E. (2021). Measuring analytic gradients of general quantum evolution with the stochastic parameter shift rule. Quantum, 5, 386.
\end{itemize}
\end{referencesbox}

\subsection{Fourier analysis and exponential encoding} 
In this section, we explore the impact of data encoding strategies on the expressive power of parameterised quantum circuits when used as function approximators. A central theme is how the choice of encoding determines the set of functions a quantum model can represent. We introduce a framework that expresses quantum models as partial Fourier series, with the frequencies directly linked to the structure of the data encoding gates. By iterating even simple encoding gates, one can enrich the frequency spectrum accessible to the model, thereby enhancing its representational capacity. We further introduce a specific exponential encoding which enables a possible quantum speedup, provided the variational part of the quantum circuit is flexible enough.

A key tool in our analysis is the representation of quantum models as Fourier-type series. We express the output of a parameterised quantum model as
\begin{align}
    f_\theta(x)&=\bra{0}U^\dag(x,\theta) O U(x,\theta) \ket{0} \\ \nonumber 
    &= \sum_{\omega \in \Omega} c_\omega(\theta) e^{i \omega \cdot x},
\end{align}
where \( \omega \cdot x \) denotes the inner product, \( \Omega \subset \mathbb{R}^N \) is a set of accessible frequencies, and the coefficients \( c_\omega(\theta) \) depend on the circuit parameters \( \theta \). 

Importantly, we will show that the frequency spectrum $\Omega$ is entirely determined by the eigenvalues of the data-encoding Hamiltonians used in the circuit. Meanwhile, the structure of the full quantum circuit governs which coefficients $c_\omega$ can be realised.

This Fourier-type representation reveals two intertwined aspects of a quantum model’s expressivity: the set of accessible basis functions $\{ e^{i \omega \cdot x} \}$, and the ability to control the corresponding coefficients $\{ c_\omega \}$. In many practical settings, the frequency set \( \Omega \) consists of integer vectors, i.e., \( \Omega \subset \mathbb{Z}^N \), and the model becomes a partial multidimensional Fourier series:
\begin{equation}
    f_\theta(x) = \sum_{n \in \Omega} c_n(\theta) e^{i n \cdot x}.
\end{equation}
Here, the functions $e^{i n\cdot x}$ form an orthogonal basis, and the term ``partial'' emphasises that only a subset of the Fourier coefficients are non-zero. This Fourier framework enables a systematic and powerful approach to understanding the functional capacity of quantum models using well-established tools from Fourier analysis.

First, we introduce our basic tool: the natural representation of a quantum model as a partial Fourier series. For simplicity, the majority of our presentation will focus on the case of univariate functions with inputs $x\in\mathbb {R}$.

We define a (univariate) quantum model $f(x)$ as the expectation value of an observable with respect to a quantum state prepared by a parameterised circuit:
\begin{align}
    f(x) = \langle 0 | U^\dagger(x,\theta) O U(x,\theta) \ket{0}, \label{eq: quantum_model}
\end{align}
where $\ket{0}$ is the initial state, $O$ is an observable, and $U(x,\theta)$ is a quantum circuit composed of data-encoding and trainable blocks. The model prediction is estimated by repeated execution and averaging over measurement outcomes.

The circuit $U(x,\theta)$ is constructed from $L$ layers, each consisting of a fixed data-encoding unitary $S(x)$ and an arbitrary trainable unitary $W(\theta)$. We use the Hamiltonian encoding via gates of the form $G(x) = e^{-\ii x H}$, where $H$ is a Hermitian operator (Hamiltonian). To isolate the effect of data encoding, we treat all trainable blocks $W(\theta)$ as general unitaries and omit the dependence on $\theta$. The circuit thus takes the form:
\begin{equation}
    U(x) = W^{(L+1)} S(x) W^{(L)} \cdots S(x) W^{(1)}. \label{eq: circuit_structure}
\end{equation}
A useful simplification is obtained by diagonalising the Hamiltonian 
$H = V^\dagger \Lambda V$, 
where $\Lambda$ is diagonal with eigenvalues $\lambda_1, \ldots, \lambda_d$. The encoding unitaries become $S(x) = V^\dagger e^{-ix\Lambda} V$, and we can absorb the $V$ and $V^\dagger$ into the surrounding unitaries to assume, without loss of generality, that $H$ is diagonal.

This decomposition allows us to isolate the data-dependent phases. The $i$-th component of the state $U(x)\ket{0}$ can be written as:
\begin{align}
    [U(x)|0\rangle]_i 
    &= \sum_{j_1, \ldots, j_L=1}^{d} e^{-i(\lambda_{j_1} + \cdots + \lambda_{j_L})x} 
    W^{(L+1)}_{i j_L} \cdots W^{(2)}_{j_2 j_1} W^{(1)}_{j_1 1}. \label{eq:component_form}
\end{align}

Introducing the multi-index $\mathbf{j} = (j_1, \ldots, j_L) \in [d]^L$, we define 
$\mu_{\mathbf{j}} = \lambda_{j_1} + \cdots + \lambda_{j_L}$, 
so that the state becomes:

\begin{align}
    [U(x)|0\rangle]_i 
    &= \sum_{\mathbf{j} \in [d]^L} e^{-i\mu_{\mathbf{j}} x} 
    W^{(L+1)}_{i j_L} \cdots W^{(2)}_{j_2 j_1} W^{(1)}_{j_1 1}. \label{eq:state_component_multiindex}
\end{align}
To compute the model output, we combine this with the measurement observable and take the real part:
\begin{align}
    f(x) 
    &= \sum_{\mathbf{k}, \mathbf{j} \in [d]^L} e^{i(\mu_{\mathbf{k}} - \mu_{\mathbf{j}})x} 
    a_{\mathbf{k}, \mathbf{j}}, \label{eq:model_fourier}
\end{align}
where the coefficients $a_{\mathbf{k}, \mathbf{j}}$ depend on the unitaries and measurement:
\begin{align}
    a_{\mathbf{k}, \mathbf{j}} 
    &= \sum_{i, i'} W^{(1)*}_{1 k_1} W^{(2)*}_{k_1 k_2} \cdots 
    W^{(L+1)*}_{k_L i} M_{i i'} 
    W^{(L+1)}_{i' j_L} \cdots 
    W^{(2)}_{j_2 j_1} W^{(1)}_{j_1 1}. \label{eq:coefficient_a}
\end{align}
By grouping terms with the same frequency $\omega = \mu_{\mathbf{k}} - \mu_{\mathbf{j}}$, we obtain the final Fourier representation:
\begin{align}
    f(x) 
    &= \sum_{\omega \in \Omega} c_\omega e^{i\omega x}, \label{eq:partial_fourier}
\end{align}
where the frequency spectrum is defined by
\begin{align}
    \Omega 
    &= \{ \mu_{\mathbf{k}} - \mu_{\mathbf{j}} \mid \mathbf{k}, \mathbf{j} \in [d]^L \}, \label{eq:spectrum}
\end{align}
and the Fourier coefficients are given by
\begin{align}
    c_\omega 
    &= \sum_{\substack{\mathbf{k}, \mathbf{j} \in [d]^L \\ \mu_{\mathbf{k}} - \mu_{\mathbf{j}} = \omega}} 
    a_{\mathbf{k}, \mathbf{j}}. \label{eq:fourier_coefficients}
\end{align}
We note several key properties of this spectrum: $0 \in \Omega$, and if $\omega \in \Omega$, then $-\omega \in \Omega$ as well. Since $c_\omega = c_{-\omega}^*$, the function $f(x)$ is real-valued. The \textbf{degree} of the spectrum is defined as 
$D = \max(\Omega)$, and we denote by $K = (|\Omega| - 1)/2$ the number of independent nonzero frequencies. If the eigenvalues $\lambda_i$ are integers, the frequency spectrum consists of integer values, and the model becomes a real-valued partial Fourier series.

These results show that a quantum model’s expressivity is determined by two independent factors: the structure of the frequency spectrum $\Omega$, governed by the eigenvalues of the encoding Hamiltonians, and the controllability of the coefficients $c_\omega$, which depends on the trainable unitaries and measurement observable. In the following example, we explore how these two aspects affect the ability of quantum models to approximate functions.

\begin{shaded}
\begin{example}\textbf{Single qubit encoding}

    As a warm-up application of the Fourier series formalism, we begin with a simple quantum model consisting of a single encoding layer ($L = 1$). Specifically, we consider a model where the input $x$ is encoded via a single-qubit gate of the form $G(x) = e^{-ixH}$, and the full circuit is given by
    \begin{align} 
    U(x) = W^{(2)} G(x) W^{(1)}. 
    \label{eq: single_layer_circuit} 
    \end{align}
    Here, $H$ is a Hermitian generator with two distinct eigenvalues $\lambda_1, \lambda_2$. Without loss of generality, we may rescale the spectrum of $H$ to $( -\gamma, \gamma )$, since global phases are unobservable in quantum mechanics. A common and important example includes Pauli rotations, where $H = \frac{1}{2} \sigma$ for $\sigma \in {\sigma_x, \sigma_y, \sigma_z}$, yielding $\gamma = \frac{1}{2}$. We aim to show that any model of the form in Eq.~\eqref{eq: single_layer_circuit} produces a function $f(x)$ of the form:
    \begin{align} 
    f(x) = A \sin(2\gamma x + B) + C, 
    \label{eq:sine_form} 
    \end{align}    
    where the constants $A$, $B$, and $C$ depend on the trainable components of the circuit but are independent of the encoded input.
    
    To demonstrate this, we first absorb the factor $\gamma$ into the input via a rescaled variable $\tilde{x} = \gamma x$. We may now assume without loss of generality that the eigenvalues of $H$ are $\lambda_1 = -1$ and $\lambda_2 = 1$. From our general Fourier formalism, the set of frequency differences $\omega = \lambda_{k_1} - \lambda_{j_1}$ for $\lambda_{k_1}, \lambda_{j_1} \in {-1, 1}$ yields:
    \begin{align} 
    \Omega = {-2, 0, 2}. 
    \end{align}
    Thus, the model output has the structure:
    \begin{align} 
    f(x) = c_{-2} e^{i2\tilde{x}} + c_0 + c_2 e^{-i2\tilde{x}}. \label{eq:single_qubit_fourier} 
    \end{align}
    
    This expression corresponds to a real-valued cosine function:
    \begin{align} 
    f(x) = c_0 + 2|c_2| \cos(2\tilde{x} - \arg(c_2)),
    \end{align}
    where $\arg(c_2)$ denotes the complex phase of $c_2$. For Pauli rotations, $\tilde{x} = x/2$, and we recover the form of Eq.~\eqref{eq:sine_form} with:
    \begin{align} 
    A = 2|c_2|, \quad B = -\pi/2 - \arg(c_2), \quad C = c_0. 
    \end{align}
    
    Importantly, this result holds regardless of the number of qubits in the system, the structure of the unitaries $W^{(1)}$, $W^{(2)}$, or the form of the measurement observable $O$. This illustrates a key insight of our analysis: even arbitrarily deep and wide quantum circuits cannot escape the fundamental expressivity limitations imposed by the encoding strategy.
\end{example}
\end{shaded}

\begin{shaded}
\begin{example}\textbf{Repeated Pauli encodings linearly extend the frequency spectrum}
    
    Given the severe limitations discussed in the previous section, a natural question arises: \textit{how can we systematically extend the frequency spectrum accessible to a quantum model?} In this section, we show that the Fourier degree can be increased linearly by repeating the data-encoding gates—either in parallel (within a single layer) or sequentially (across multiple layers).
    
    We analyse two cases:
    \begin{enumerate}
        \item \textbf{Parallel repetition}: A single-layer model ($L = 1$) where the encoding gate is repeated $r$ times in parallel across different qubits.
        \item \textbf{Sequential repetition}: A multi-layer model ($L = r$), where a single-qubit encoding gate is applied repeatedly across $r$ layers.
    \end{enumerate}
    Both techniques are common in practical quantum machine learning applications, and this analysis provides a theoretical foundation for their effectiveness.
    
    \subsection*{1. Parallel repetition of Pauli encodings}
    We first consider a model with $L = 1$ and an encoding gate applied in parallel across $r$ qubits:
    \begin{align}
        S(x) &= e^{-i \frac{x}{2} \sigma_r} \otimes \cdots \otimes e^{-i \frac{x}{2} \sigma_1}, \label{eq:parallel_encoding}
    \end{align}
    where each $\sigma_j \in \{\sigma_x, \sigma_y, \sigma_z\}$. Since the rotations act on different qubits, they commute, and the full encoding unitary can be written as:
    \begin{align}
        S(x) &= V^\dag \exp\left(-i \frac{x}{2} \sum_{q=1}^{r} \sigma_z^{(q)} \right) V = V^\dag e^{-ix \Lambda} V,
        \label{eq:parallel_diag}
    \end{align}
    where $\sigma_z^{(q)}$ denotes a Pauli-Z operator acting on qubit $q$, and $\Lambda$ is diagonal with eigenvalues:
    \begin{align}
        \lambda_p = \frac{p - r}{2}, \quad p \in \{0, 1, \dots, r\}. \label{eq:eigenvalues_parallel}
    \end{align}
    These values arise from all possible sums of $r$ terms $\pm \frac{1}{2}$. The spectrum of the model consists of differences between any two of these eigenvalues:
    \begin{align}
        \Omega_{\text{par}} &= \left\{ \lambda_{k_1} - \lambda_{j_1} \mid k_1, j_1 \in \{1, \dots, 2^r\} \right\} \\
        &= \left\{ p - p' \mid p, p' \in \{0, \dots, r\} \right\} \\
        &= \{-r, -r+1, \dots, 0, \dots, r-1, r\}. \label{eq:parallel_spectrum}
    \end{align}
    Hence, a quantum model with $r$ parallel Pauli encodings yields a \textbf{truncated Fourier series of degree $r$}.
    
    \subsection*{2. Sequential repetition of Pauli encodings}
    
    Now consider a circuit where a single-qubit Pauli encoding gate is repeated across $r$ layers:
    \begin{align}
        U(x) = W^{(r+1)} e^{-i \frac{x}{2} \sigma_r} W^{(r)} \cdots W^{(2)} e^{-i \frac{x}{2} \sigma_1} W^{(1)}. \label{eq:sequential_model}
    \end{align}
    Assume all $\sigma_j$ are equal and act on the same qubit. As before, we diagonalize each encoding gate to get $\Lambda = \frac{1}{2} \sigma_z$, and the full spectrum consists of differences of sums of $r$ eigenvalues:
    \begin{align}
        \Omega_{\text{seq}} = \left\{ \left(\sum_{l=1}^r \lambda_{k_l} \right) - \left( \sum_{l=1}^r \lambda_{j_l} \right) \;\middle|\; k_l, j_l \in \{1, 2\} \right\}. 
        \label{eq: sequential_spectrum}
    \end{align}
    This again results in:
    \begin{align}
        \Omega_{\text{seq}} = \{-r, -r+1, \dots, r-1, r\}. \label{eq: sequential_spectrum_final}
    \end{align}
    Thus, the \textbf{frequency spectrum grows linearly} with the number of repetitions $r$, whether applied in parallel or sequentially.
    
    \subsection*{Summary}
    Both parallel and sequential repetitions of Pauli-rotation encodings enable a quantum model to express a larger class of functions via an extended Fourier spectrum. Specifically, the model output can be described as a \textbf{truncated Fourier series of degree $r$}, and this scaling mechanism is independent of other details like the choice of trainable unitaries or measurement observable.
\end{example}
\end{shaded}

\begin{shaded}
\begin{example}\textbf{Exponential encoding and frequency spectrum}
    
    We can further improve the set of available frequencies by an exponential encoding strategy where input data $x \in [-\pi,\pi)$ is embedded into a quantum circuit using a sequence of $N$ single-qubit $Z$-rotations with integer weights $\beta_j$:
    \begin{equation}
        S(x) = \bigotimes_{j=1}^{N} e^{-\ii\,\frac{x\,\beta_j}{2}\,\sigma^{\rm z}}. 
        \label{eq: expenc_circuit}
    \end{equation}
    Each $\beta_j$ defines the frequency contribution of the $j$-th encoding gate. The total unitary $S(x)$ acts diagonally in the computational basis, such that the encoded state accumulates a phase depending on the sum of the weighted bit values of the computational basis states.
   
    The frequency spectrum $\Omega^{(N)}$ of the encoded model consists of all integer differences between accumulated phases of computational basis states:
    \begin{equation}
        \Omega^{(N)} = \left\{ \sum_{j=1}^N (k_j' - k_j)\beta_j \;\middle|\; k_j, k_j' \in \{0,1\} \right\}.
    \end{equation}
    The set $\Omega^{(N)}$ can be built recursively. Given $\Omega^{(N-1)}$, we have:
    \begin{equation}
        \Omega^{(N)} = \Omega^{(N-1)} \cup (\Omega^{(N-1)} + \beta_N) \cup (\Omega^{(N-1)} - \beta_N). \label{eq:recursive_omega}
    \end{equation}

    To ensure a \textbf{dense}, \textbf{non-degenerate} spectrum, we choose the $\beta_j$ recursively such that each new frequency added is larger than twice the maximum existing frequency:
    \begin{equation}
        \beta_j = 2 \sum_{i=1}^{j-1} \beta_i + 1.
    \end{equation}
    Starting with $\beta_1 = 1$, this recurrence yields:
    \begin{equation}
        \beta_j = 3^{j-1}, \quad \text{for } j=1,\ldots,N.
    \end{equation}    
    Thus, the resulting circuit is:
    \begin{equation}
        S(x) = \bigotimes_{j=1}^{N} e^{-\ii\,x\frac{3^{j-1}}{2}\,\sigma^{\rm z}}, 
        \label{eq: expenc_final}
    \end{equation}
    and generates a frequency spectrum:
    \begin{equation}
        \Omega = \left\{ \sum_{j=1}^N b_j 3^{j-1} \;\middle|\; b_j \in \{-1, 0, 1\} \right\}.
    \end{equation}
    This forms a symmetric, dense, integer-valued spectrum of size $|\Omega| = 3^N$.

    We can generalise the exponential encoding by using $\beta_j = l^{j-1}$ for some $l > 1$. Larger $l$ creates a sparser spectrum with potentially higher maximum frequencies. This trade-off allows one to:
    \begin{itemize}
        \item Increase expressivity by expanding the range of representable frequencies,
        \item Or control overfitting by limiting the number of distinct frequency components.
    \end{itemize}    
    Moreover, to target a specific maximum frequency $\beta^*$, one can select weights $\beta_j$ such that:
    \begin{equation}
        \sum_{j=1}^{N} \beta_j = \beta^*,
    \end{equation}
    which may result in degeneracies (repeated frequencies) and can be leveraged to concentrate representational power around specific frequencies.

    Exponential encoding provides a flexible and powerful mechanism to modulate the Fourier spectrum of quantum models. By carefully selecting the $\beta_j$ weights, one can tailor the expressivity and generalisation behaviour of the quantum model to match the desired function class.
\end{example}
\end{shaded}

\begin{referencesbox}
\begin{itemize}
  \item Schuld, M., Sweke, R., \& Meyer, J. J. (2021). Effect of data encoding on the expressive power of variational quantum-machine-learning models. Physical Review A, 103(3), 032430.
  \item Shin, S., Teo, Y. S., \& Jeong, H. (2023). Exponential data encoding for quantum supervised learning. Physical Review A, 107(1), 012422.
\end{itemize}
\end{referencesbox}

\subsection{Quantised classical models}
The main goal of quantum machine learning discussed so far is to obtain some computational speedup with respect to the best possible classical algorithms. This can be done by transferring the critical parts of the classical algorithms to the quantum devices. Though predominant, quantum speedup is not the only goal of quantum machine learning research. This section will examine genuine quantum models and what insight they can offer for machine learning. Many questions regarding the usefulness of genuine quantum distributions/models are still open. The training properties of quantum models and their generalisation and robustness are also not fully understood. Concrete, quantised classical models can perhaps provide some answers that might be applicable more generally. 

This section will discuss two quantised approaches, the quantum Boltzmann machines and the qboost algorithm. 

\paragraph{Quantum Boltzmann machines}
We start our discussion with a short review of the classical recurrent networks, specifically Hopfield models and then Boltzmann machines as their generalisation.
Hopfield models and Boltzmann machines are recurrent neural networks with $G$ binary variables $s=s_1s_2\ldots s_G$, where $s_j\in\{-1,1\}$. We can separate the variables $s$ into the hidden $h_j=s_j$ (for $j=1,2\ldots,N_h$) and visible variables $v_j=s_{j+N_h}$ for $j=1,2\ldots N_v$, where $N = N_h + N_v$. An important object in both the Hopfield networks and the Boltzmann machines is the energy function
\begin{align}
    E(s) = -\sum_{i,j=1}^Nw_{i,j}s_is_j-\sum_{i=1}^Nb_is_i,
    \label{eq:ising energy}
\end{align}
with real parameters $w_{i,j}$ and $b_i$. In Hopfield models, the energy function is symmetric $w_{i,j}=w_{j, i}$ and without self connections $w_{i, i}=0$ and biases $b_i=0$. The state of each variable in the Hopfield model is then updated according to the perceptron rule
\begin{align}
    s^{t+1}_i={\rm sign}\left(\sum_{j=1}^Nw_{i,j}s_j\right).
\end{align}
The perceptron update does not increase the Ising energy \eref{eq:ising energy}. Therefore, we are guaranteed to arrive at a local minimum. The perceptron dynamics can be interpreted as an associative recall of patterns saved in the local minima. The problem is that we can store only around 15 different states with 100 bits, which is far from all possible $2^{100}$ different binary strings. 

The recall algorithm is similar to the zero-temperature Monte Carlo algorithm. The Monte Carlo algorithm obtains a new configuration of variables $s'$ by flipping one or more variables of the old configuration $s$. At zero temperature, the change is accepted if $E(s')<E(s)$. 

In this framework, a generalisation of the Hopfield model is straightforward. We associate to each configuration a probability $p(s)=\ee^{-E(s)/T}/Z$, where $Z=\sum_{s}\ee^{-E(s)/T}$. The introduced distribution is known as the Boltzmann distribution. If the zero temperature Monte Carlo is related to the Hopfield model, the finite temperature Monte Carlo results in a classical Boltzmann machine. The Boltzmann machines are a variational generative model. The parameters of the Ising energy should be chosen such that the Boltzmann distribution is as close as possible to a real dataset distribution. If this is achieved, we can generate new samples by sampling from the model using the finite temperature Monte Carlo algorithm.

Here, we are interested in the quantum Boltzmann machine. To arrive at the quantum generalisation of the Boltzmann machine, we first have to determine the quantum Hamiltonian associated with the classical Ising energy \eref{eq:ising energy}. We arrive at the quantum formulation of the energy \eref{eq:ising energy} by applying the following identities
\begin{align}
    \bra{0}\sigma^{\rm z}\ket{0}=1, \quad \bra{1}\sigma^{\rm z}\ket{1}=-1,\quad \bra{i,j}A\otimes B\ket{i,j}=\bra{i}A\ket{i}\bra{j}B\ket{j}.
\end{align}
Hence, we can write 
\begin{align}
    E(s)=\bra{(1-s)/2}(-\sum_{i,j}w_{i,j}\sigma^{\rm z}_i\sigma^{\rm z}_i -\sum_{i}b_i\sigma^{\rm z}_i)\ket{(1-s)/2} =\bra{(1-s)/2} H_{\rm Ising} \ket{(1-s)/2}.
    \label{eq:ising H}
\end{align}
The Hamiltonian is the Ising Hamiltonian, a quantum formulation of the classical Ising energy. The Hamiltonian is diagonal and contains all possible values of the classical Ising energy. We can now introduce quantum effects by adding non-commuting terms to the Hamiltonian \eref{eq:ising H}. A typical quantum extension is the transverse field Ising model (TFIM)
\begin{align}
    H_{\rm TFIM} = H_{\rm Ising} + \sum_{i}c_i\sigma^{\rm x}_i.
\end{align}

Now we can introduce the quantum Boltzmann machines as the Gibbs distribution with respect to the TFIM, namely
\begin{align}
    \rho_{\rm TFIM}=\ee^{H_{\rm TFIM}/T}/Z,\quad Z=\tr{\ee^{H_{\rm TFIM}/T}}.
\end{align}

For a projectors $\Lambda_s$ the quantum Boltzmann machine introduces a classical probability distribution by $p(s)=\tr{\Lambda_s\rho_{\rm TFIM}}$, where $\Lambda_s=\ketbra{s}{s}$. It is an open problem to describe the properties of this kind of "quantum" distribution with regard to generative machine learning tasks. An open problem is finding an efficient training algorithm for quantum Boltzmann machines. We currently have an algorithm that minimises an upper bound of the desired Log-likelihood objective. However, the upper bound can be very far from the actual values; hence, the algorithm often performs poorly. 

The described approach to the "quantisation" of the Boltzmann machines encapsulates the main general steps:
\begin{enumerate}
    \item Write the classical model in the framework of quantum probability
    \item Add "quantum" terms, and define the appropriate quantum probability distribution, which reduces to the classical in the case of diagonal operators.
    \item Study the properties of the quantised model. How is it different from the classical? Can we efficiently exploit this difference? Applications on NISQ devices...
\end{enumerate}

Besides the quantum Boltzmann machines, there are many other probabilistic quantum models: quantum Hopfield models, quantum hidden Markov models, quantum graphical models (quantum belief propagation), and quantum causal models.

\paragraph{Qboost}
The final quantum model we will discuss is the Qboost algorithm, one of the first quantum machine learning algorithms. It is a "quantisation" of the classical ensembling algorithms, such as Adaboost or XGboost. The ensembling problem is formulated as follows. Given a labeled dataset $\{x^j,y^j\}$ and $K$ models $h_m(x)$ we want to find a linear combination 
\begin{align}
    F_K(w,x)=\sum_{k=1}^K w_k h_k(x)
\end{align}
that performs better than the individual models. What is the best ensembling method strategy is still an open problem. 

A simple classical approach to the problem is the AdaBoost algorithm. In this algorithm, the ensemble is increased sequentially, and the weights are derived from the exponential loss $L_c= \sum_i \ee^{y_iF_{m}(x_i)}$. The problem with this algorithm is that it does not use any regularisation. Hence, a model is included irrespective of its utility or even if it is repeated many times. 

Our strategy to obtain a quantum ensembling algorithm will be to define a quadratic loss, which we will then write as an Ising model. We can directly apply a quantum annealing algorithm or the QAOA algorithm from there. The loss that we will minimise is the mean-square error of the ensemble prediction
\begin{align}
    \min_{W}\left[\frac{1}{N}\sum_{i=1}^N(\sum_{k=1}^Kw_kh_k(x_i)-y_i)^2+\lambda ||w||_0\right],
\end{align}
where $||w||_0=\sum_{k=1}^Kw_k^0$. We can rewrite the above expression as 
\begin{eqnarray}
    &\min_w\left[\frac{1}{N}\sum_{i=1}^N\left[(\sum_{k=1}^Kw_kh_k(x_i))^2-2\sum_{k=1}^K(w_kh_k(x_i)y_i)+y_i^2\right]+\lambda ||w||_0\right]&\\ \nonumber
    &\downarrow&\\ \nonumber
    &\min_w\left[\frac{1}{N}\sum_{k,l=1}^Kw_lw_k(\sum_{i=1}^Nh_k(x_i)h_l(x_i))-2\sum_{k=1}^Kw_k (\sum_{i=1}^Nh_k(x_i)y_i)+\lambda ||w||_0\right]&
\end{eqnarray}
The last expression is similar to the classical Ising model. The first term in the brackets is the interaction energy; the last two translate to the bias term. The main difference is that the variables $w_i$ are continuous. Therefore, we only have to use a finite bitwidth/precision. For example we represent each $w_k$ with only three bits $w_k=0.w_k^1w_k^2w_k^3$, where $w_k^j\in \{0,1\}$. With this modification, the minimisation objective is exactly the classical Ising model. We can now use a quantum annealer or a QAOA to optimise our objective, including a regularisation term. Therefore, we can tune the balance between the ensemble's accuracy and complexity/size. An early Kaggle evaluation of the boosting model on realistic datasets showed that the classical ensembling methods outperform the quantum version.

\begin{referencesbox}
\begin{itemize}
  \item Schuld, M., \& Petruccione, F. (2021). Machine learning with quantum computers (Vol. 676, pp. 163-169). Berlin: Springer. (Chapter: Fault-Tolerant Quantum Machine Learning)
  \item \href{https://lucidalu.github.io/Quantum-algorithms-A-survey-of-applications-and-end-to-end-complexities/areas-of-application/machine-learning-with-classical-data/quantum-machine-learning-via-quantum-linear-algebra}{Quantum algorithms: quantum machine learning via linear algebra}
\end{itemize}
\end{referencesbox}

\subsection{Quantum kernels}
In this section, we will show that many of the quantum models we have discussed can be represented in the framework of kernel methods. In particular, we will show that
\begin{enumerate}
    \item quantum models are linear models in the feature vectors $\rho(x)$
    \item Quantum models that minimise typical machine learning cost functions have measurements that can be written as "kernel expansions in the data" $A=\sum_{m}\alpha_m \rho(x_m)$.
    \item The problem of finding the optimal measurement for typical machine learning cost functions trained with $M$ data samples can be formulated as an $M$-dimensional convex optimisation problem.
\end{enumerate}

Before we start our discussion of the quantum kernels, we will briefly review the classical kernel methods. 

\paragraph{Kernel methods}
Kernels can be considered similarity functions between data points of the input space $\mathcal{X}$. More formally, given a non-empty input space $\mathcal{X}$, with $x_1x_2,\ldots x_N\in \mathcal{X}$, and complex numbers $c_1,c_2,\ldots,c_N\in\mathds{C}$ a function $\kappa: \mathcal{X}\times\mathcal{X}\rightarrow \mathds{C}$ is a kernel if 
\begin{align}
    \sum_{i,j=1}^Nc_i^*c_{j}\kappa(x_i,x_j)\geq 0.
    \label{eq: kernel positivity}
\end{align}
Defining a Gram matrix $K_{i,j}=\kappa(x_i,x_j)$ is also useful. Then the above condition translates to positive-semidefiniteness of the Gram matrix $K\geq0$. The positivity condition \eref{eq: kernel positivity} implies that $\kappa(x,x)\geq 0$ and $\kappa(x,y)=\kappa(y,x)^*$, where the star denotes the complex conjugation.

The kernel can be used to define a canonical feature map $\phi$ from the input set to the complex-valued functions on the input set
\begin{align}
    \Phi: \mathcal{X}\rightarrow \mathds{C}^{\mathcal{X}},\quad x\xrightarrow{\Phi}\kappa(\bullet,x).
    \label{eq: canonical map}
\end{align}
The image of the map is a function which maps elements of the input space to complex numbers. The space of these functions is a vector space. Two different elements of that vector space can be expressed as 
\begin{align}
    f(\bullet) = \sum_{i=1}^I\nu_i\kappa(\bullet,x_i),\quad \nu_i\in\mathds{R}, \\ \nonumber
    g(\bullet) = \sum_{j=1}^J\mu_j\kappa(\bullet,x_j),\quad \mu_j\in\mathds{R}. \\ \nonumber
\end{align}
Given arbitrary elements $g, f$ of the vector space defined by the canonical feature map, we can define an inner product on that space as
\begin{align}
    \langle f,g\rangle = \sum_{i=1}^I\sum_{j=1}^J\nu_i^*\mu_j\kappa(x_i,x_j).
\end{align}
That this is a valid product follows from the positivity condition of the kernel \eref{eq: kernel positivity}. An inner product between two canonical feature maps of different elements of the input space is then
\begin{align}
    \langle\Phi(x),\Phi(y)\rangle =\langle\kappa(\bullet,x),\kappa(\bullet,y)\rangle=\kappa(x,y).
    \label{eq: kernel distance}
\end{align}
The above expression in \eref{eq: kernel distance} solidifies the statement that the kernels are similarity or distance measures between points of the input space $\mathcal{X}$. Moreover, the vector space obtained by the canonical feature map (also called the feature space and denoted by $\mathcal{F}$) can be extended to a Hilbert space by including all limit points of Cauchy series, where the norm is defined by the inner product defined in \eref{eq: kernel distance}.

We can also make a reverse statement. Given a feature map from the input space to some Hilbert space $\phi:\mathcal{X}\rightarrow\mathcal{H}$, we can define a kernel on the input space as the inner product between feature maps of the input elements
\begin{align}
    \kappa(x,y)=\langle\phi(x),\phi(y)\rangle.
    \label{eq: scalar product kernel}
\end{align}
This is a valid kernel, since the inner product is linear and positive-semidefinite. Using the expression \eref{eq: kernel distance} on the kernel \eref{eq: scalar product kernel} we see that the Hilbert space $\mathcal{H}$ embedding of the input dataset and the canonical feature space embedding of the dataset have the same geometry, i.e. $\langle\phi(x),\phi(y)\rangle=\langle\Phi(x),\Phi(y)\rangle$ for $x,y\in\mathcal{X}$.

We have shown the main relation between the Hilbert space of the canonical feature map and kernels. Given a kernel, we can find a map from the input space to a Hilbert space, where the inner product between feature maps reproduces the kernel. Conversely, given a map to a Hilbert space, the inner product defines a kernel on the input space. Since the inner product on the Hilbert space obtained by the canonical map \eref{eq: canonical map} reproduces the kernel, we call the obtained feature space the Reproducing Kernel Hilbert Space (RKHS).

\paragraph{The representer theorem}
The representer theorem is one of kernel theory's most important classical theorems. We want to solve a supervised problem on an input domain $\mathcal{X}$ and a dataset $\mathcal{D}=(x_m,y_m)\in\mathcal{X}\times\mathds{R}$, where $m=1,2,\ldots, M$. We consider a class of model functions $f:\mathcal{X}\rightarrow\mathds{R}$ which we can write as an expansion over the canonical feature maps
\begin{align}
f(x)=\sum_{l=1}^\infty\mu_i\kappa(x,x_i),\quad x_i\in\mathcal{X}.
\label{eq: kernel model class}
\end{align}
Further, we assume that the cost function $\mathcal{C}:\mathcal{X}\rightarrow\mathds{R}$ quantifies the quality of the predictions and includes a monotonically increasing regularisation term. The representer theorem guarantees that the model minimising the empirical risk (the cost function over the training dataset $\mathcal{R}$) is of the form
\begin{align}
    f_{\rm opt}(x)=\sum_{m=1}^M\nu_m\kappa(x,x_m).
    \label{eq: representer theorem}
\end{align}
The main difference between \eref{eq: kernel model class} and \eref{eq: representer theorem} is that in the first equation we have an infinite sum over $x_j\in\mathcal{X}$ that are not necessarily in the dataset. In contrast, the second sum is only over the canonical maps of the input set elements of the dataset. The representer theorem has many important consequences also in the quantum setting as we shall see in the following sections.

\paragraph{Quantum kernel}
Before introducing the quantum kernel, we introduce a Hilbert space of operators. Since we are concerned only with the finite-dimensional case, we can consider density matrices and quantum operators as finite-dimensional matrices. The space of matrices is a vector space. If we further introduce the Hilbert-Schmidt inner product
\begin{align}
    \langle\langle A,B\rangle\rangle = \tr{A^\dag B},
\end{align}
the space of matrices and hence also the space of operators becomes a Hilbert space $\mathcal{H}$. Therefore, any encoding 
\begin{align}
\phi: \mathcal{X}\rightarrow\mathcal{H},\quad x\rightarrow\rho(x), \quad x\in\mathcal{X},\quad \rho>0,\quad \tr{\rho}=1    
\label{eq: quantum map}
\end{align}
maps the input elements to a feature Hilbert space $\mathcal{H}$. The data-encoding feature map $\phi$ and the Hilbert-Schmidt inner product in $\mathcal{H}$ then define a quantum kernel 
\begin{align}
    \kappa(x,y)=\tr{\rho(x)\rho(y)}.
    \label{eq: quantum kernel}
\end{align}
The quantum kernel has an associated canonical feature map \eref{eq: canonical map} that maps the input examples into the RKHS $\mathcal{F}$. In the following, we will study the relation between the operator Hilbert space $\mathcal{H}$ and the RKHS $\mathcal{F}$.

\begin{shaded}
\begin{example}
Let us consider common feature encodings (some of which we discussed in the previous section) and check what types of kernels they produce. In all cases we will have $x\rightarrow \rho(x)=\ketbra{\phi(x)}{\phi(x)}$
\begin{itemize}
    \item \textit{Basis encoding:} In this case we consider $x=i\mathds{N}$. We have
    \begin{align}
        \rho(i)=\ketbra{i}{i},
    \end{align}
    where $\ket{i}$ denotes the $i$-th basis state. Since the basis states are orthogonal $\ketbra{i}{j}=\delta_{i,j}$ we have
    \begin{align}
        \kappa(i,j)=|\braket{i}{k}|^2=\delta_{i,j}^2=\delta_{i,j}.
    \end{align} 
    The basis encoding reproduces the Kronecker delta kernel.
    \item \textit{Amplitude encoding:} Where we consider normalised complex vectors $x\in\mathds{C}^{2^n}$ with $|x|_2=1$. The amplitude encoding leads to the feature map
    \begin{align}
        \rho(x)=\ketbra{x}{x}=\sum_{i,j=1}^{2^n}x_ix_j^*\ketbra{i}{j}
    \end{align}
    The corresponding kernel is 
    \begin{align}
        \kappa(x,y)=|\braket{x}{y}|^2=\sum_{i,j=1}^{2^n}|x_i^*y_j|^2\delta_{i,j}=|x^\dag y|_2^2
    \end{align}
    The amplitude encoding reproduces the quadratic kernel.
    \item \textit{Repeated amplitude encoding:} We can also repeat the amplitude encoding as follows
    \begin{align}
        \rho(x)=\ketbra{x}{x}\otimes\ketbra{x}{x}\otimes\ldots\otimes \ketbra{x}{x}.
    \end{align}
    For $r$ repetitions, this leads to the polynomial kernel
    \begin{align}
        \kappa(x,y)=|x^\dag y|^{2r}
    \end{align}
    \item \textit{Phase encoding:} In the repeated polynomial encoding, we map elements of a vector $x\in[0,2\pi]^n$ to angles of the Bloch sphere as
    \begin{align}
        \rho(x_i) = \ketbra{\phi(x_i)}{\phi(x_i)},\quad \ket{\phi(x_i)}=\cos(x_i)\ket{0}+\sin(x_i)\ket{1}.
    \end{align}
    The phase encoding of the whole vector is then
    \begin{align}
        \rho(x)=\rho(x_1)\otimes\rho(x_2)\otimes\ldots\otimes\rho(x_n).
    \end{align}
    The associated kernel is 
    \begin{align}
        \kappa(x,y)=& \prod_{i=1}^n|\braket{\phi(x_i)}{\phi(y_i)}|^2=\prod_{i=1}^n|\sin(x_i)\sin(x_i)+\cos(x_i)\cos(x_i)|^2\\ \nonumber
        =&\prod_{i=1}^n|\cos(x_i-y_i)|^2.
    \end{align}
    We reproduced the cosine distance kernel. 
\end{itemize}
\end{example}
\end{shaded}

\paragraph{Quantum models are linear models in embeddings (feature vectors) $\rho(x)$}
We begin by rewriting variational quantum models in the operator space formalism. A variational quantum model calculates an expectation value of the form 
\begin{align}
    f_\theta(x)=\bra{\psi_0}U^{\dag}(x)A_\theta U(X)\ket{\psi_0},
    \label{eq: variational circuit}
\end{align}
where $\ket{\psi_0}$ is the initial state, $U(x)$ is the data encoding procedure, $A_\theta$ is the variational circuit that is merged with the measurement, and $\theta$ are the variational parameters. We now write the function $f$ by using the operator formalism as
\begin{align}
    f_\theta(x)=\tr{A_\theta U(x)\ketbra{\psi_0}{\psi_0}U^\dag(x)} = \tr{A_\theta\rho(x)},\quad \rho(x)=U(x)\ketbra{\psi_0}{\psi_0}U^\dag(x).
    \label{eq: linear quantum model}
\end{align}
From \eref{eq: linear quantum model} we immediately see that any quantum circuit of the form \eref{eq: variational circuit} is a linear model in the operator Hilbert space. Moreover, since $A_\theta$ is Hermitian it can be decomposed as $A_\theta=\sum_{\mu}a_\mu\ketbra{\mu}{\mu}$, where $\ketbra{\mu}{\mu}$ are projectors on the eigenstates of $A$ and $a_\mu$ are the corresponding eigenvalues. Since projectors $\ketbra{\mu}{\mu}$ are also density matrices we have 
\begin{align}
    A_\theta=\sum_{x_l\in\mathcal{X}}\mu_l\rho(x_l).
\end{align}
Therefore, we can associate with each quantum model $A_\theta$ an element of the RKHS via the map
\begin{align}
    A_\theta\rightarrow \sum_{x_l\in\mathcal{X}}\mu_l\Phi(x_l).
\end{align}

\paragraph{The quantum RKHS and the space of quantum models are equivalent}
The previous statement shows that the quantum RKHS $\mathcal{F}$ includes linear models. We now show that functions in the quantum RKHS are quantum models. A general function in the quantum RKHS, $f\in\mathcal{F}$, can be written as
\begin{align}
    f(x)=\sum_k\gamma_k\kappa(x_k,x)
\end{align}
where $\kappa$ is now a quantum kernel and $\gamma_k\in\mathds{R}$. We now use the linearity of the quantum kernel (trace of the embeddings) and express the function $f(x)$ as
\begin{align}
    f(x) =& \sum_k\gamma_k\tr{\rho(x_k)\rho(x)}\\ \nonumber
    =&\tr{\sum_{k}\gamma_k}\rho(x_k)\rho(x)\\ \nonumber
    =&\tr A\rho(x)
    \label{eq:rkhs linear model}
\end{align}
where $A=\sum_{k}\gamma_k\rho(x_k)$ is a Hermitian operator. The final expression in \eref{eq:rkhs linear model} is exactly the definition of a variational quantum model in the operator space formalism \eref{eq: linear quantum model}. 

We have now shown that the space of variational quantum models of the form \eref{eq: linear quantum model} is equivalent to the quantum RKHS. This equivalence holds for the chosen embedding procedure $\rho(x)$.

We can now invoke the representer theorem and write the model that minimises the regularised empirical risk as a linear combination of the dataset embeddings
\begin{align}
    f_{\rm opt} =& \sum_{m=1}^M\alpha_m\tr{\rho(x_m)\rho(x)}\\ \nonumber
    =&\tr{\sum_{m=1}^M\alpha_m\rho(x_m) \rho(x)} = \tr{A_{\rm opt}\rho(x)}.
\end{align}
The optimal measurement/variational circuit can be expressed as a linear combination of the input space embeddings, i.e. $A_{\rm opt} = \sum_{m=1}^M\alpha_m\rho(x_m)$.

\paragraph{Optimising quantum kernels}
The kernel method viewpoint on the variational circuits guarantees that the global optimum of the variational problem is linear in dataset feature maps $\rho(x_m)$. Now we will show that the problem of finding the corresponding coefficients $\alpha_m$ is a simple, convex optimisation problem. 

We consider a dataset $\mathcal{D}=\{(x_m,y_m),m=1,2,\ldots,M\}$, and minimise regularised loss function $L:\mathcal{X}\times\mathcal{Y}\times\mathds{R}\rightarrow [0,\infty)$, i.e. we want to solve the following optimisation problem
\begin{align}
    f_{\rm opt} = \inf_{f\in\mathcal{F}}\lambda |f|_F^2+\mathcal{R}_L(f).
    \label{eq: optimisation problem}
\end{align}
where $\lambda$ is the regularisation parameter, $|\bullet|_F^2$ denotes the Frobenious norm in the RKHS, and $\mathcal{R}_L(f)$ calculates for a given function $f$ the average loss over the dataset. The representer theorem guarantees that the optimal model has the form
\begin{align}
    f_{\rm opt} = \sum_{\alpha=1}^M\alpha_m\tr\rho(x_,)\rho(x).
\end{align}
Our optimisation problem now reduces to the problem of determining the linear expansion coefficients $\alpha_m$. Let us first consider the functional $\mathcal{R}_L(f)$ which can be written as
\begin{align}
    \mathcal{R}_{L}(f)=\frac{1}{M}\sum_{m=1}^ML(x_m,y_m,\sum_{l=1}^M\alpha_l\tr\rho(x_m)\rho(x_l)).
\end{align}
If we now assume that $L$ is a convex function of the network outputs (which is true for typical loss functions), we see that $\mathcal{R}_L(\alpha)$ is also a convex function of $\alpha_m$. Moreover, $\mathcal{R}_L$ has only $M$ degrees of freedom. 

We can express the second part of the optimisation function as
\begin{align}
    |f|_F^2=&\sum_{m,l}\alpha_m\alpha_l\tr{\rho(x_m)\rho(x_l)}\\ \nonumber
    =&\sum_{m,l}\alpha_m\alpha_l\kappa(x_m,x_k)=\alpha^{\rm T} K \alpha.
\end{align}
Since the Gram matrix $K$ is positive-semidefinite, the function of the Frobenius norm is a convex function of $\alpha$. Therefore, the optimisation problem \eref{eq: optimisation problem} in the RNKS reduces to an $M$-dimensional convex optimisation problem 
\begin{align}
    \inf_{\alpha\in\mathds{R}^M}\frac{1}{M}\sum_{m=1}^ML\left(x_m,y_m,\sum_{l=1}^M\alpha_l\kappa(x_l,x_m)\right)+\lambda \alpha^{\rm T}K\alpha.
    \label{eq: convex optimisation problem}
\end{align}

\begin{figure}[!htb]
    \centering
    \includegraphics[width=0.8\textwidth]{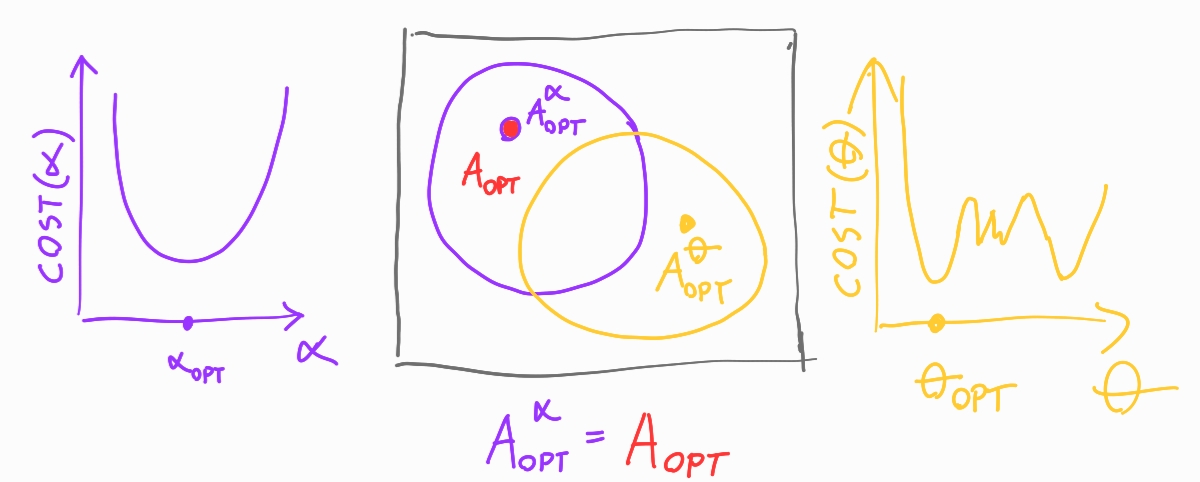}
    \caption{Schematic representation of the optimisation problems in the RKHS setting (left) and the variational setting (right). The RKHS optimisation problem is convex, i.e., it has only one global/local minimum. The variational problem is higher dimensional (generally $2^N$ where $N$ is the number of qubits) and is a non-convex optimisation problem.}
    \label{fig: kernel optimisation}
\end{figure}
We have transformed a potentially hard $2^N$ dimensional optimisation problem into an $M$-dimensional convex optimisation problem. Moreover, the solution of the convex problem is guaranteed to be globally optimal (see \fref{fig: kernel optimisation}). To find a solution to this problem is still quadratic in $M$ and is hence not practical for large datasets. Still, the theory provides new tools to construct practically relevant quantum machine learning models. The discussed results also determine which types of embedding procedures will not lead to significant quantum speedups.

\begin{referencesbox}
\begin{itemize}
  \item Schuld, M., \& Petruccione, F. (2021). Machine learning with quantum computers (Vol. 676, pp. 163-169). Berlin: Springer. (Chapter: Quantum Models as Kernel Methods)
  \item Schuld, M., \& Killoran, N. (2019). Quantum machine learning in feature Hilbert spaces. Physical review letters, 122(4), 040504.
  \item Schuld, M. (2021). Supervised quantum machine learning models are kernel methods. arXiv preprint arXiv:2101.11020.
\end{itemize}
\end{referencesbox}

\subsection{Barren Plateaus in Random Parameterised Quantum Circuits}

In previous lectures, we introduced parameterised quantum circuits (PQCs) and their use in hybrid quantum-classical algorithms such as the Variational Quantum Eigensolver (VQE) and Quantum Approximate Optimisation Algorithm (QAOA). These methods are especially relevant in the context of Noisy Intermediate-Scale Quantum (NISQ) devices, which can execute shallow quantum circuits but remain limited by decoherence and gate fidelity.
The core idea of these hybrid algorithms is to optimise a parameterised quantum circuit with respect to a cost function, often expressed as the expectation value of a Hamiltonian or an overlap with a target quantum state. This process is reminiscent of training classical neural networks, where parameters are updated to minimise a loss function.  
When physical insight is available, the circuit ansatz can be tailored to the structure of the problem. For example, the Unitary Coupled Cluster ansatz is often used in quantum chemistry. However, a structured ansatz may not be feasible in many machine learning applications or under severe hardware constraints. Instead, a generic or random parameterised circuit—sometimes called a “hardware-efficient ansatz”—is used.
While random PQCs are easy to implement and hardware-friendly, they introduce a major challenge: the \emph{barren plateau} problem. In such circuits, the optimisation landscape often features large regions where the cost function gradient is exponentially small in the number of qubits. This makes gradient-based optimisation methods ineffective. 

This phenomenon can be understood through the lens of concentration of measure. In high-dimensional spaces, most points are close to the mean value of any smooth function. In the quantum setting, this implies that expectation values and gradients of random observables over random states concentrate near their average. This effect has been formalised using tools such as Levy's lemma.

Importantly, the flat regions of the cost landscape do not correspond to useful local minima, but to large, featureless regions with values close to the global average (such as that of the maximally mixed state). Numerical and analytical results show that even relatively shallow circuits, with depth scaling as $O(n^{1/d})$ on a $d$-dimensional layout, can enter this barren regime.

In this section, we will analyse the mathematical structure behind this gradient vanishing phenomenon and explore how it influences the trainability of PQCs. In the following section, we will discuss practical strategies to mitigate barren plateaus, particularly problem-inspired ansätze and adiabatic optimisation.

To explicitly verify the vanishing gradient and its exponential suppression in variational quantum circuits (RPQCs), we examine the expectation value and variance of the gradient under design assumptions.

We will discuss random parameterised quantum circuits (RPQCs) 
\begin{align}
U(\vec{\theta}) = U(\theta_1, ..., \theta_L) = \prod_{l=1}^L U_l(\theta_l) W_l
\end{align}
where $U_l(\theta_l) = \exp \left(-i \theta_l V_l \right)$, $V_l$ is a hermitian operator, and $W_l$ is a generic unitary operator that does not depend on any angle $\theta_l$.  Circuits of this form are a natural choice due to a straightforward evaluation of the gradient with respect to most objective functions. We have discussed the evaluation of the gradients in previous sections. Consider an objective function $E(\theta)$ expressed as the expectation value over some hermitian operator $H$,
\begin{align}
E(\vec{\theta}) =  \bra{0}U(\vec{\theta})^\dagger H U(\vec{\theta})\ket{0}.
\end{align}
When the RPQCs are parameterised in this way, the gradient of the objective function takes a simple form:
\begin{align}
\partial_k E \equiv \frac{\partial E(\vec{\theta})}{\partial \theta_k} 
= i \bra{0} 
U_{-}^\dagger \left[V_k, U_{+}^\dagger H U_{+} \right] U_{-} \ket{0} 
\end{align}
where we introduce the notations $U_{-} \equiv \prod_{l=0}^{k-1} U_l(\theta_l) W_l$, $U_{+} \equiv \prod_{l=k}^{L} U_l(\theta_l) W_l$, and henceforth drop the subscript $k$ from $V_k \rightarrow V$ for ease of exposition.  Finally, we will define our RPQCs $U(\vec{\theta})$ to have the property that for any gradient direction $\partial_k E$ defined above, the circuit implementing $U(\vec{\theta})$ is sufficiently random such that either $U_{-}$, $U_+$, or both match the Haar distribution up to the second moment. The circuits $U_-$ and $U_+$ are independent.

We will use the properties of the Haar measure on the unitary group $d\mu_{\text{Haar}}(U) \equiv d\mu(U)$, which is the unique left- and right-invariant measure such that
\begin{align}
\int_{U(N)}\!\!\!\!\!\!\!\!\!\! d\mu(U) f(U) = \int\! d\mu(U) f(VU) = \int\! d\mu(U) f(UV)
\end{align}
for any $f(U)$ and $V \in U(N)$, where the integration domain will be implied to be $U(N)$ when not explicitly listed.  While this property is valuable for proofs, quantum circuits that exactly achieve this invariance generically require exponential resources.  This motivates the concept of unitary $t$-designs, which satisfy the above properties for restricted classes of $f(U)$, often requiring only modest polynomial resources.  Suppose $\{p_i, V_i\}$ is an ensemble of unitary operators, with unitary $V_i$ being sampled with probability $p_i$.  The ensemble $\{p_i, V_i\}$ is a $k$-design if
\begin{align}
\sum_i p_i V_i^{\otimes t} \rho (V_i^\dagger)^{\otimes t} =
\int d\mu(U) U^{\otimes t} \rho (U^\dagger)^{\otimes t}.
\end{align}
This definition is equivalent to the property that if $f(U)$ is a polynomial of at most degree $t$ in the matrix elements of $U$ and at most degree $t$ in the matrix elements of $U^*$, then averaging over the $t$-design $\{p_i, V_i\}$ will yield the same result as averaging over the unitary group with the respect to the Haar measure.

The average value of the gradient is a concept that requires additional specification because, for a given point, the gradient can only be defined in terms of the circuit that led to that point.  We will use a practical definition that leads to the value we are interested in, namely
\begin{align}
\average{\partial_k E} = \int dU p(U) \partial_k \bra{0}U(\vec{\theta})^\dagger H U(\vec{\theta})\ket{0}
\end{align}
where $p(U)$ is the probability distribution function of $U$. 

\paragraph{Expectation Value of the Gradient} By the RPQC definition, the full unitary $U = U_+ U_-$ is composed of two independently distributed unitaries, with either $U_+$ or $U_-$ forming at least a 2-design. The independence implies the joint distribution factorises as:
\begin{align}
p(U) = \int dU_+ \, p(U_+) \int dU_- \, p(U_-) \, \delta(U_+ U_- - U).
\end{align}

Using this, the average gradient becomes:
\begin{align}
\average{\partial_k E} = i \int dU_- \, p(U_-) \, \tr \left\{ \rho_- \int dU_+ \, p(U_+) \left[ V, U_+^\dagger H U_+ \right] \right\}.
\end{align}

Assuming Haar-randomness up to the first moment (1-design), we utilise the identity:
\begin{align}
\int d\mu(U) \, U O U^\dagger = \frac{\tr O}{N} I,
\end{align}
where $N = 2^n$ and $O$ is any operator.

\paragraph{Case 1: $U_-$ is a 1-design}

Define $\rho_- = U_- \ket{0}\bra{0} U_-^\dagger$. Then:
\begin{align}
\average{\partial_k E} &= i \int d\mu(U_-) \, \tr \left\{ \rho_- \left[ V, \int dU_+ \, p(U_+) \, U_+^\dagger H U_+ \right] \right\} \notag \\
&= \frac{i}{N} \tr \left\{ \left[ V, \int dU_+ \, p(U_+) \, U_+^\dagger H U_+ \right] \right\} = 0.
\end{align}

\paragraph{Case 2: $U_+$ is a 1-design}

\begin{align}
\average{\partial_k E} &= i \int dU_- \, p(U_-) \, \tr \left\{ \rho_- \int d\mu(U_+) \left[ V, U_+^\dagger H U_+ \right] \right\} \notag \\
&= i \frac{\tr H}{N} \int dU_- \, p(U_-) \tr \left\{ \rho_- \left[ V, I \right] \right\} = 0.
\end{align}

\paragraph{Variance of the Gradient}

The variance is defined by:
\begin{align}
\text{Var}[\partial_k E] = \average{(\partial_k E)^2},
\end{align}
given that $\average{\partial_k E} = 0$. For Haar averages up to the second moment (2-design), we use the integration formula:
\begin{align}
\int d\mu(U) \, U_{i_1 j_1} U_{i_2 j_2} U^*_{i'_1 j'_1} U^*_{i'_2 j'_2} = & \frac{\delta_{i_1 i'_1} \delta_{i_2 i'_2} \delta_{j_1 j'_1} \delta_{j_2 j'_2} + \delta_{i_1 i'_2} \delta_{i_2 i'_1} \delta_{j_1 j'_2} \delta_{j_2 j'_1}}{N^2 - 1} \\ & - \frac{\delta_{i_1 i'_1} \delta_{i_2 i'_2} \delta_{j_1 j'_2} \delta_{j_2 j'_1} + \delta_{i_1 i'_2} \delta_{i_2 i'_1} \delta_{j_1 j'_1} \delta_{j_2 j'_2}}{N(N^2 - 1)}.
\end{align}

We evaluate three distinct scenarios:

\paragraph{Case 1: $U_-$ is a 2-design}
\begin{align}
\text{Var}[\partial_k E] = \frac{2 \tr(\rho^2)}{N^2} \tr \left\langle H_u^2 V^2 - (H_u V)^2 \right\rangle_{U_+} = - \frac{\tr(\rho^2)}{2^{2n}} \tr \left\langle \left[ V, H_u \right]^2 \right\rangle_{U_+},
\end{align}
where $H_u = u^\dagger H u$.

\paragraph{Case 2: $U_+$ is a 2-design}
\begin{align}
\text{Var}[\partial_k E] = \frac{2 \tr(H^2)}{N^2} \tr \left\langle \rho_u^2 V^2 - (\rho_u V)^2 \right\rangle_{U_-} = - \frac{\tr(H^2)}{2^{2n}} \tr \left\langle \left[ V, \rho_u \right]^2 \right\rangle_{U_-},
\end{align}
with $\rho_u = u \rho u^\dagger$.

\paragraph{Case 3: Both $U_-$ and $U_+$ are 2-designs}
\begin{align}
\text{Var}[\partial_k E] = 2 \tr(H^2) \tr(\rho^2) \left( \frac{\tr(V^2)}{2^{3n}} - \frac{\tr(V)^2}{2^{4n}} \right).
\end{align}

All three cases reveal exponential decay in $n$, highlighting the severity of barren plateaus in high-dimensional quantum circuits.

\subsubsection{Contrast with Gradients in Classical Deep Networks}

Finally, we contrast our results with the vanishing (and exploding) gradient problems observed in classical deep neural networks. Two key differences emerge in the quantum case: (i) the scaling behaviour of the vanishing gradients, and (ii) the complexity involved in estimating expectation values.

In classical networks, gradients typically vanish (or explode) exponentially with the number of \emph{layers}. In contrast, in quantum circuits, gradients decay exponentially with the number of \emph{qubits}, which is generally a much more severe form of decay for fixed-depth architectures. This arises due to the structure of the parameter landscape. In classical deep networks, the gradient for a particular parameter depends on the sum over exponentially many paths connecting input to output neurons, where cancellation occurs due to random initialisation. Similarly, in quantum circuits, the gradient depends on interference among exponentially many computational paths, and the amplitude signs fluctuate randomly. Due to normalisation, this leads to gradient magnitudes that saturate at exponentially small values in the number of qubits.

Moreover, gradient \emph{estimation} differs significantly. In classical training, the batch gradient is estimated numerically and is limited by machine precision, scaling logarithmically as $O(\log(1/\epsilon))$ in the desired error $\epsilon$. Small gradients are still trackable across batches, allowing for eventual convergence. In contrast, quantum gradients must be estimated via repeated circuit evaluations, and the estimation error scales as $O(1/\epsilon^\alpha)$, with $\alpha \geq 1$ depending on the estimator. If the number of measurements is insufficient to resolve the exponentially small signal, the result is effectively indistinguishable from noise, leading to a random walk in parameter space.

By the concentration of measure, such a walk will have an exponentially small probability of escaping the barren plateau. Hence, without additional algorithmic strategies (e.g., tailored ansätze, parameter initialisation schemes, or alternative cost functions), gradient descent in quantum machine learning cannot overcome this obstacle in polynomial time. This highlights a fundamental barrier to naively porting classical training paradigms to quantum systems.

\begin{referencesbox}
\begin{itemize}
  \item McClean, J. R., Boixo, S., Smelyanskiy, V. N., Babbush, R., \& Neven, H. (2018). Barren plateaus in quantum neural network training landscapes. Nature communications, 9(1), 4812.
\end{itemize}
\end{referencesbox}

\subsection{Quantum Convolutional Neural Networks}
In the previous section, we have seen that in general, random PQC exhibit a barren plateau problem. In this section, we will consider a specific PQC architecture, which avoids the problem of barren plateaus and performs well on a variety of supervised machine learning tasks. 

We take inspiration from classical machine learning. Convolutional neural networks (CNNs) have demonstrated remarkable success in classical machine learning tasks such as image classification, leveraging architectural elements like convolution and pooling to extract hierarchical and translationally invariant features from data. In a typical CNN, feature maps evolve through layers of local convolutions—weighted sums over small neighbourhoods—and pooling, which reduces dimensionality by downsampling key activations. Nonlinear activation functions are used between layers to increase expressiveness, and the final classification is made via a fully connected layer. The number of learnable parameters is typically independent of input size, enabling efficient scaling.

Inspired by this structure, we introduce a quantum analogue: the quantum convolutional neural network (QCNN), which adapts the core principles of CNNS to quantum systems. The QCNN is composed of alternating convolution and pooling layers followed by a fully connected unitary layer, and is designed to act on an $n$-qubit quantum state $\rho_{\rm in}$. 

In this quantum model, each convolution layer applies a translationally invariant pattern of two-qubit unitary gates (W), akin to sliding convolution kernels. The pooling layers implement a reduction of degrees of freedom by performing measurements on specific qubits, with outcomes conditioning unitaries (I) on neighbouring qubits, thus introducing a nonlinearity. This sequence is repeated for $L$ layers until only a few qubits remain. A fully connected unitary $F$ is then applied, and a final observable $O$ is measured on the resulting state $\rho_{\rm out}$.

Like classical CNNs, the QCNN architecture is parameter-efficient: only $O(\log n)$ parameters are required for an $n$-qubit input, representing an exponential reduction relative to generic quantum circuits. This efficiency makes QCNNs particularly suitable for scalable quantum learning tasks.

Given a training dataset of labelled quantum states $\{(\ket{\psi_\alpha}, y_\alpha)\}_{\alpha=1}^M$ with $y_\alpha \in \{0,1\}$, the QCNN can be trained to minimise a loss function such as the mean squared error:
\begin{equation}
\label{eq:mse}
\text{MSE} 
= \frac{1}{2M} \sum_{\alpha = 1}^M (y_\alpha - f_{\{W,I,F\}}(\ket{\psi_\alpha}))^2,
\end{equation}
where $f_{\{W,I,F\}}(\ket{\psi_\alpha})$ denotes the expectation value of the output measurement.

To understand the theoretical underpinnings of QCNNs, we can relate their structure to two powerful concepts in quantum information and tensor networks (discussed in the following sections): the multiscale entanglement renormalisation ansatz (MERA) and quantum error correction (QEC). MERA efficiently represents many-body quantum states (Vidal, 2007; Aguado and Vidal, 2008; Pfeifer et al., 2009) through alternating layers of unitaries and isometries, building long-range entanglement hierarchically. The QCNN reverses this structure, effectively collapsing entanglement while preserving relevant features.

Furthermore, the measurement and feedback mechanism in pooling layers parallels QEC: measurement outcomes can be interpreted as syndromes, with corresponding unitaries acting to correct localised errors (Preskill, 1998). In this sense, QCNNs implement a hybrid between MERA and nested QEC protocols, allowing them to detect and correct deviations from target states or phases.

This synergy enables QCNNs to classify quantum phases and detect quantum phase transitions with exponentially fewer measurements than traditional observables. For instance, a QCNN trained to recognise a representative ground state $\ket{\psi_0}$ of a given phase can map perturbed versions of $\ket{\psi_0}$ back toward it via repeated correction, effectively simulating a renormalisation-group flow. As a result, QCNNs offer a promising architecture for both practical quantum machine learning and foundational studies of quantum many-body systems.

Importantly, it has been shown that the variance of the derivatives of the cost function vanishes only polynomially with the system size. However, more recent studies show that due to the same reason, the QCNN can be trained and evaluated classically. This is in agreement with a more general result, which argues that current PQCs without barren plateaus are classically simulable and can not provide any benefit with respect to classical machine learning methods. 

In the next section, we will discuss a model-agnostic approach to combat barren plateaus, which will combine the insights from adiabatic quantum computing and parametrised quantum circuits.

\begin{referencesbox}
\begin{itemize}
  \item Cong, I., Choi, S., \& Lukin, M. D. (2019). Quantum convolutional neural networks. Nature Physics, 15(12), 1273-1278.
\end{itemize}
\end{referencesbox}

\subsection{Variational Ground-State Adiabatic Theorem and Barren Plateaus in PQCs}
\begin{figure}[!htb]
    \centering
    \includegraphics[width=0.5\columnwidth]{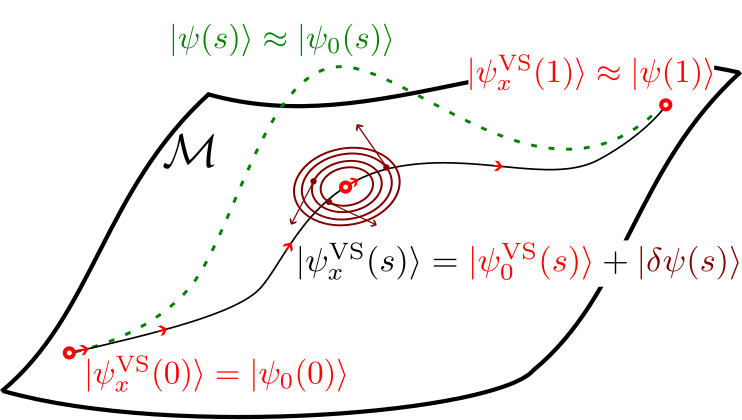}
    \caption{Schematic representation of the variational ground-state quantum adiabatic theorem. The dashed green line represents the exact quantum adiabatic evolution $\ket{\psi(s)} \approx \ket{\psi_0(s)}$, visiting high-entanglement regions. The solid black line corresponds to the instantaneous variational ground state $\ket{\psi^{\rm VS}_0(s)}$, with the small red arrows indicating that it changes slowly with time. The time-dependent variational state $\ket{\psi_x^{\rm VS}(s)}$ is separated into $\ket{\psi^{\rm VS}_0(s)}$ and their difference $\ket{\delta\psi(s)}$. The latter can be decomposed in the eigenbasis of the linearised evolution map, whose eigenvalues $\omega_l$ determine the frequencies of the elliptic trajectories around the instantaneous variational ground state. Since $1/T \ll \omega_l$, the fast dynamics of $\ket{\delta\psi(s)}$ (denoted by long, dark-red arrows) averages to zero, and the time-dependent variational state $\ket{\psi_x^{\rm VS}(s)}$ follows the instantaneous variational ground state $\ket{\psi^{\rm VS}_0(s)}.$ 
    When the target ground state of the final Hamiltonian lies within the variational manifold, the time-dependent variational state will converge to it as the annealing time $T$ becomes large.}
    \label{fig: manifold volution}
\end{figure}
Parameterised quantum circuits (PQCs) have emerged as a leading framework for variational quantum algorithms (VQAs), such as the variational quantum eigensolver (VQE) and the quantum approximate optimisation algorithm (QAOA). Despite their versatility, PQCs suffer from the \emph{barren plateau} problem—an exponentially vanishing gradient in the cost landscape—which severely limits the scalability of variational methods. This phenomenon has been analysed in depth by McClean et al. (2018) and Cerezo et al. (2021), where it was shown that overly expressive or random ansätze tend to generate states close to the Haar measure, leading to vanishing gradients.

A promising approach to mitigating this issue is rooted in the \emph{variational ground-state quantum adiabatic theorem}. This strategy emulates adiabatic quantum computation (AQC) by evolving a parameterised quantum state slowly along a path determined by a time-dependent Hamiltonian $H(s)$, where $s=t/T$ is the normalised time. The conventional adiabatic theorem, discussed by Albash and Lidar (2018), guarantees that a system prepared in the ground state of $H_0$ will remain in the ground state of the instantaneous Hamiltonian $H(s)$ if the evolution is sufficiently slow.

In gate-based quantum circuits, we can approximate this continuous evolution within a variational manifold $\mathcal{M}$ defined by a PQC ansatz. The resulting variational trajectory $\ket{\psi_x^{\mathrm{VS}}(s)}$, governed by the time-dependent variational principle (TDVP), evolves in a physically meaningful subspace that avoids the random sampling responsible for barren plateaus.

As illustrated in Fig.~\ref{fig: manifold volution}, the variational state follows the instantaneous variational ground state $\ket{\psi_0^{\mathrm{VS}}(s)}$ with high fidelity if the initial and final ground states both lie within $\mathcal{M}$. The deviation $\ket{\delta\psi(s)} = \ket{\psi_x^{\mathrm{VS}}(s)} - \ket{\psi_0^{\mathrm{VS}}(s)}$ is of order $\mathcal{O}(1/T)$, provided that the annealing time $T$ is large compared to the inverse of the lowest excitation frequencies in the variational subspace (see Hackl et al., 2020).

\begin{figure}[htb!]
    \centering
    \includegraphics[width=\textwidth]{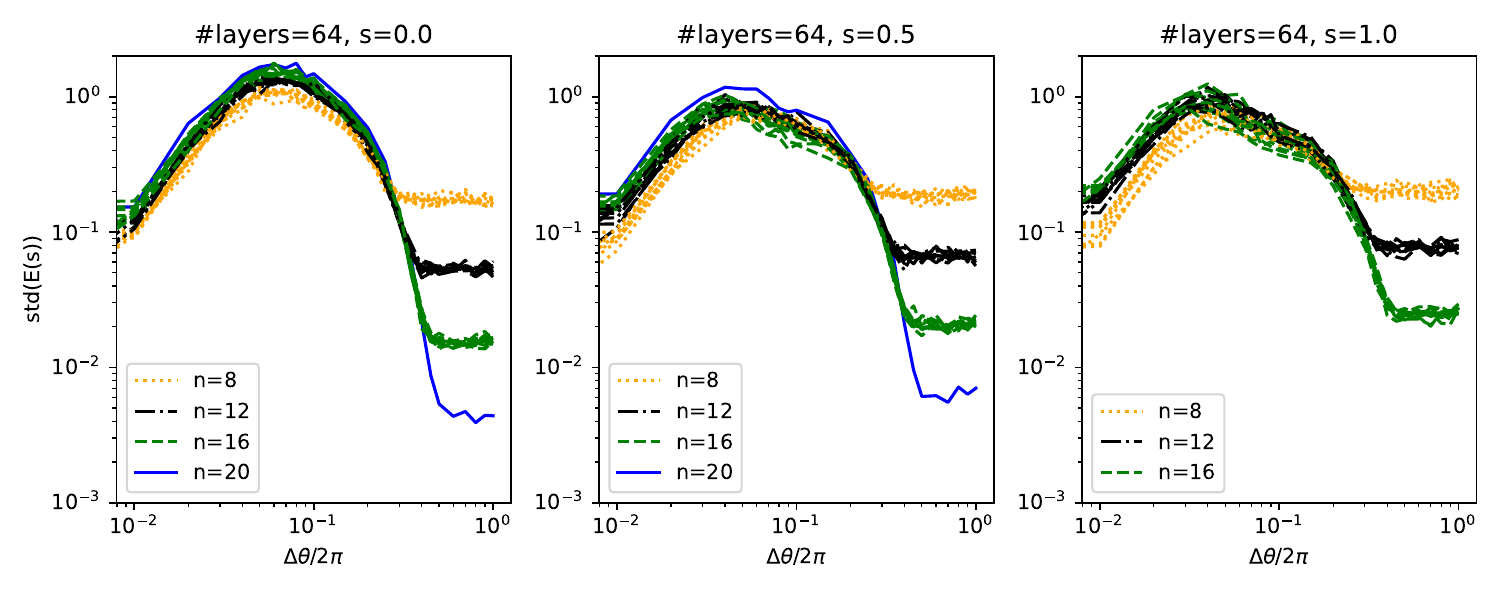}
    \caption{Scaling of the standard deviation of the loss (in this case, the energy of the model) close to the variational ground-state obtained by a time-dependent gradient descent method. In a finite region shrinking as a square root of the number of parameters, the standard deviation of the loss function and hence the gradient, is independent of the number of qubits. The PQC used to produce these results exhibits the barren plateaus problem. However, by following the adiabatic path, we are able to follow a particular local minimum. $\Delta\theta$ is the size of the hypercube around the variational ground state in which the standard deviation is calculated. If $\Delta\theta$ is very small, we are in a linear regime close to the variational ground state, where the standard deviation grows quadratically, and if $\Delta\theta$ is $2\pi$, we essentially choose a random point in the parameter space.}
    \label{fig: qvat bp}
\end{figure}

This variational approach can be adopted with classical and quantum ans\"{a}tze. Here we will consider quandum PQCs. The main question in this context is what happens with the gradients in the vicinity of the variational ground states. In this regard, we consider a hardware-efficient ansatz exhibiting barren plateaus and use it to solve an Ising spin glass problem, which is computationally demanding. We first calculate the variational ground state by using a variant of gradient descent, and then calculate the energy standard deviation around the obtained variational minimum. The \fref{fig: qvat bp} shows the result at different points in the annealing procedure. In all cases, we find that close to the variational ground state, the standard deviation remains independent of the system size, while at random points $\Delta\theta\approx 2\pi$ it vanishes exponentially.

This simple example demonstrates that we can circumvent the barren plateau problem by following the adiabatic path.

\begin{referencesbox}
\begin{itemize}
  \item Unpublished work
  \item Žunkovič, B., Torta, P., Pecci, G., Lami, G., \& Collura, M. (2025). Variational ground-state quantum adiabatic theorem. Physical Review Letters, 134(13), 130601.
\end{itemize}
\end{referencesbox}

\subsection{Power of data in quantum machine learning}
In the previous sections, we considered specific training problems with quantum machine learning models that arise from the concentration of measure in expressive quantum ansatzes. In this section, we will discuss the effect of data in the context of quantum simulation and how this translates to quantum machine learning. We will see that data itself can elevate classical models to rival or even surpass quantum approaches without data, even when the underlying quantum circuits are classically intractable. Consequently, the key to understanding quantum advantage in machine learning lies not only in computational complexity but also in the geometric and statistical properties of the data. This perspective motivates a principled framework to assess prediction advantages, using tools such as kernel methods and geometric tests, to quantify when quantum models offer meaningful improvements over classical counterparts.

We consider a supervised learning task with a set of $N$ training examples $\{(x_i, y_i)\}$, where $x_i$ are input vectors sampled i.i.d. from a distribution $\mathcal{D}$ and $y_i \in \mathbb{R}$ are real-valued outputs generated by a quantum model. Each input $x_i$ is mapped into a quantum state using a continuous data-encoding unitary $U_{\text{enc}}(x_i)$ acting on the $n$-qubit initial state $\ket{0}^{\otimes n}$, resulting in the state $\ket{x_i} = U_{\text{enc}}(x_i) \ket{0}^{\otimes n}$, with density matrix $\rho(x_i)$.

A quantum neural network (QNN), defined by a parametrised unitary $U_{\text{QNN}}(\theta)$, is applied to the encoded state, followed by a measurement of an observable $O$. The expected value gives the output label:
\[
f(x_i) = \bra{x_i} U_{\text{QNN}}^\dagger O U_{\text{QNN}} \ket{x_i}.
\]

While computing $f(x)$ can be classically intractable in general, as we now demonstrate, the presence of training data changes the problem substantially.

\paragraph{Hardness Without Data.}  
Suppose we use amplitude encoding for classical vectors $x_i \in \mathbb{R}^p$ with $\|x_i\|_2 = 1$, yielding quantum states:
\[
\ket{x_i} = \sum_{k=1}^p x_i^k \ket{k}.
\]
For a general QNN defined by time evolution under a many-body Hamiltonian, computing $f(x)$ is known to be classically hard and has been discussed in the context of Hamiltonian simulation. More precisely:
\begin{proposition}\label{prop:BQP=BPP}
If a classical algorithm without training data can efficiently compute $f(x)$ for arbitrary $U_{\text{QNN}}$ and $O$, then $\mathrm{BPP} = \mathrm{BQP}$.
\end{proposition}

\paragraph{Ease With Data.}  
Despite this hardness in the absence of data, training a classical model to approximate $f(x)$ may be straightforward when training data is available. Expanding the expectation, we find:
\begin{align}
    f(x_i) &= \left( \sum_{k=1}^p x_i^{k*} \bra{k} \right) U_{\text{QNN}}^\dagger O U_{\text{QNN}} \left( \sum_{l=1}^p x_i^l \ket{l} \right) \notag \\
           &= \sum_{k=1}^p \sum_{l=1}^p B_{kl} x_i^{k*} x_i^l,
\end{align}
where $B_{kl} = \bra{k} U_{\text{QNN}}^\dagger O U_{\text{QNN}} \ket{l}$ are fixed coefficients. Thus, $f(x)$ is a quadratic function in the input components.

Using classical regression techniques, one can fit such a function with high accuracy. 

\paragraph{Testing Quantum Advantage} \label{sec: tests}
Continuing our exploration of quantum machine learning, we now develop a general framework to assess the potential for quantum prediction advantage in supervised learning tasks. This approach builds on the geometry of learning spaces and results in practical tests for quantum superiority.

We begin by considering a quantum model defined by the function
\[
f(x) = \Tr( O^U \rho(x)), \quad \text{with } O^U = U_{\text{QNN}}^\dagger O U_{\text{QNN}},
\]
where $\rho(x)$ is the input state and $O$ is an observable. Suppose we are given $N$ training samples $\{(x_i, y_i = f(x_i))\}$.

After training on this data, there exists an ML algorithm that outputs $h(x) = w^\dagger \phi(x)$, where $\phi(x)$ is a feature map and $k(x_i, x_j) = K_{ij}=\phi(x_i)^\dagger \phi(x_j)$ is the kernel, which has a prediction error bounded by:
\[
\mathbb{E}_{x \sim \mathcal{D}}|h(x) - f(x)| \leq c \sqrt{\frac{s_K(N)}{N}},
\]
for some constant $c > 0$, where $s_K(N)$ quantifies the model complexity. The model complexity is defined as:
\[
s_K(N) = \sum_{i,j=1}^N (K^{-1})_{ij} \Tr(O^U \rho(x_i)) \Tr(O^U \rho(x_j)).
\]
After training we have $s_K(N)= \|w\|^2$. Smaller $s_K(N)$ implies better generalisation, reflecting how well the kernel geometry aligns with the structure of the data labels. The derivation of this result is out of the scope of the lectures (for now). We will discuss only the implications.

\textbf{Computational Aspects:} Given measurements of $\Tr(O^U \rho(x_i))$, $s_K(N)$ can be computed by inverting the $N \times N$ kernel matrix $K$, requiring $\mathcal{O}(N^3)$ time classically.

\textbf{Regularization:} To prevent overfitting, a regularization term $\|w\|^2$ is often added during training. This ensures $s_K(N)$ remains bounded even if the model does not perfectly fit the training data.

\paragraph{Implications for Quantum Advantage}
If a classical model achieves a small $s_K(N)$ on a quantum dataset, then it can predict $f(x)$ accurately, even if $f(x)$ is hard to compute classically. Hence, the potential for quantum advantage depends on minimising $s_K(N)$ over classical models.

To assess this, we define an asymmetric geometric quantity:
\[
g_{12} =g(K^1||K^2)= \sqrt{\|\sqrt{K^2} (K^1)^{-1} \sqrt{K^2}\|_\infty},
\]
where $K^1$ and $K^2$ are the kernel matrices of two models, and the spectral norm $\|\cdot\|_\infty$ captures maximal directional discrepancy. This leads to the inequality:
\[
s_{K^1} \leq g_{12}^2 s_{K^2},
\]
indicating how well model $K^1$ can perform relative to model $K^2$.

\textbf{Classical vs Quantum Models:} Define $g_{\mathrm{CQ}} = g(K^\mathrm{C} \| K^\mathrm{Q})$. If $g_{\mathrm{CQ}} \approx 1$, classical ML can match quantum ML performance. If $g_{\mathrm{CQ}} \gg 1$, quantum methods may yield significantly better predictions.

\textbf{Constructive Insight:} When $g_{\mathrm{CQ}}$ is large, one can explicitly construct a dataset on which quantum models outperform classical ones. In practice, $g_{\mathrm{CQ}}$ is minimised over a suite of classical methods to find the best classical "adversary."

\subsubsection*{Role of Effective Dimension in Quantum Kernel Methods}

Consider quantum kernel methods with $k^{\mathrm{Q}}(x_i, x_j) = \Tr(\rho(x_i)\rho(x_j))$. Let $\mathrm{vec}(X)$ denote the vectorization of matrix $X$, then:
\[
s_Q = \mathrm{vec}(O^U)^T P_Q \mathrm{vec}(O^U),
\]
where $P_Q$ is the projector onto the subspace spanned by $\{\mathrm{vec}(\rho(x_i))\}_{i=1}^N$.

Define the effective dimension:
\[
d = \mathrm{rank}(K^{\mathrm{Q}}) = \dim(P_Q) \leq N.
\]
Since $P_Q$ is a projector:
\[
s_Q \leq \min(d, \Tr(O^2)),
\]
and the prediction error becomes:
\[
\mathbb{E}_{x \sim \mathcal{D}} |h(x) - f(x)| \leq c \sqrt{\frac{\min(d, \Tr(O^2))}{N}}.
\]

\textbf{Computational Note:} Both $d$ and $g_{\mathrm{CQ}}$ can be computed classically via singular value decomposition on the $N \times N$ kernel matrices.

\subsubsection*{Concluding Remarks on Predictive Advantage}

When both $g_{\mathrm{CQ}}$ and $\min(d, \Tr(O^2))$ are small, we conclude that classical models can learn the target function effectively, regardless of the quantum circuit depth.

Finally, quantum advantage arises only when $s_C \gg s_Q$—that is, when quantum models align well with the function $f(x)$ but classical models do not. The geometric and model-complexity tests described provide a principled, data-driven methodology for assessing this possibility.

\begin{referencesbox}
\begin{itemize}
  \item Huang, H. Y., Broughton, M., Mohseni, M., Babbush, R., Boixo, S., Neven, H., \& McClean, J. R. (2021). Power of data in quantum machine learning. Nature communications, 12(1), 2631.
\end{itemize}
\end{referencesbox}



\section{Quantum-inspired machine learning}
In this section, we will introduce two frameworks/methods originating from quantum mechanics and apply them to machine learning problems. First, we will discuss tensor networks, which are an efficient way to represent low-entanglement many-body quantum states variationally. Then we will look at dequantisation methods and algorithms that assume sample and query access to input vectors.

\subsection{Tensor networks}
A tensor network is a set of tensors which are contracted to form a new tensor. Common operations like matrix multiplication and scalar product can be interpreted as tensor networks where we contract two tensors in order to get a new one. In this context also a number is a zero-dimensional tensor. 

\paragraph{Diagramatic notation} A standard notation that helps us to visualise operations on tensors and keep track of indices is the diagramatic notation. The tensors are diagramatically represented as shapes (circles, squares, triangles) with legs. The number of legs determines the dimension of the tensor (see table \ref{tab: diagram notation}). For example, a number can be represented as a circle without legs and a vector as a circle with one leg.
\begin{table}[!htb]
    \centering
    \begin{tabular}{c|c|c}
        Object & Index & Diagrammatic  \\
        \hline
        \hline
        Number & a & \includegraphics[width=0.08\textwidth]{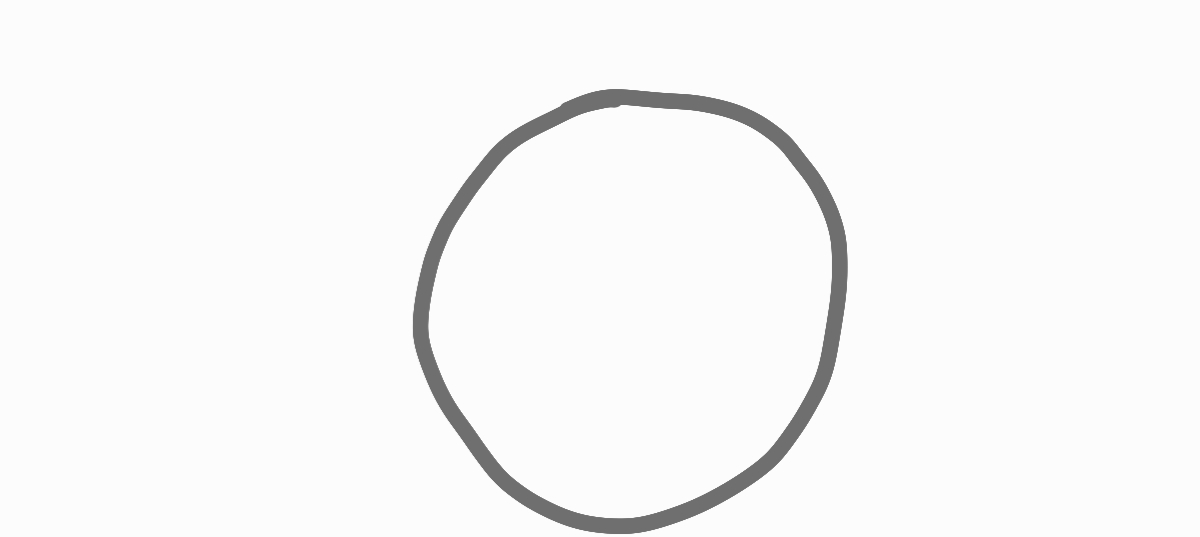}\\
        Vector & $a_i$ & \includegraphics[width=0.08\textwidth]{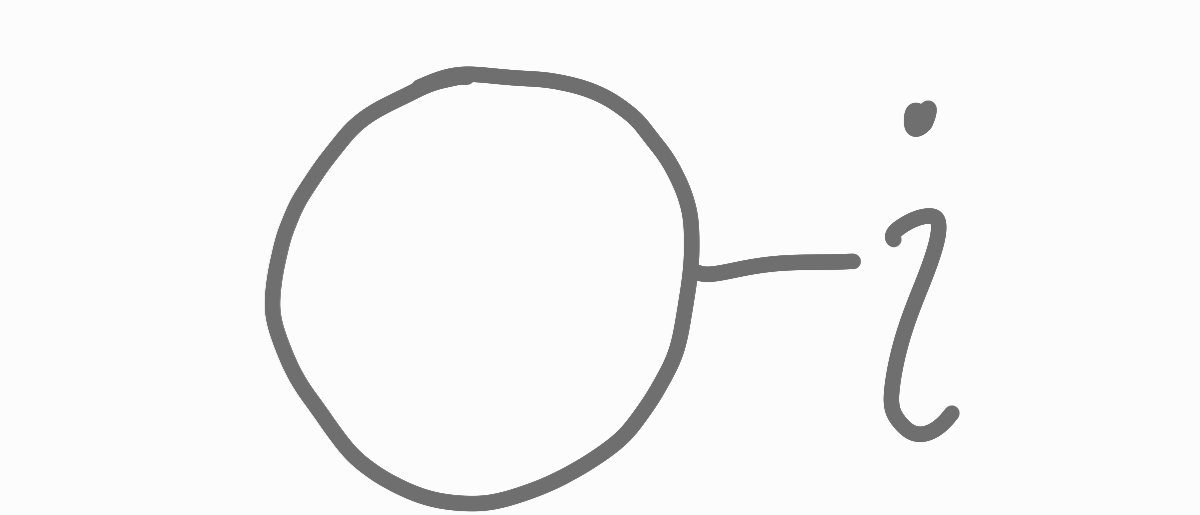}\\
        Matrix & $M_{i,j}$ & \includegraphics[width=0.08\textwidth]{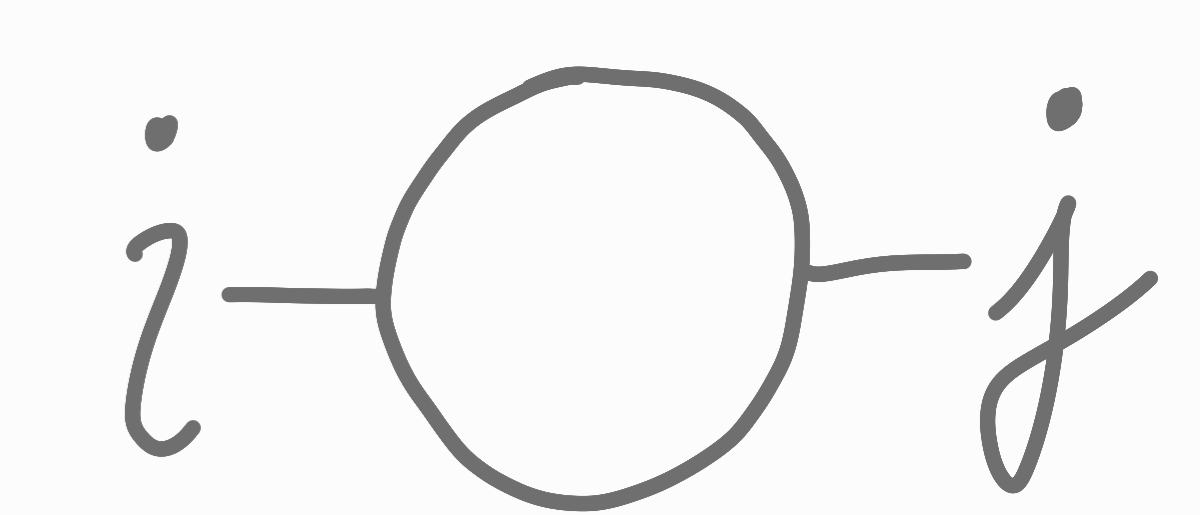}
        \\
        3rd order Tensor & $T_{i,j,k}$ & \includegraphics[width=0.08\textwidth]{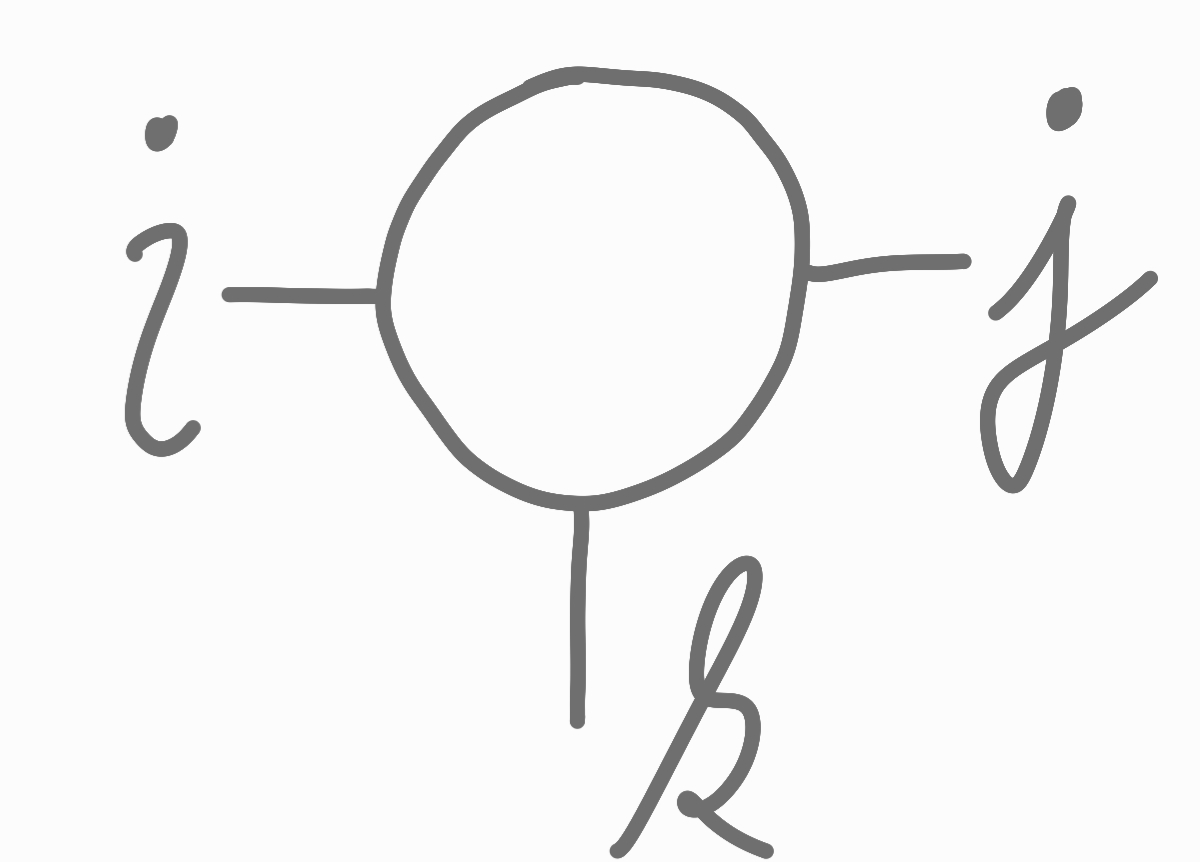}
    \end{tabular}
    \caption{Comparison of diagrammatic and index notation of low-order tensors.}
    \label{tab: diagram notation}
\end{table}

Typically, we do not write indices and names of the tensors in the diagrammatic notation, since this is clear from the context/drawing and does not provide any new information. Also, the orientation of the legs does not matter unless we use a convention, e.g., the legs pointing up in a vector means a complex conjugate of the legs pointing down. In most literature, no such rule is assumed.

The primary operations we can do on tensors are tensor product, contractions, trace, and grouping/splitting of indices. 

\paragraph{Tensor product} is the most common operation when we are dealing with many-body quantum systems. The reason is that quantum mechanics has a tensor product structure. If one system is described with a quantum probability vector $a\in \mathcal{H}_1$ and the other with a vector $b\in \mathcal{H}_1$, then the composite system is described with a tensor product $c=a\otimes b\in \mathcal{H}_{1}\otimes \mathcal{H}_2$. Tensor product of tensors $A$ and $B$ is defined as 
\begin{align}
    [A\otimes B]_{i_1,\ldots,i_N,j_1,\ldots j_M}=A_{i_1,\ldots i_N}B_{j_1,\ldots j_M}.
\end{align}
In the diagrammatic notation, the tensor product of tensors is represented as tensors drawn one after another (without any legs connected)
\begin{align}
    \includegraphics[width=0.4\textwidth]{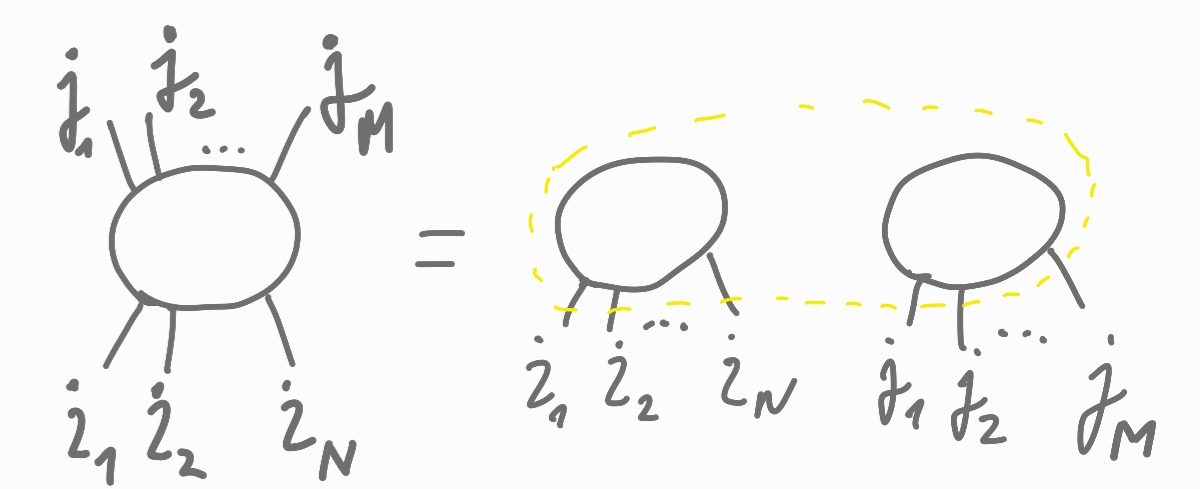}.
\end{align}
The yellow dashed line is drawn for convenience and is usually omitted.  We also omit the indices of legs since they do not provide any new information.

\paragraph{Contractions} A generalisation of the tensor product is a contraction. We not only construct a new tensor by multiplying elements from different tensors, but also sum over contracted indices. A simple example is matrix multiplication
\begin{align}
    A_{ij}=\sum_{k}B_{ik}C_{kj}.
\end{align}
In the diagrammatic notation, the contraction is represented by joining the corresponding legs of different tensors. The matrix multiplication example is then
\begin{align}
    \includegraphics[width=0.4\textwidth]{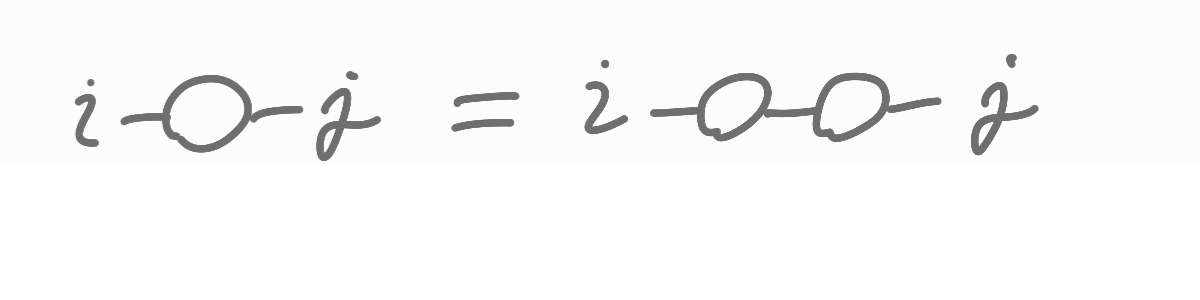}.
\end{align}

\paragraph{Trace} A contraction of indices on the same tensor is called a trace. For example, a trace of a matrix is a sum of diagonal elements $\tr{A}=\sum_{i}A_{i,i}$. In the diagrammatic notation, this is 
\begin{align}
    \includegraphics[width=0.4\textwidth]{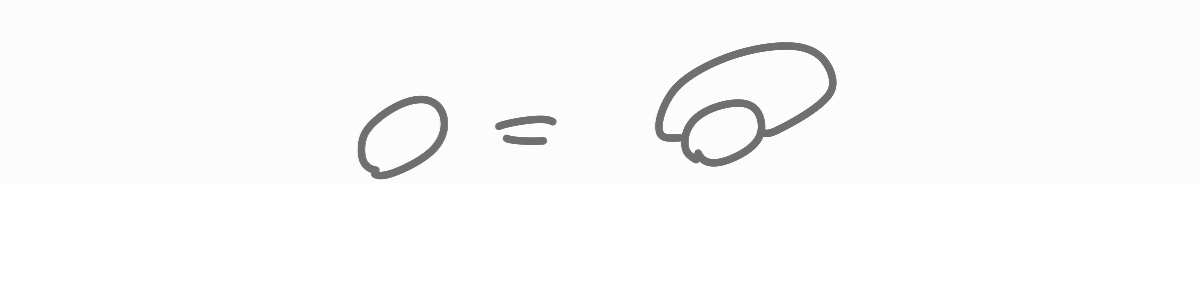}.    
\end{align}
Generally, we can also take a trace in tensors with more than two indices. In this way, the result is not a number but a tensor with a lower dimension. For example, if we trace a 3rd-order tensor, we obtain a vector 
\begin{align}
    a_i=\sum_{j}T_{i,j,j}
\end{align}
or in the diagrammatic notation
\begin{align}
    \includegraphics[width=0.4\textwidth]{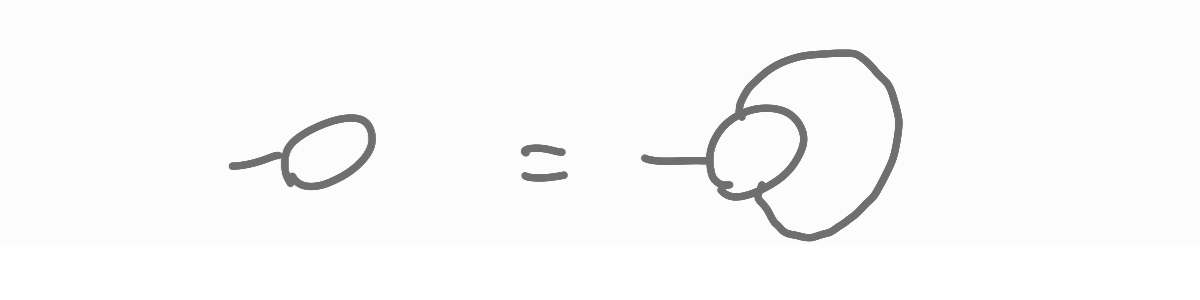}.    
\end{align}

\paragraph{Reshape: grouping and splitting of indices} (\textit{vector = matrix = tensor})
Critical operations that we can apply to tensors are grouping and splitting of indices. In other words, the reshaping of tensors. For example, a matrix with elements $A_{i,j}$ can be represented as a vector $v_\alpha$ as $v_(i M + j)=A_{i,j}$, where $i=0,\ldots N-1$ and $j=0,\ldots M-1$. This vectorisation is not unique. We could equally well vectorise the matrix $A$ as $v_(jN + i)=A_{i,j}$. These two options are commonly referred to as column major and row major vectorisations. In the diagrammatic notation, we group indices by simply drawing them close together or with a bold leg
\begin{align}
    \includegraphics[width=0.4\textwidth]{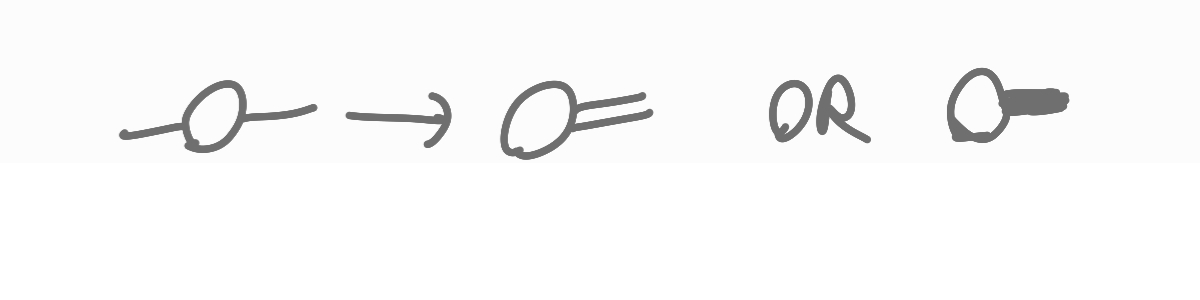}.   
\end{align}
The size of the new index is equal to the product of the sizes of the grouped indices. In the above matrix-to-vector example, this means that $\alpha=0,\ldots MN-1$. 

Grouping of indices can also be applied only on a part of the tensor indices, and together with another grouping of indices. For example, a five-dimensional tensor can be transformed to a matrix by grouping two and three indices 
\begin{align}
    \includegraphics[width=0.4\textwidth]{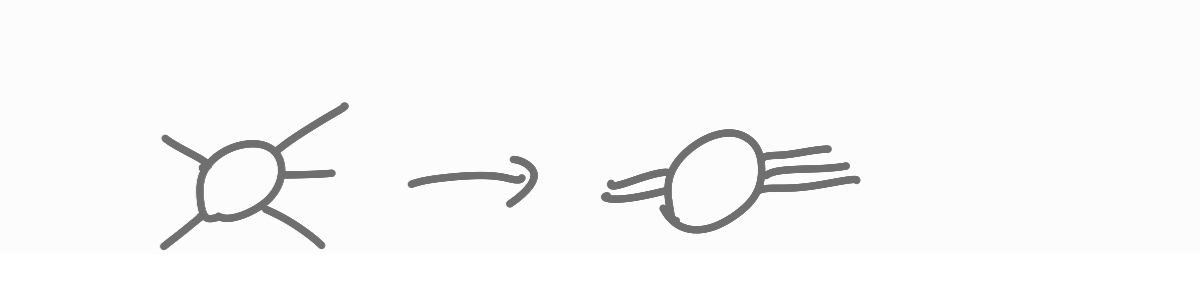}.   
\end{align}

Whereas we can always group two indices, one index can not be split into an arbitrary number of indices. This is clear since the reshaped tensor must have the same number of elements. For example, a vector with $NM$ elements can be split into an $N\times M$ matrix. 

\paragraph{Singular value decomposition on tensors}
Tensor reshaping allows us to reshape any tensor to a matrix (assuming the tensor is not a number or a vector with a prime number of elements). This allows us to apply operations that are typically reserved for matrices to general tensors. The most important operation in this context is the singular value decomposition (SVD). Given a complex matrix $A$, the SVD results in two unitaries $U$ and $V$ and a real positive diagonal matrix $S$ such that $A=USV^\dag$. If we keep all the singular values, the decomposition is exact. However, if we keep only the most significant $D$ elements, the decomposition results in a best approximation (measured in Hilbert-Schmidt norm) of the original matrix by a matrix of rank $D$. In fact, the SVD decomposition decomposes a matrix into a tensor network. If we apply this decomposition repeatedly, we can decompose a high-dimensional tensor as a tensor network of low-dimensional tensors. The most common tensor network that is obtained is the matrix product state (see \exref{ex: matrix product state})

\begin{shaded}
\begin{example}
\label{ex: matrix product state}
Let us consider a tensor $T$ with $N$ legs of size 2. The tensor $T$ has $2^N$ elements. Our aim is to represent the $N$ dimensional tensor as a contraction of two two-dimensional and $N-2$ three-dimensional tensors $A^{j}$, $j=1,\ldots N$
$$
    \includegraphics[width=0.4\textwidth]{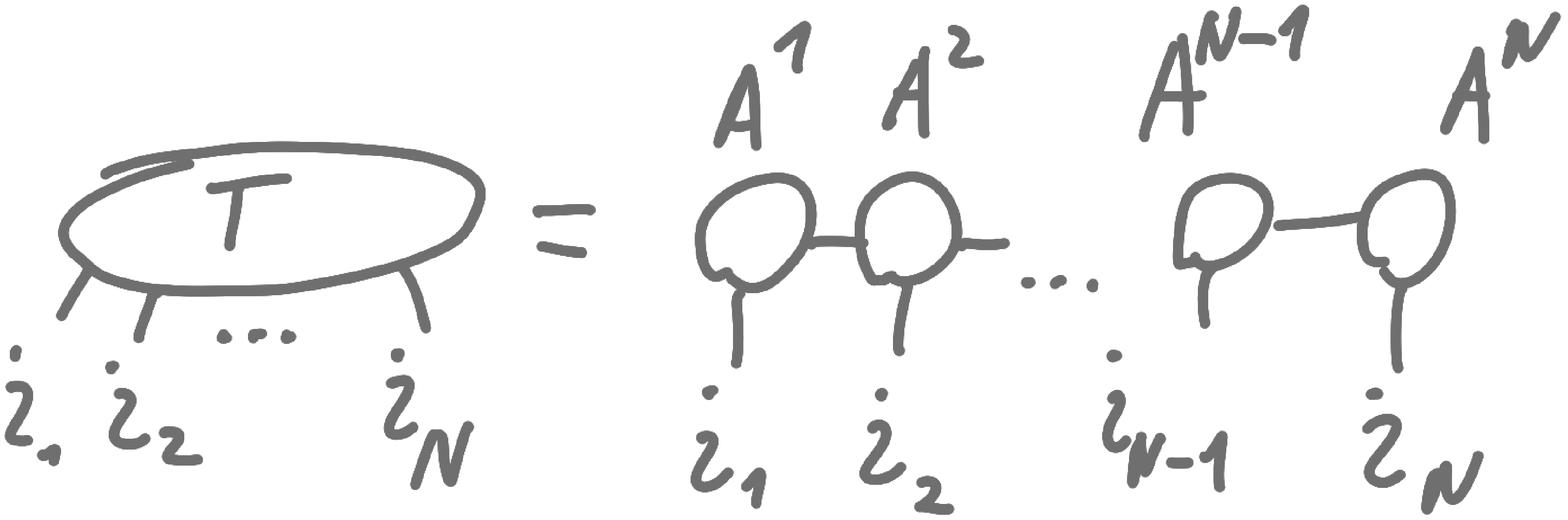}. 
$$
To get the above decomposition, we will successively reshape the largest tensor in the decomposition and apply the SVD. First, we reshape the tensor $T$ into a matrix by grouping the indices $i_2,\ldots, i_N$. Then we use the SVD on the matrix. And finally, we contract the matrices $S$ and $V$ from the SVD decomposition. In the diagrammatic notation,n these steps are 
$$
    \includegraphics[width=0.4\textwidth]{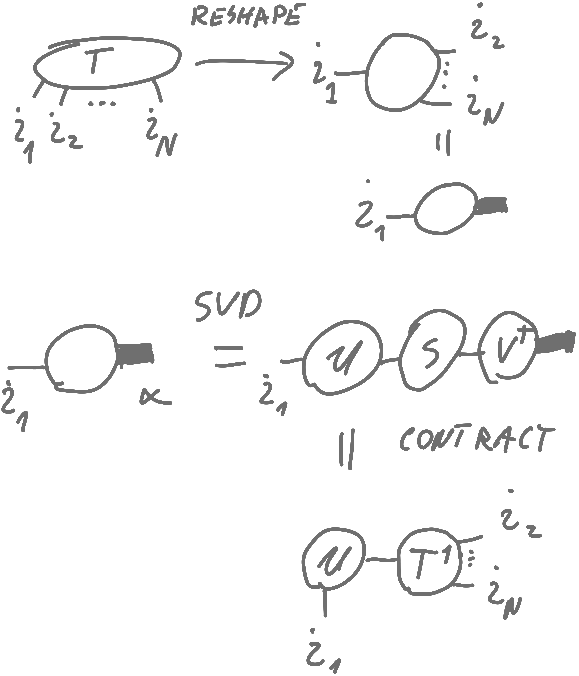}. 
$$
We already obtained the first tensor, namely $A^1=U$, from the above decomposition. To get the next tensor $A^2$, we perform similar transformations on $T^1$, i.e. a matrix obtained by multiplying $S$ and $V^\dag$ from the first SVD decomposition. First, we reshape the matrix $T^1$ by grouping the left index and the index $i_2$ in the left index of the reshaped matrix and indices $i_3,\ldots, i_N$ in the right index of the reshaped matrix. Then we perform the SVD, contraction of $S$ and $V$, and reshaping to finally get the tensors $A^2$ and $T^2$. We write these steps diagramatically as
$$
    \includegraphics[width=0.4\textwidth]{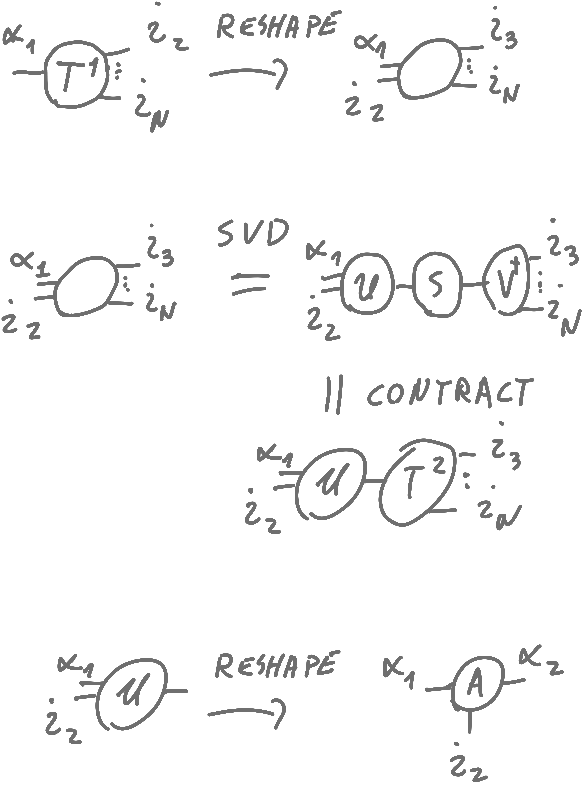}. 
$$
To get the tensor $A^3$, we perform the same steps as above but on the tensor $T^2$; similarly, for the remaining tensors up to the last. The last tensor $A^N$ is equal to $T^{N-1}$ since it has only one index on the right side and has therefore already the correct shape.  

By keeping all the singular values in the SVD decomposition, we arrive at an exact matrix product state representation of the original tensor. However, if we keep only the first $D$ singular values in each SVD decomposition, we obtain an approximation. These approximations are very common in many-body quantum systems, since they describe physically important quantum probability distributions. In particular, ground states of gapped one-dimensional Hamiltonians.
\end{example}
\end{shaded}

\paragraph{Tensor network contraction} Contracting a general tensor network is an NP-hard problem. The most straightforward argument to see this is to transform the NP-hard graph colouring problem into a tensor network contraction problem. 

\textit{The graph colouring problem}: Given a graph, find the number of different colourings with $d$ colours such that all connected vertices have different colours.

\begin{figure}[!htb]
    \centering
    \includegraphics[width=0.6\textwidth]{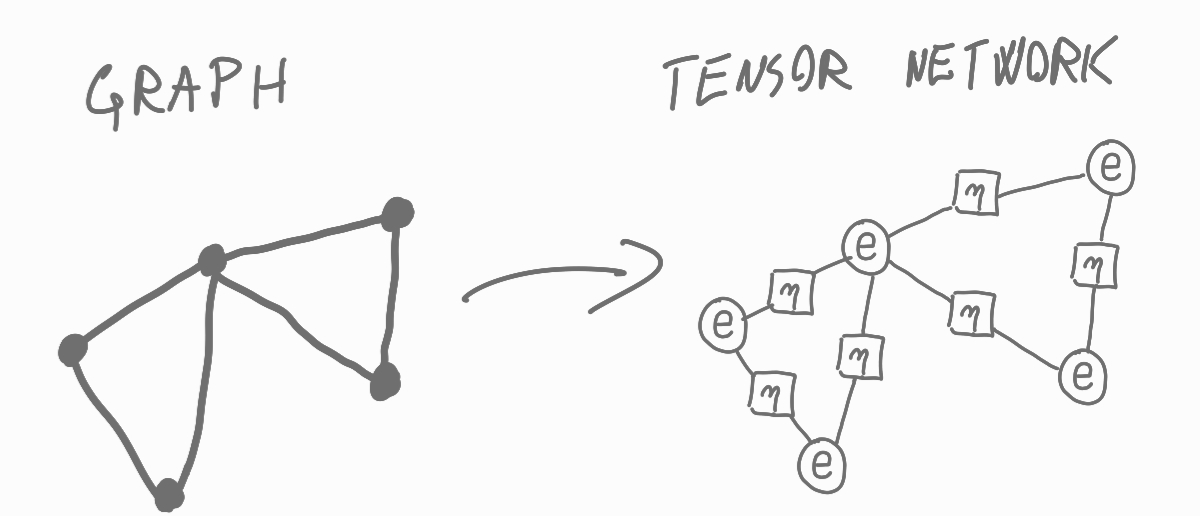}
    \caption{Example of the transformation from the graph to a tensor network contraction, which results in the number of different colourings with $d$ colours. The size of each contracted leg in the tensor network diagram is equal to the number of colours $d$.}
    \label{fig: coloring problem}
\end{figure}

We transform this problem into a tensor network contraction problem by introducing two types of tensors
\begin{align}
    e_{i_1,\ldots i_K}=&\begin{cases}1&i_1=i_2=\ldots i_K\\0& \mathrm{else}\end{cases} \\ \nonumber
    \eta_{i,j}=&\begin{cases}1&i\neq j\\ 0& i=j\end{cases},
\end{align}
where $i,j=1,\ldots d$.
To each vertex, we attach the tensor $e$, and to each edge, we attach the tensor $\eta$. The tensor $e$ ensures that the contraction of the obtained tensor network will have non-vanishing contributions only if all outgoing indices have the same value (corresponding to the colour). On the other hand, the tensor $\eta$ ensures that all the connected edges will have a different colour. Hence, the contraction of the obtained network gives exactly the number of allowed colourings with $d$ colours. For an example see \fref{fig: coloring problem}

\begin{shaded}
\begin{example}
\label{ex:mps contractions}
As we have seen, the tensor network contraction problem is NP-hard. However, particular tensor networks have efficient contractions. In this example, we will look at one inefficient and two efficient contractions of a matrix product state norm given by the tensor network diagram 
\begin{align}
    \includegraphics[width=0.13\textwidth]{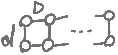}.
    \label{eq:mps norm}
\end{align}

We get an inefficient contraction of the tensor network \eref{eq:mps norm} if we first contract the inner bonds of the matrix product states and obtain an exponentially large tensor in the number of legs $N$. The final contraction is then obtained by calculating the 2-norm of that tensor (see \eref{eq:mps norm inefficient}). 
\begin{align}
    \includegraphics[width=0.4\textwidth]{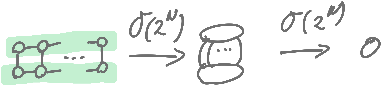}.
    \label{eq:mps norm inefficient}
\end{align}

The first efficient contraction can be done in parallel. First, we calculate the contractions corresponding to the same tensors in the MPS (denoted by green in the first step in \eref{eq:mps norm efficient 1}) to obtain $N$ matrices of size $D^2\times D^2$. The parallel complexity of this step is $\mathcal{O}(D^4d)$ since all the contractions can be done in parallel. In the successive steps, we multiply together pairs of neighbouring matrices $\log_2(N)$ times. In each step, all the calculations can be done in parallel with complexity $\mathcal{O}(D^6)$. The total parallel complexity of this algorithm is hence $\mathcal{O}(\log_2(N)D^6+D^4d)$.
\begin{align}
    \includegraphics[width=0.5\textwidth]{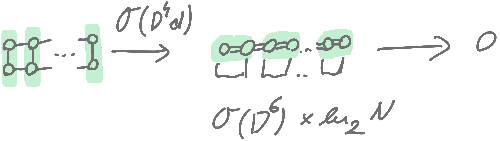}.
    \label{eq:mps norm efficient 1}
\end{align}

The final, most efficient, sequential algorithm is given by the steps shown in \eref{eq:mps norm efficient 2}
\begin{align}
    \includegraphics[width=0.5\textwidth]{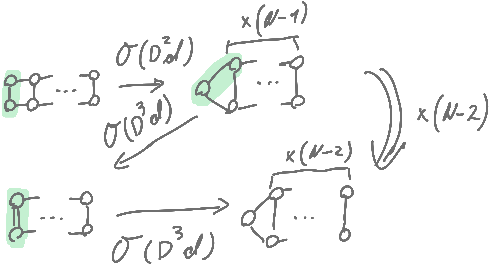}.
    \label{eq:mps norm efficient 2}
\end{align}
We contract from left to right. In the initial step, performed only once, we contract the leftmost tensors -- $\mathcal{O}(dD^2)$. Then we perform the following two contractions $N-2$ times. First, we contract the leftmost tensor with the top remaining matrix product state tensor -- $\mathcal{O}(dD^3)$. Then we contract the obtained tensor with the leftmost tensor of the bottom MPS tensor -- $\mathcal{O}(D^3d)$. The final two contractions are similar to the main $N-2$ contractions, with the difference that the remaining MPS tensors do not have the right bond index. The total complexity of this algorithm is hence $\mathcal{O}(NdD^3)$.
\end{example}
\end{shaded}

\begin{figure}[!htb]
    \centering
    \includegraphics[width=0.6\textwidth]{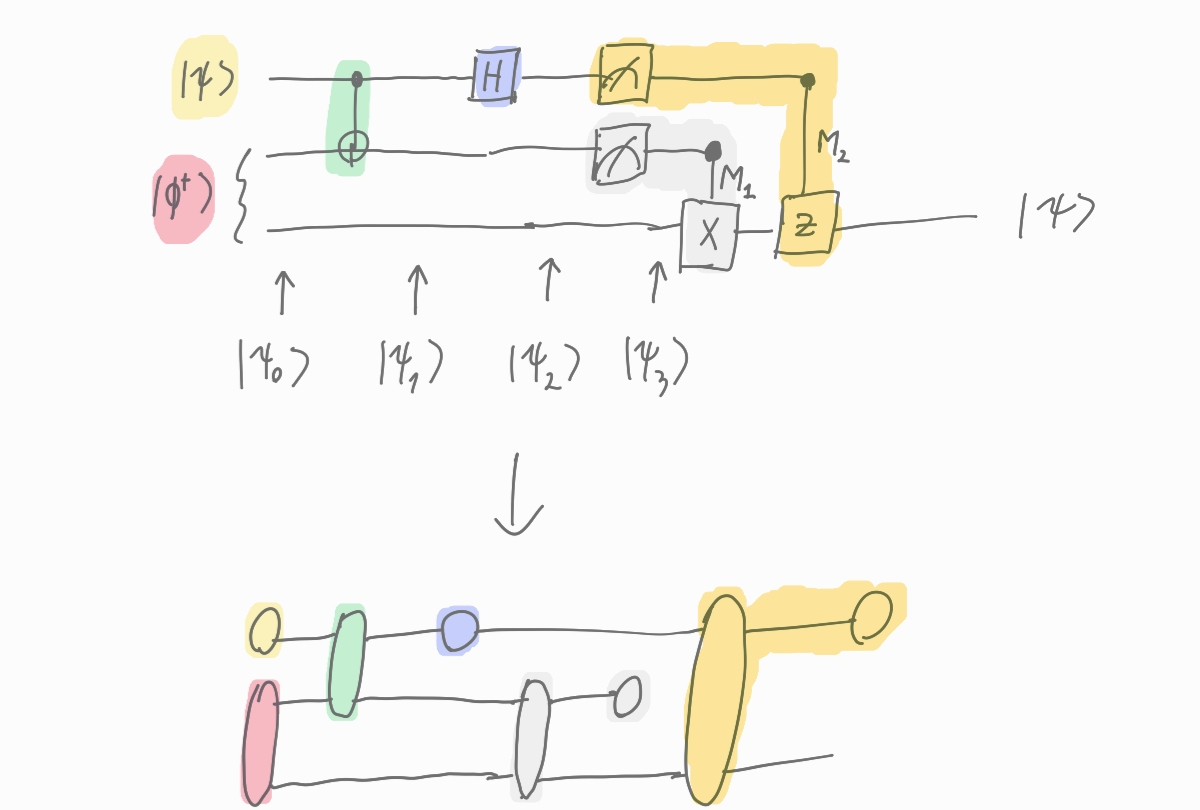}
    \caption{Interpreting the teleportation circuit as a tensor network contraction. The initial state is a tensor product of a vector with a 2-dimensional tensor. The following operations act on one or two qubits. The corresponding tensors are two and four-dimensional. The measurement is simulated by a contraction with an appropriate state.}
    \label{fig: teleportation tn}
\end{figure}

\paragraph{Tensor network contractions and quantum computation} The circuit model of quantum computation can be interpreted as a contraction of a tensor network with appropriate unitary constraints. A teleportation example is shown in \fref{fig: teleportation tn}. Hence, efficient classical simulation of quantum circuits is intimately related to tensor network contractions. In order to determine if a given quantum circuit is hard to calculate classically, we have to find its best contraction scheme, which is a hard problem in general.


\paragraph{Tensor networks in machine learning}
Tensor networks have been used as a special type of multi-linear model similar to variational quantum circuits and as a compact representation of weight matrices in deep neural networks. They have been applied to many supervised, unsupervised, and reinforcement learning problems. Most applications of tensor networks to machine learning problems are proof of principle. So far, the only exceptions are the anomaly detection problem discussed in \exref{ex: anomaly detection} and its extension to positive unlabelled learning.

\begin{figure}[!htb]
    \centering
    \includegraphics[width=0.6\textwidth]{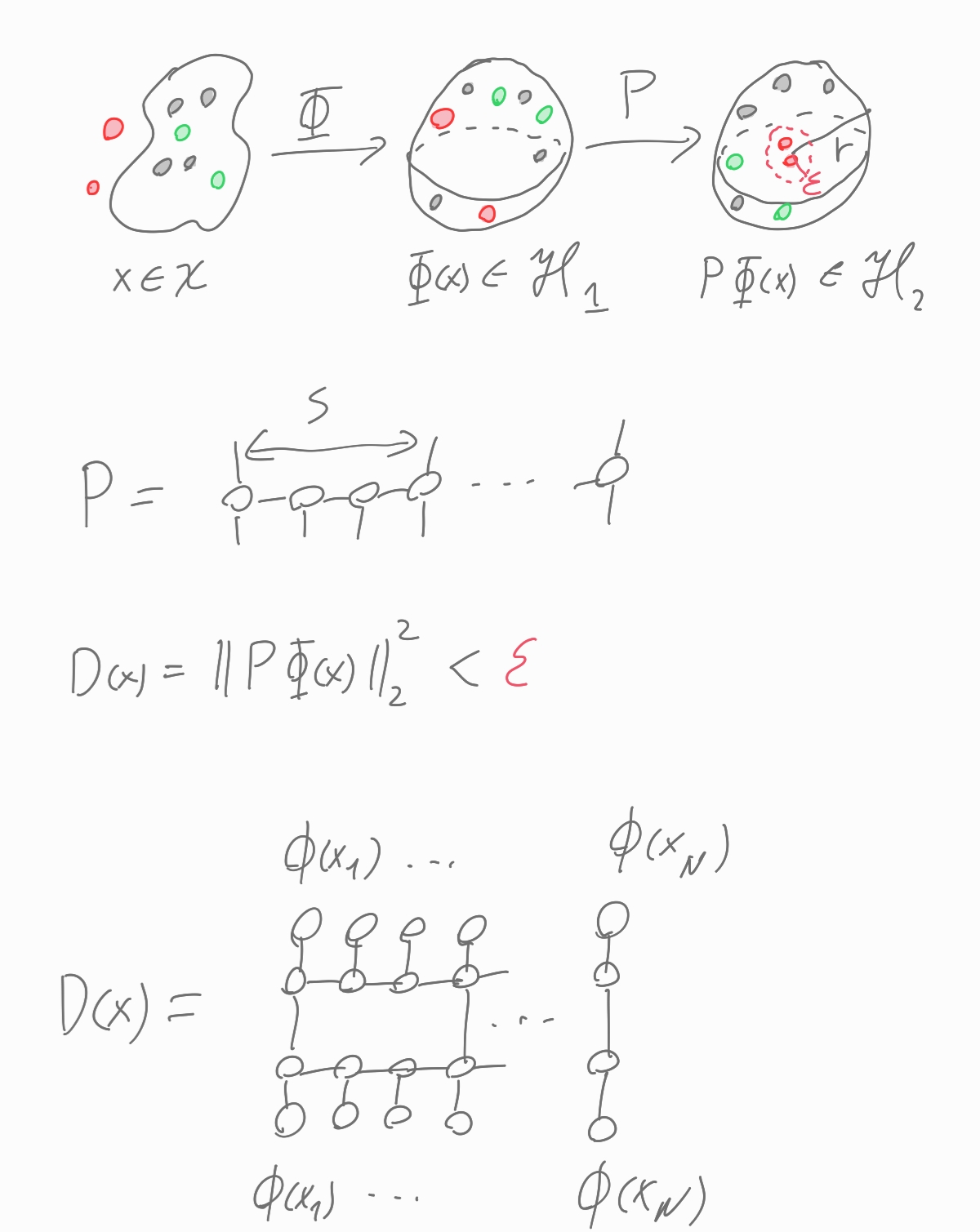}
    \caption{The figure shows the idea of anomaly detection with tensor networks. An example is first mapped to a Hilbert $\mathcal{H}_1$ space and then projected to the Hilbert space $\mathcal{H}_2$. If the norm of the projected vector is smaller than $\epsilon$, we have an anomaly (red circles). The projector $P$ has a tensor network structure with a large kernel. The norm of the final projected vector can be evaluated efficiently.}
    \label{fig: anomaly detection}
\end{figure}

\begin{shaded}
\begin{example}
\label{ex: anomaly detection} The anomaly detection problem is formulated as follows. Given a dataset, the task is to determine if a new, test example is from the same distribution as the training examples. The difference with respect to the standard classification problem is that we do not have out-of-distribution examples for training. 

The tensor network solution to this problem is summarised in \fref{fig: anomaly detection}. 
\begin{enumerate}
    \item First, we embed the examples in a high-dimensional Hilbert space $\mathcal{H_1}$ by using a data encoding procedure discussed in Section \ref{sec: embeddings}, $\Phi(x)\in\mathcal{H}_1$. 
    \item Then we project the encoded example into a new Hilbert space $P\Phi(x)\in H_2$. If the norm of the projected vector is smaller than $\epsilon$, we have an anomaly.
    \item The projector is chosen such that its kernel is large. For the choice in \fref{fig: anomaly detection}, the kernel has size $d^{N-\lfloor N/S\rfloor}$, where $d$ is the embedding dimension, $N$ the number of examples, and $S$ determines the size of the Hilbert space $\mathcal{H}_2$.
    \item We can use standard gradiend descend based training with the loss $\mathcal{L}=\frac{1}{M}\sum{i=1}^M||\log(D(x))-1||_2+\alpha \log(||P||_F)$. The loss has two contributions. The first decreases if the in-distribution training examples have a norm close to $\sqrt{e}$ after the projection. The second decreases if the kernel of the projector is large. The balance between the two contributions is controlled by a hyperparameter $\alpha>0$.
    \item Finally, we have to ensure that the calculation of the projected norm is efficient. This is clear since after the contractions of the embeddings with the projectors, we obtain a standard matrix product state norm calculation (see \exref{ex:mps contractions}).
\end{enumerate}
\end{example}
\end{shaded}

\begin{referencesbox}
    \begin{itemize}
        \item Fernández, Y. N., Ritter, M. K., Jeannin, M., Li, J. W., Kloss, T., Louvet, T., ... \& Waintal, X. (2024). Learning tensor networks with tensor cross interpolation: new algorithms and libraries. arXiv preprint arXiv:2407.02454.
        \item https://github.com/google/TensorNetwork (With few tutorials, references and video lectures)
        \item https://www.math3ma.com/blog/matrices-as-tensor-network-diagrams
        \item https://arxiv.org/abs/1306.2164
        \item https://arxiv.org/abs/1708.00006
        \item https://arxiv.org/abs/1603.03039
        \item https://doi.org/10.1137/090752286
        \item Wang, J., Roberts, C., Vidal, G., \& Leichenauer, S. (2020). Anomaly detection with tensor networks. arXiv preprint arXiv:2006.02516.
    \end{itemize}
\end{referencesbox}


\subsection{Dequantisation}
Quantum computation consists of three steps: 1) data encoding 2) quantum algorithm 3) measurement (readout of the result). When calculating the runtime of quantum algorithms, we implicitly assume that the data encoding step and the measurement step can be performed efficiently. However, this is a very strong assumption. In fact, only data with a special structure can be encoded exponentially fast, which is typically required to get the exponential speedup in many quantum algorithms. In particular, the algorithms relying on qBLAS. To complicate matters even further, the data encoding complexity highly depends on the exact architecture of the quantum device. Therefore, it seems reasonable to abstract away the data encoding and readout procedures when comparing quantum and classical computation. In quantum algorithms, we already assume efficient state preparation. Hence, we have to equip classical algorithms with an equally strong assumption and then check if we can improve their runtime. 

The classical equivalent to the efficient quantum state preparation assumption is the sample and query access assumption.

\begin{figure}[!htb]
    \centering
    \includegraphics[width=0.6\textwidth]{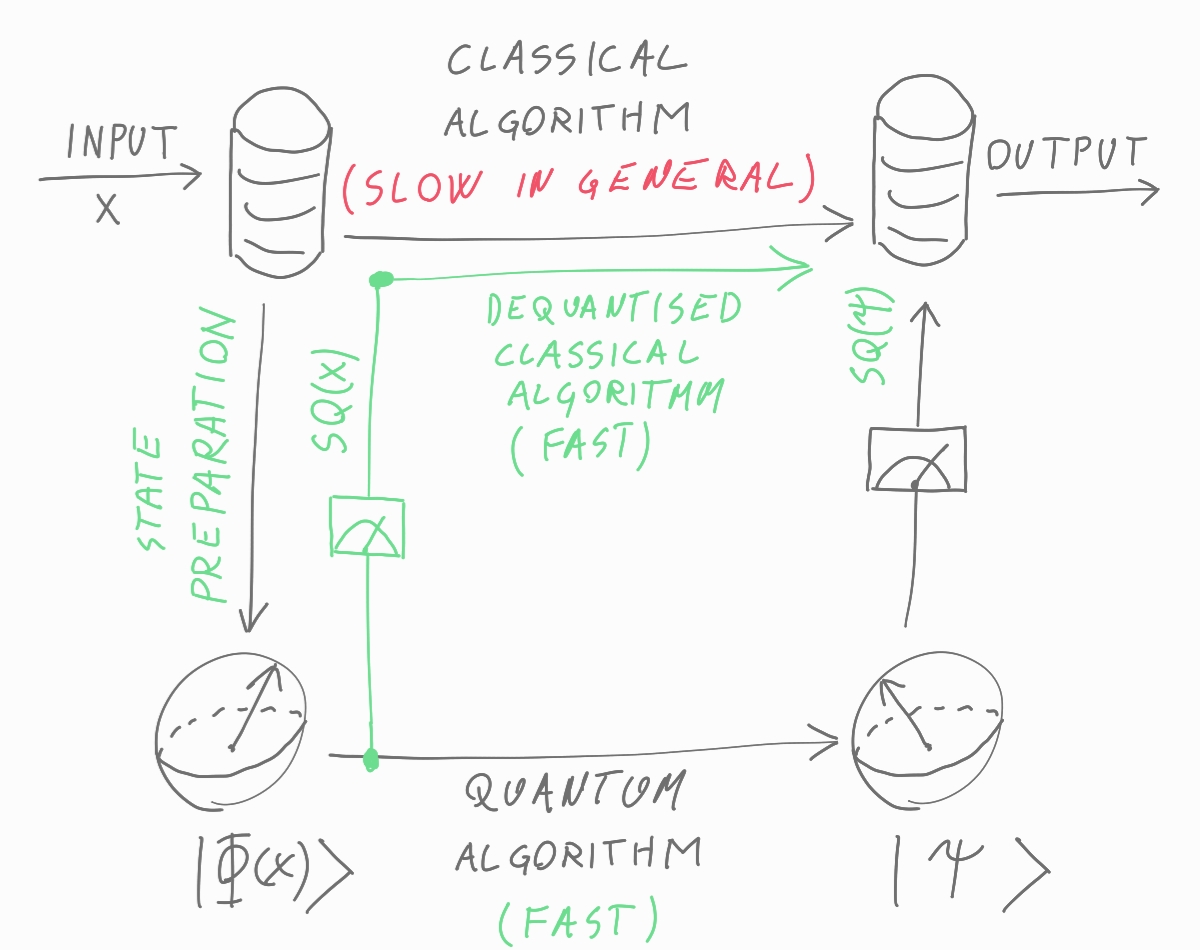}
    \caption{Intuitive representation of the dequantisation procedure and sample and query access assumption. We dequantise the quantum algorithm (marked with blue) if we find a polynomially equivalent classical algorithm having access to the initial quantum state (marked with green). This procedure is interesting only if a classical algorithm without the $SQ(x)$ assumption is slow. Slow in this context means exponentially slower in the size of the input compared to the quantum algorithm.}
    \label{fig: dequantisation}
\end{figure}

\paragraph{Sample and query access} We have a sample and query access to $x\in\mathcal{C}^N$, denoted by $SQ(x)$, iff we can query an index $i\in 1,2,\ldots N$ for its value $x_i$ and quary for $||x||_2$. With $SQ^\nu(x)$ we denote $SQ(x)$ with an access to an approximate norm $\bar{x}=(1\pm\nu)||x||_2$. 
We get an intuitive understanding of the sample and query access assumption as follows. We want a fair comparison between the quantum and classical algorithms. Suppose we assume an efficient quantum state preparation procedure for the quantum algorithm. In that case, we have to give the classical algorithm access to measurements of the initial state in the computational basis (see \fref{fig: dequantisation}). In other words, we start with a quantum state that we can access by measurements, but can only perform classical processing. In this way, the quantum and classical algorithms are given the same resources. Since the norm of the quantum state is always one, and we can determine efficiently (in time $\mathcal{O}(T)$, for some fixed $T$) the elements $x_i$, the efficient state preparation procedure implements the $SQ(x)$.

\paragraph{Dequantisation} We say we dequantize a quantum protocol $\mathcal{S}: \ket{\phi_1}\ket{\phi_2}\ldots\ket{\phi_N}\rightarrow \ket{\psi}$ if we find a classical algorithm of the form $SQ(\phi_1,\phi_2,\ldots,\phi_N)\rightarrow SQ(\psi)$. In other words, given a sample and query access to the inputs, we have to efficiently prepare a sample and query access for the output of the desired calculation (see \fref{fig: dequantisation}). The fact that Born-type measurements significantly speed up several machine learning protocols and randomised linear lagebra is a well known fact in numerical methods literature, however, a connection to quantum machine learning has been recognised only recently.
The dequantised algorithms provide a clean framework for the abstraction of the preparation procedures and strong bounds on the exponential speedup of the quantum algorithms. Moreover, this framework can be a pillar of new types of classical-quantum algorithms, where the quantum part provides the sample and query access to the input of the classical algorithm. 

\begin{shaded}
\begin{example} \textbf{Nearest centroid classification}
The nearest centroid classification assigns a class based on the distance to the centroids of the training examples. To do that, we need to calculate the scalar product of the training examples of each class and the test example. Classically, this is done in $\mathcal{O}(N)$ steps, where $N$ is the size of the input. In the following, we will first consider the quantum algorithm and then the dequantised classical algorithm with complexity $\mathcal{O}(T\frac{1}{\epsilon}\log\frac{1}{\delta})$.

The complexity of the nearest centroid algorithm is determined by the complexity of the calculation of the scalar product between two vectors $x,y\in\mathds{C}^N$, i.e. $x\cdot y$.

\textbf{Quantum algorithm} 
In the quantum case, we assume that we can efficiently prepare the states $\ket{x}$ and $\ket{y}$ corresponding to amplitude-encoded vectors x and y. The algorithm giving the overlap $\braket{x}{y}$ is 
\begin{align}
    \includegraphics[width=0.35\textwidth]{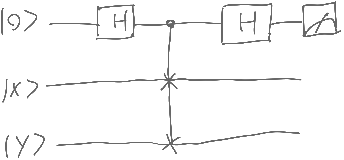}.
\end{align}
The probability value that the auxiliary qubit is in the state $\ket{0}$ is $P(0)=\frac{1}{2}(1+|\ketbra{x}{y}|^2)$. The overlap can be calculated to precision $\epsilon$ and with probability $1-\delta$ in time $\mathcal{O}(T\frac{1}{\epsilon^2}\log\frac{1}{\delta})$.

\textbf{Dequantised classical algorithm}
The classical algorithm calculating the scalar product $x\cdot y$ given the sample and query access to $x$ and query access to $y$ is given below
\begin{enumerate}
    \item $s=54\frac{1}{\epsilon^2}\log\frac{2}{\delta}$
    \item Collect measurements $i_1,\ldots i_s$ from $\ket{x}$ ; requires sample access
    \item Calculate $z_j=x_{i_j}y_{i_j}\frac{||x||_2^2}{|x_{i_j}|}$ ; requires query access to $x$, $||x||_2$, and $y$
    \item Separate $z_j$'s into $6\log \frac{2}{\delta}$ buckets of size $\frac{9}{\epsilon^2}$ and take the mean of each bucket.
    \item The output of the algorithm is the component-wise median of the means.
\end{enumerate}

\textit{Scetch of the proof} The number of samples $s$ gives the time complexity and is trivially $\mathcal{O}(\frac{T}{\epsilon^2}\log\frac{1}{\delta})$, where $T$ is the time necessary to determine $x$ in the querry access (similar to determining one expectation value in the quantum algorithm). We now have to bound the error and success probability.

First we note that $z_j$ is a random variable with mean $x\cdot y$ and bounded variance $\mathrm{var}(z_j)\leq ||x||_2^2||y||_2^2$. The variance of the mean of the copies in one bucket is hence bounded by $\frac{\epsilon^2}{9}||x||_2^2||y||_2^2$. Using the Chebishev inequality, we observe that the probability that the mean is $\frac{\epsilon}{\sqrt{2}}||x||_2||x||_2$ away from the desired result is bounded from above by $\frac{2}{9}$. Finally, we use the Chernoff-Hoeffding inequality to show that the probability that the median of the means of the buckets is with probability $1-\delta$ inside the desired result $x\cdot y\pm \epsilon||x||_2||y||_2$.  

The quantum case can be sped up quadratically by using the amplitude amplification algorithm with the final runtime complexity $\mathcal{O}(\frac{T}{\epsilon}\log\frac{1}{\delta})$. Nevertheless, the discussed dequantised classical algorithm is polynomially equivalent to the best quantum algorithm. 
\end{example}
\end{shaded}
\begin{referencesbox}
    \begin{itemize}
        \item Tang, E. (2023). Quantum machine learning without any quantum. University of Washington.
    \end{itemize}
\end{referencesbox}

\end{document}